\documentclass[11pt]{article}
\usepackage{amsmath,amsfonts,amsbsy,amssymb,array,accents,dsfont}
\usepackage{enumerate,array,latexsym,graphicx,mathrsfs,verbatim,psfrag}
\usepackage{bm} %math bold symbols
\usepackage[normalem]{ulem}%strikethrough
\usepackage{booktabs} %some alignment stuff
\usepackage[usenames]{color}
\usepackage[utf8]{inputenc}
\usepackage[english]{babel}
\usepackage{fancybox}

\usepackage{datetime}

\usepackage[nosort]{cite}
\usepackage{chngpage} % allows for temporary adjustment of side margins

\usepackage{enumitem}
\setlist{itemsep=0pt}

\usepackage[colorlinks=true,      linkcolor=darkblue,      urlcolor=darkblue,      
            filecolor=darkblue,      citecolor=darkblue,       pdfstartview=FitH,     
						pdfpagemode=UseNone,      bookmarksopen=true]{hyperref}  % LINKS
\usepackage[all]{hypcap}     %should be loaded AFTER hyperref, which should otherwise be loaded last.

\newcommand{\captionfonts}{\small}
\makeatletter  % Allow the use of @ in command names
\long\def\@makecaption#1#2{%
  \vskip\abovecaptionskip
  \sbox\@tempboxa{{\captionfonts #1: #2}}%
 \ifdim \wd\@tempboxa >\hsize
    {\captionfonts #1: #2\par}
  \else
    \hbox to\hsize{\hfil\box\@tempboxa\hfil}%
  \fi
  \vskip\belowcaptionskip}
\makeatother   % Cancel the effect of \makeatletter

\DeclareMathSymbol{\medhatsym}{\mathord}{largesymbols}{"62} % basic symbol

\DeclareMathSymbol{\medtildesym}{\mathord}{largesymbols}{"65}% basic symbol

\newcommand{\comm}[1]{} %for commenting out blocks of text

\def\IC{\mathbb{C}}

\def\IR{\mathbb{R}}

\def\IT{\mathbb{T}}

\def\({\left(}
\def\){\right)}
\def\[{\left[}
\def\]{\right]}

\def\half{\frac12}
\def\coeff#1#2{{\textstyle \frac{#1}{#2}}}

\def\One{{\hbox{ 1\kern-.8mm l}}}

\def\barray{\begin{array}}
\def\earray{\end{array}}
\def\be{\begin{equation}}
\def\ee{\end{equation}}
\def\bea{\begin{eqnarray}}
\def\eea{\end{eqnarray}}
\def\bal{\begin{align}}
\def\eal{\end{align}}

\numberwithin{equation}{section} % replaces the hack below

\makeatletter
\g@addto@macro\bfseries{\boldmath}
\makeatother

\definecolor{cardinal}{rgb}{0.6,0,0}
\definecolor{darkgreen}{rgb}{0,0.4,0}
\definecolor{purple}{rgb}{0.5, 0, 0.5}
\definecolor{golden}{rgb}{0.92, 0.7, 0}
\definecolor{midnight}{rgb}{0, 0, 0.5}
\definecolor{darkblue}{rgb}{0, 0, 0.8}

\def\flux{\Pi}
\def\oneone{\rlap 1\mkern4mu{\rm l}}
\def\IC{\mathbb{C}}
\def\Neql#1{{\cal N}\!=\!{#1}}
\def\coeff#1#2{\relax{\textstyle {#1 \over #2}}\displaystyle}

\def\IR{\mathds{R}}
\def\ZZ{\mathds{Z}}
\def\flux{\Pi}
\def\cA{{\cal A}}
\def\cB{{\cal B}}

\def\cI{{\cal I}}

\def\cL{{\cal L}}
\def\cM{{\cal M}}
\def\cN{{\cal N}}
\def\cO{{\cal O}}

\def\cQ{{\cal Q}}

\def\cS{{\cal S}}

\def\cX{{\cal X}}

\def\nBPS#1{$\frac{1}{#1}$-BPS}

\def\th{$^{\rm th}$\ }
\def\combox#1{\bigskip \framebox{\parbox{5.8 in}{{\bf \underline{Comment}:\ \ }{\it #1}}}\bigskip }
\def\exbox#1{\bigskip \framebox{\parbox{6 in}{{\bf \underline{Exercise}:\ \ }{\it #1}}}\bigskip }
\def\exboxx#1{\bigskip \framebox{\parbox{4 in}{{\bf \underline{Exercise}:\ \ }{\it #1}}}\bigskip }
\begin{document}

%\begin{titlepage}

\begin{flushright}
%
% PHT-T17/134\\
%
\end{flushright}

\vspace{14mm}

\begin{center}

{\huge \bf{Lectures on Microstate Geometries}} \medskip \\

\vspace{13mm}

%\bigskip\bigskip
\centerline{{\bf Nicholas P. Warner$^{}$}}
\bigskip
\bigskip
\vspace{1mm}

\centerline{Institut de Physique Th\'eorique,}
\centerline{Universit\'e Paris Saclay, CEA, CNRS, }
\centerline{Orme des Merisiers,  F-91191 Gif sur Yvette, France}
\bigskip
\centerline{Department of Physics and Astronomy,}
\centerline{University of Southern California,} \centerline{Los
Angeles, CA 90089-0484, USA}
\bigskip
\centerline{Department of Mathematics,}
\centerline{University of Southern California,} \centerline{Los
Angeles, CA 90089, USA}

\vspace{4mm}

%
%{\small\upshape\ttfamily  ~ejmartin @ uchicago.edu,  walkerra @ usc.edu, ~warner @ usc.edu} \\

\vspace{15mm}
%\bigskip\bigskip\bigskip
 
\textsc{Abstract}
\vspace{8mm}
\begin{adjustwidth}{10mm}{10mm} % to adjust the L and R margins
 %
%\begin{abstract}
These are notes for some introductory lectures about microstate geometries and their construction.  The first lecture considers BPS black holes in four dimensions as a way to introduce what one should expect from the BPS equations.  The second lecture discusses the  ``no solitons without topology'' theorem.  The subsequent lectures  move on to the construction and properties of bubbled microstate geometries in five dimensions. Since these are graduate lectures, they involve a strongly computational 
narrative intended to build some proficiency and understanding of supersymmetric solitons in five-dimensions. The narrative also regularly ``plays tourist''  pointing to specific features that are more general or have broader impact.  The last sections contain  brief comments on the larger setting of microstate geometries and describe some of the more recent developments.
These lectures were given as a mini-course at the IPhT, Saclay in May, 2019. 

\vspace{3mm}
\noindent
%
 
%
%\end{abstract}
\end{adjustwidth}

\end{center}

%\end{titlepage}

\thispagestyle{empty}

\newpage

%%%%%%%%%%%%%%%%%%%%%%%%%%%%%%%%%%%%%

\baselineskip=14pt
\parskip=2pt

\tableofcontents

% \newpage

\baselineskip=15pt
\parskip=3pt

%%%%%%%%%%%%%%%%%%%%%%%%%%%%%%%%%%%%%
\section{Introduction}
\label{Sect:introduction}
%%%%%%%%%%%%%%%%%%%%%%%%%%%%%%%%%%%%%

 Solitons are loosely defined to be smooth, stable, localized, finite-energy solutions of some particular system of equations. They have played a remarkably rich role in physics and typically arise in non-linear systems when there is a stable balance between an attractive, cohesive force  and some form of dispersive phenomenon.  The most celebrated, early recording of this phenomenon was  John Russell's horseback pursuit of a solitary wave along two miles of the Union Canal.  This was followed by the modeling such phenomena via the Korteweg-de Vries (KdV) equation.  The 20\th century led to the deeper understanding, and solution of, a number of non-linear, solitonic systems using integrable hierarchies and inverse scattering methods.   
 
In physics, there are many beautiful examples of the role of solitons in the semi-classical description of phase transitions.  In one phase there might be natural, low-energy perturbative excitations that give a fundamental description of the system, and there may be some high energy, non-linear solitonic excitations.   As a coupling constant, or the energy changes, a new phase can emerge in which the solitons  become light while the original perturbative degrees of freedom becomes massive.  The fundamental  description of the system must change dramatically in the new phase because it is usually simpler to describe the system in terms of its light degrees of freedom.    In the Ising model at low temperature, the light fields are the localized spins while the solitons are ``massive'' domain walls between domains of different spin.  At the critical temperature, the domain walls become light and the entire system has a simple description in terms of a massless Majorana fermion.  In supersymmetric Yang-Mills theories,  the standard phase is described in terms of gluons and quarks.  However, at strong coupling the light degrees of freedom can be monopoles or dyons.

In string theory, the vast array of dualities between apparently very different string/membrane theories (IIA, IIB, heterotic, M-theory ... ) may be viewed in terms of precisely such phase transitions in which solitonic degrees of freedom in one theory become light and thus provide a better effective description of the theory in terms of a ``dual'' form of string/M-theory.  Such  phase transitions are often driven by changes of topology.  For example, the solitons might be branes wrapping cycles, and if the cycle collapses, the corresponding soliton becomes massless.  In this way, the W-bosons of the heterotic string emerge from M2 branes wrapping 2-cycles in M-theory. 

Since solitons are generically smooth configurations of fields, while particles are typically singular ($\delta$-function) sources, the discovery of solitons led to the idea that perhaps they could lead to better descriptions of fundamental particles in Nature.  The classically infinite self-energy of the electron was unsatisfactory and the hope was that, at some scale, it would resolve into a smooth solitonic ``lump.''   There was also a desire that electromagnetism should be ``unitary,'' not in the sense of quantum mechanics, but meaning that electromagnetism should be self-consistent and complete in itself:  Rather than having the electron as a separate fundamental particle, the hope was to describe it as a lump made out of electromagnetic fields.  This required a ``non-linear'' electromagnetism and this was one of the driving forces behind the development of Born-Infeld theory.  While this idea is no longer a serious contender in describing the electron, Born-Infeld theories have had a renaissance in the description of field theories on branes in string theory.

With the advent of General Relativity, and particularly because of the non-linearities of the equations of motion, there was deep interest in finding solitonic solutions and seeing whether such solitons could play a role in describing the end-states of stars, or other collapsed objects.   Unfortunately, it  became clear by the 1960's, and particularly through the work of Lichnerowicz, that there are no smooth solitonic solutions to General Relativity in $(3+1)$-dimensions.    Indeed, the singularity theorems of Penrose and Hawking showed that singularities were, in fact, an essential part of Nature, as described by General Relativity.  The scientific community thus began to embrace singularities, so long as they are safely inside horizons. Indeed, the triumphs of LIGO in the last few years have confirmed the accuracy of General Relativity in  extreme, non-linear regimes and  black hole and their horizons appear to be an essential part of Nature.

The study of solitons in General Relativity continued, but was re-shaped by the singularity theorems.  The idea was  to weaken the definition of solitons and  look for stationary (time-independent) lumps that can have singularities so long as the  singularities are safely inside horizons. Indeed, theorems were proved to show that this was the only option and the   mantra ``No Solitons without Horizons'' became lore in General Relativity, even when coupled to all sorts of different matter systems.   

Ideally, a soliton in General Relativity should be not only classically stable, but stable within quantum mechanics.  Hawking's discovery of black-hole evaporation meant that solitons must have vanishing Hawking temperature and thus be extremal.  Indeed, to ensure the stability, the solitons must be BPS, that is, they must be the lowest energy state in the sector of a theory with a given set of charges.  This implies that there is no other state into which they can decay.  Therefore, the first step was to find BPS solitons, which, as we will discuss, are typically supersymmetric.  Once one finds this class of solitons, one can try to be more adventurous and find solitons that are, perhaps, classically stable, or meta-stable, but decay by tunneling, or perhaps gravitational radiation.  The BPS solitons thus become important islands of stability from which one can explore more physical, time dependent ``lumps and their quantum descendants.''

There are indeed, simple, solitonic black-hole solutions in Einstein-Maxwell gravity, and this is where we will start this course. They are invaluable for illustrating the structure of BPS solutions and for highlighting the Faustian Bargain of supersymmetry.

So, in the face of the triumph of General Relativity in describing everything seen by LIGO, why should we be concerned about General Relativity and its description of black holes?  Why should we revisit gravitational solitons? The answer is ``The Black-Hole Information Problem.''   

Because Hawking radiation originates from vacuum polarization  just above the horizon, the uniqueness of black holes in General Relativity implies that Hawking radiation is universal, thermal and (almost) featureless. In particular, it is independent of how, and from what, the black hole formed. Semi-classical back-reaction of this Hawking radiation also implies that the black hole will evaporate, albeit extremely slowly.   It is, therefore, impossible to reconstruct the interior state of a black hole (apart from mass, charge and angular momentum) from the exterior data, and thus from the final state of the Hawking radiation.  The evaporation process cannot, therefore, be represented through a unitary transformation of states in a Hilbert space.  Hence black-hole evaporation, as predicted by General Relativity and Quantum Mechanics, is inconsistent with a foundational postulate of Quantum Mechanics.  

Based on its horizon area, the black hole at the core of the Milky Way should have about $e^{10^{90}}$  microstates.  From the outside, black-hole uniqueness implies that its state is unique, as would be the state of its Hawking radiation were it to evaporate.  The problem is therefore vast: $e^{10^{90}} \ne 1$!

However, the evaporation time-scale of a Schwarzschild black hole is also vast:   a one solar mass black hole evaporates into a zero-temperature reservoir in about $10^{67}$ years. Moreover the evaporation time is  proportional to $M^3$.  For the black hole at the core of the Milky Way, this evaporation time is about $10^{87}$ years.  This led to the idea that, over 
 such vast time-scales, the microstate data of a black hole  should  be able to dribble out with only tiny (quantum?) corrections to General Relativity.  In 2009, Mathur used strong sub-additivity of quantum information to show that this idea was wrong and that solving the Information Problem would require corrections of order $1$ to the physics at the horizon scale.
 
Because of the information problem, there is  a growing consensus that new physics must emerge at the horizon scale of a black hole.  The challenge is to find a classical mechanism to support such horizon-scale structure, and if one can do this, then one can  investigate how black-hole microstructure can be encoded in the horizon-scale structure.  We thus come back to the issue of gravitational solitons.    To support horizon-scale microstructure we need {\it smooth, horizonless} geometries that closely approximate the geometries of black holes up where the horizon would form.   This is the definition of a {\it Microstate Geometry}.   Instead of having a horizon, such geometries cap-off smoothly at very high red-shift.  The absence of horizon is not only essential to solving the information problem but also to smoothness:  singularity theorems tell us that singularities must form inside horizons.  

This would be a very short course if Microstate Geometries did not exist.  Examples of such geometries were first constructed almost 15 years ago (largely because the creators of the geometries were ignorant of the many theorems that forbade their existence).  It is thus extremely interesting to chart how the Microstate Geometries dodged the ``No Go'' theorems.  The answer is a combination of higher-dimensional gravity theories, topology and Chern-Simons interactions.  We  discuss this in the second lecture. 

Subsequent lectures will involve the construction of Microstate Geometries and an investigation of their properties.  As I indicated above, the explicit examples that we will discuss here will be supersymmetric/BPS.  This part of the story is now very well-developed.  The systematic construction of generic, non-supersymmetric Microstate Geometries remains a major unsolved problem: we have sporadic, very exotic examples, but, as yet, we do not have Microstate Geometries that match the properties of astrophysical black holes.  This is a problem we hope to solve in the not too-distant future. 

%%%%%%%%%%%%%%%%%%%%%%%%
\subsection{Background}
\label{ss:Background}
%%%%%%%%%%%%%%%%%%%%%%%%

In this course I will assume that you are familiar with
\begin{itemize}
\item The basics of general relativity:  that you can work with metrics, compute connections, curvatures and geodesics and know about isometries and Killing vectors.  I will assume you know action principles and how to compute energy-momentum tensors and obtain Einstein's equations.
\item Differential forms:  You should be familiar with differential forms, exterior derivatives  and  Hodge duals.  If not, it would be important to review the relevant section of any source like \cite{Nakahara:2003nw} Sections 5.4 and 7.9  before Lecture 2.  
\item Lorentz spinors in four dimensions.  While we will use Lorentz spinors in a generic space-time and in both four and five dimensions, it will not be a major part of the course and you will be fine if you have only seen the Dirac equation in Minkowski space.  
\end{itemize}
%

%%%%%%%%%%%%%%%%%%%%%%%%%%%%%%%%%%%%%
\section{BPS Black Holes in Four Dimensions}
\label{Sect:4DBPS}
%%%%%%%%%%%%%%%%%%%%%%%%%%%%%%%%%%%%%

%%%%%%%%%%%%%%%%%%%%%%%%
\subsection{The Reissner-Nordstr\"om black hole}
\label{ss:RNBH}
%%%%%%%%%%%%%%%%%%%%%%%%

The most general, spherically symmetric, time-independent metric\footnote{I do not need to assume time-independence to arrive at the Reissner-Nordstr\"om black hole. One can prove a generalization of Birkhoff's theorem that shows that spherical symmetry implies time-independence.  I am taking a short-cut here to simplify the pedagogy.}   has the form
\begin{align}
ds^2&  ~=~ -  e^{2\, \alpha(r)} \, dt^2  ~+~  e^{2\, \beta(r)} \, dr^2 ~+~  e^{2\, \gamma(r)} \, dr \,dt   ~+~  e^{2\, \delta(r)} \, \big(d\theta^2 + \sin^2 \theta\,d \phi^2)\\
 & ~\to~  -  e^{2\, \alpha(r)} \, dt^2  ~+~  e^{2\, \beta(r)} \, dr^2 ~+~  e^{2\, \gamma(r)} \, dr \,dt   ~+~  r^2 \, \big(d\theta^2 + \sin^2 \theta\,d \phi^2) \,.
\end{align} 
where the  possible re-definition $r \to \tilde r(r)$ has been fixed by setting the  areas of the concentric spheres to $4 \pi r^2$. 

One is also free to define $t =  \tilde t + f(r)$ and so  $dt^2  = (d \tilde t^2 + 2 f'(r) dr d\tilde t + f'(r)^2 dr^2)$.   This can be used to gauge away the $e^{2\, \gamma(r)} \, dr \,dt$ term, and so, without loss of generality we may take:
\begin{equation}
ds^2 ~=~ -  e^{2\, \alpha(r)} \, dt^2  ~+~  e^{2\, \beta(r)} \, dr^2  ~+~  r^2\, \big(d\theta^2 + \sin^2 \theta\,d \phi^2) \,.
\label{gen4met}
\end{equation} 

Choosing  $(x^0,x^1,x^2,x^3) = (t,r,\theta, \phi)$, the non-zero components of the affine connections are:
\begin{align}
\Gamma^{0}_{01}&   ~=~  \Gamma^{0}_{10}  ~=~ \alpha'(r) \,, \qquad \Gamma^{1}_{00} ~=~ e^{2\,( \alpha -\beta)} \, \alpha'(r)  \,, \qquad \Gamma^{1}_{11} ~=~ \beta'(r)  \,,  \nonumber \\ 
\Gamma^{1}_{22} & ~=~  -r \,e^{-2 \, \beta}  \,, \qquad   \Gamma^{1}_{33}  ~=~  -r \, \sin^2 \theta \, e^{-2 \, \beta} \,, \qquad \Gamma^{2}_{12} ~=~  \Gamma^{2}_{21}  ~=~   \frac{1}{r} \,, \nonumber  \\
 \Gamma^{2}_{33}  & ~=~  - \sin \theta \, \cos \theta \,, \qquad \Gamma^{3}_{13} ~=~  \Gamma^{3}_{31}  ~=~   \frac{1}{r} \,, \qquad \Gamma^{3}_{23}  ~=~  \Gamma^{3}_{32}  ~=~ \cot \theta  \,.
\end{align} 
The independent, non-zero components of the Riemann tensor are:
\begin{align}
{R^{0}}_{101} &   ~=~   \alpha'(r) \,  \beta'(r) -\alpha''(r)  -(\alpha'(r) )^2\,, \qquad {R^{0}}_{202}  ~=~-r \,  e^{-2\,\beta} \, \alpha'(r)  \,, \qquad {R^{0}}_{303}  ~=~  {R^{0}}_{303}  \,   \sin^2 \theta  \,,  \nonumber \\ 
{R^{1}}_{212} &   ~=~   r \,  e^{-2\,\beta} \, \beta'(r) \,, \qquad {R^{1}}_{313}  ~=~{R^{1}}_{212}  \,   \sin^2 \theta    \,, \qquad {R^{2}}_{323}  ~=~  \big(1 -  e^{-2\,\beta} \big)\,   \sin^2 \theta      \,.
\end{align} 
The non-zero components of the Ricci tensor are:
\begin{align}
R_{00}  &   ~=~  e^{2\,(\alpha-\beta)} \, \bigg(\alpha''(r)  + (\alpha'(r) )^2 -\alpha'(r) \,  \beta'(r)  +\frac{2}{r}\,  \alpha'(r) \bigg)\,,   \nonumber \\ 
R_{11}  &   ~=~ - \bigg(\alpha''(r)  + (\alpha'(r) )^2 -\alpha'(r) \,  \beta'(r)  -\frac{2}{r}\,  \beta'(r) \bigg)\,,  \nonumber \\
R_{22}  &   ~=~  e^{-2\,\beta} \, \big(r\, ( \beta'(r) - \alpha'(r)) -1 \big)~+~ 1\,, \qquad R_{33}    ~=~ R_{22}  \,  \sin^2 \theta   \,.
\end{align} 

The most-general, spherically symmetric, time-independent electromagnetic field is given by
\begin{equation}
F ~=~  E(r)  \, dt \wedge dr   ~+~ p\, \sin \theta \, d\theta \wedge d \phi \,.
\label{FForm1}
\end{equation} 
Closure of $F$, that is, $d F =0$, implies that $p$ is a constant.  The non-trivial Maxwell equation is 
\begin{equation}
d \star_4 F ~=~ 0 \qquad  \Leftrightarrow \qquad \frac{d}{dr}\, \big(r^2 \,  e^{- \alpha-\beta}  \, E(r)   \big) ~=~0  \,.
\label{Max1}
\end{equation} 

The energy-momentum tensor is given by
\begin{equation}
T_{\mu \nu}    ~=~   \frac{1}{4 \pi} \, ({F_\mu}^\rho \, F_{ \nu \rho} ~-~ \coeff{1}{4}\, g_{\mu \nu} \,  F^{\rho \sigma}  F_{\rho \sigma}) \,,
\label{TMax}
\end{equation} 
whose components are
\begin{align}
T_{00}  &   ~=~ \frac{1}{8 \pi} \, \Big( e^{-2\, \beta} \, \big( E(r)  \big)^2  ~+~ e^{2\, \alpha} \,  \frac{p^2}{r^4} \Big)\,,   \qquad  T_{11}   ~=~ -  \frac{1}{8 \pi} \, \Big(e^{-2\, \alpha} \, \big( E(r) \big)^2  ~+~ e^{2\, \beta} \,  \frac{p^2}{r^4}\Big)\,,  \nonumber \\
 T_{22}  &~=~ \frac{1}{8 \pi} \, \Big(  r^2 \, e^{-2\, (\alpha+\beta)} \, \big( E(r) \big)^2   ~+~ \frac{p^2}{r^2}\Big) \,,     \qquad T_{33}    ~=~ T_{22}  \,  \sin^2 \theta   \,.
\end{align} 
Note that this is traceless.

Note:  The factor of $\frac{1}{4 \pi}$ in (\ref{TMax}) comes from taking CGS/Gaussian units in which $\epsilon_0 = \frac{1}{4 \pi}$ and  $\mu_0 = 4 \pi$.  This makes the equations simpler if one takes $G=1$.  This is probably the last time I will get the units correct.  

Einstein's equations are: 
\begin{equation}
R_{\mu \nu} ~=~   8 \pi G \, ( \, T_{\mu \nu}  - \coeff{1}{2} \, T \, g_{\mu \nu}  ) ~=~    8 \pi G \, T_{\mu \nu} \,.
\end{equation} 
since $T \equiv g^{\mu \nu} T_{\mu \nu} =0$.  

The first equation (taking $G=1$) is:
\begin{equation}
R_{11} ~+~   e^{2\, (\beta-\alpha)}  \,R_{00}~=~ 8 \pi  \, (   T_{11} ~+~   e^{2\, (\beta-\alpha)} \, T_{00}) \,.
\end{equation} 
and is equivalent to
\begin{equation}
\frac{2}{r}\,  (\alpha'(r) + \beta'(r) )   ~=~ 0\,.
\end{equation} 
Hence
\begin{equation}
\alpha  ~=~   -\beta ~+~ {\rm constant}\,.
\end{equation} 
One can absorb the constant of integration into a rescaling of $t$ and so, without loss of generality, one can take:
\begin{equation}
\alpha  ~=~   -\beta \,.
\end{equation} 

This means that the Maxwell equation (\ref{Max1}) becomes 
\begin{equation}
\frac{d}{dr}\, \big(r^2 \,   E(r)   \big) ~=~0  \qquad \Leftrightarrow \qquad E(r)~=~ \frac{q}{r^2} \,,
\label{Max2}
\end{equation} 
for some constant, $q$.  If the metric is asymptotically Minkowskian, then (\ref{FForm1}) tells us that $q$ is the electrostatic charge.

The remaining Einstein equations are 
\begin{align}
e^{-2\, \alpha}   \bigg(\alpha''(r)  + 2(\alpha'(r) )^2  + \frac{2}{r}\,  \alpha'(r) \bigg)   &   ~=~  \big( E(r) \big)^2 ~+~  e^{2\, (\alpha+\beta)} \,  \frac{p^2}{r^4} ~=~   \frac{p^2+ q^2}{r^4}  \,,   \nonumber \\
e^{2\,\alpha} \, \big( 2\, r\,\alpha'(r) + 1 \big)  ~-~ 1&~=~ - r^2 \, \big( E(r) \big)^2  - \frac{p^2}{r^2}   ~=~  - \frac{p^2+ q^2}{r^2}     \,.
\label{Ein1}
\end{align} 
The second equation is equivalent to
\begin{equation}
\frac{d}{dr} \,\big( r\, e^{2\,\alpha} \big) ~=~ 1  - \frac{p^2+q^2}{r^2}    \qquad \Leftrightarrow \qquad   e^{2\,\alpha} ~=~ 1~-~  \frac{2\,m}{r} ~+~ \frac{p^2 + q^2}{r^2}   \,,
\end{equation} 
for some constant of integration, $m$.  The other equation in (\ref{Ein1}) is automatically satisfied (as it must be because of the Bianchi identities for $R_{\mu \nu}$ and conservation of the energy-momentum tensor).

Hence the solution is given by the Reissner-Nordstr\"om metric:
\begin{equation}
ds^2 ~=~ -  \Big(1~-~  \frac{2\,m}{r} ~+~ \frac{p^2 + q^2}{r^2}  \Big)  \, dt^2  ~+~  \Big(1~-~  \frac{2\,m}{r} ~+~  \frac{p^2 + q^2}{r^2}  \Big)^{-1} \, dr^2  ~+~  r^2\, \big(d\theta^2 + \sin^2 \theta\,d \phi^2) \,.
\label{RNmet}
\end{equation} 

In the Newtonian approximation for asymptotically flat spaces, one has 
\begin{equation}
g_{00} ~\approx~ -  \Big(1~+~  \frac{2 \, \phi(\vec x)}{c^2}  \Big)   \,,
\label{Newton1}
\end{equation} 
where $\phi(\vec x)$ is the Newtonian gravitational potential.   This means that we can identify the parameter $m$ with the Keplerian mass of the system.  The electric field is given by 
\begin{equation}
F ~=~  dA ~=~  \frac{q}{r^2}  \, dt \wedge dr   ~+~ p\, \sin \theta \, d\theta \wedge d \phi \,, \qquad A ~=~ \Phi(r) \, dt ~+~ p\, \cos \theta \, d \phi \,, \qquad 
\Phi~=~ \frac{q}{r}  \,,
\label{FForm2}
\end{equation} 
and $p$ and $q$ represent the magnetic and electric charges (in units with  $\epsilon_0 = \frac{1}{4 \pi}$ and  $\mu_0 = 4 \pi$).

For simplicity I will, henceforth set $p=0$.

%%%%%%%%%%%%%%%%%%%%%%%%
\subsection{Extreme Reissner-Nordstr\"om as an exemplar of branes}
\label{ss:extRN}
%%%%%%%%%%%%%%%%%%%%%%%%

The horizons of a simple black hole are identified by seeking the null hypersurfaces defined by constant $r$.  That is, one solves $g^{rr}=0$, which yields:  
\begin{equation}
r ~=~  r_\pm ~=~ m ~\pm~ \sqrt{ m^2 - q^2} \,.
\label{horlocation}
\end{equation} 
The physical range of Reissner-Nordstr\"om metrics have  $m \ge q$ (if $m < q$ then the metric has a naked singularity).   

The metrics with $m=q$ are known as extremal Reisnner-Nordstr\"om and have the form: 
\begin{align}
ds^2 ~=~ & -  H(r)^{-2}  \, dt^2  ~+~ H(r)^2 \, dr^2  ~+~  r^2\, \big(d\theta^2 + \sin^2 \theta\,d \phi^2)  \nonumber \\
~=~ & -  H(r)^{-2}  \, dt^2  ~+~ H(r)^2 \, \Big[ d\rho^2  ~+~  \rho^2\, \big(d\theta^2 + \sin^2 \theta\,d \phi^2) \Big]    \,.
\label{extRNmet}
\end{align} 
where
\begin{equation}
H ~\equiv~ \Big(1~-~  \frac{q}{r}  \Big)^{-1} ~=~1 +\frac{q}{\rho}\,, \qquad  \rho  ~\equiv~  r~-~ q \,.
\label{Hrhatdefn}
\end{equation} 
The inner and outer horizon merge into a single, extremal horizon at $\rho =0$ or $r=q$.

There are several things to note here:  
\begin{itemize}
\item  The spatial sections of the metric, defined by $(\rho, \theta, \phi)$, have the flat, Euclidean metric on $\IR^3$.
\item  $H(\rho)$ is a harmonic function on the flat $\IR^3$.
\item  The function, $H^{-1}$, is an electric potential for the Maxwell field, $E(r)$ in (\ref{Max2}).  
\end{itemize}
None of these facts are an accident.  

Indeed, a generic BPS $p$-brane has a metric of the form 
\begin{equation}
 (H(\vec y))^\alpha  \, \eta_{\mu \nu}  \, dx^\mu dx^\nu    ~+~ (H(\vec y))^\beta   \, d\vec y  \cdot d\vec y   \,,
\label{branemet}
\end{equation} 
where $\alpha,\beta \in \IR$, $\eta_{\mu \nu}$ is the metric on a Minkowski space, $\IR^{1,p}$, parallel to the brane, the metric,  $d\vec y  \cdot d\vec y$, transverse to the brane is flat, and $H(\vec y)$, is harmonic.  Moreover, $H(\vec y)$ is the electrostatic potential for the Maxwell field sourced by the brane.  A generic brane also sources a dilaton field, $\phi$, that is also entirely determined in terms of $H(\vec y)$.

Returning to the Reissner-Nordstr\"om metric, there are some other general threads that I wish to draw out.  First, note that as $r \to \infty$, the metric  (\ref{RNmet}) goes to flat Minkowski space:
\begin{equation}
ds^2 ~=~ -   \, dt^2  ~+~  dr^2  ~+~  r^2\, \big(d\theta^2 + \sin^2 \theta\,d \phi^2) \,.
\label{MinkMet1}
\end{equation} 
However, as one approaches the horizon of the extremal Reissner-Nordstr\"om metric,  $\rho \to 0$, 
\begin{align}
ds^2 ~\sim~ &  -  \Big(\frac{\rho}{q}\Big)^2    \, dt^2  ~+~ \Big(\frac{\rho}{q}\Big)^{-2}  \, \big( d\rho^2  ~+~ \rho^2\, (d\theta^2 + \sin^2 \theta\,d \phi^2) \big)\\
~=~ &   q^2  \, \bigg[- \frac{1}{q^{4}} \, \rho^2\, dt^2  ~+~  \, \frac{d\rho^2}{\rho^2}  \bigg]~+~ q^2\, \big(d\theta^2 + \sin^2 \theta\,d \phi^2 \big) \,.
\label{RBmet}
\end{align} 
This is a product space: AdS$_2 \times  S^2$.  (Historically it is known as the Robinson-Bertotti metric.)  The ``cosmological constant'' of the AdS$_2$ arises because the electromagnetic flux becomes $q^{-1}$ times the volume form of AdS$_2$ in the near-horizon region:
\begin{equation}
F ~=~ \frac{q}{r^2}  \, dt \wedge dr  ~\sim~ \frac{1}{q}  \, dt \wedge dr ~=~\frac{1}{q} \, \bigg(\frac{1}{q} \, \rho \, dt \bigg) \wedge \bigg(q \,\frac{d\rho}{\rho}\bigg)  \,.
\label{RBMax}
\end{equation} 
This is also a feature of many extremal brane geometries: The near-horizon limit of (\ref{branemet}) typically yields something of the form
\begin{equation}
q^2\,  \bigg[\frac{d\rho^2}{\rho^2} ~+~q^{-2}\,  \rho^2\, \eta_{\mu \nu}  \, dx^\mu dx^\nu \bigg]   ~+~ q^2\, d\Omega_n^2   \,,
\label{nearbranemet}
\end{equation} 
where $d\Omega_n^2$ is the metric of the unit sphere in the flat space described by the Cartesian coordinates, $\vec y$.  For a $p$-brane, this the metric (of a Poincar\'e slice) of AdS$_{p+2}$ $\times S^n$.

It is also useful to note that to go from the asymptotically-flat metric (\ref{MinkMet1}) to the AdS$_2$ $\times S^2$ metric in (\ref{RBmet}), one simply drops the $1$ in the harmonic function, $H(\rho)   = 1 +\frac{q}{\rho}$.  In more general geometries whose asymptotic form is governed by harmonic functions, one often finds that the transition from asymptotically-flat to asymptotically-AdS is achieved in exactly this way, and the mathematical process of going between such classes of asymptotics is sometimes referred to as ``Throwing away the 1's.''

One should also be mindful in using metrics like  (\ref{nearbranemet}): they look ``complete'' and smooth for $0 < \rho < \infty$, but we know they are not complete and that the apparent singularity at $\rho =0$ is a coordinate singularity.  One should remember that $\rho =0$, or $r =q$, is the {\it horizon} and that the patch covered by the metric (\ref{nearbranemet}) is geodesically incomplete.  In particular, the interior of the black hole corresponds to $0 < r <q$, or $-q < \rho < 0$.  One can use infalling null rays to set up Eddington-Finkelstein coordinates that continue  (\ref{nearbranemet}) across the horizon to the actual, physical singularity at $\rho = -q$, where $H$ vanishes. We will tend to use metrics like  (\ref{nearbranemet}) and one should remember that when we do this, we are neglecting the black-hole interior. 

That said, in microstate geometries there is no ``interior'' because the geometry caps off smoothly above where the black-hole horizon would form. 

%%%%%%%%%%%%%%%%%%%%%%%%
\subsection{Hawking temperature}
\label{ss:HT}
%%%%%%%%%%%%%%%%%%%%%%%%

The most direct way to compute the Hawking temperature of a static black hole is to use the fact that thermal ensembles are periodic in imaginary time with period 
\begin{equation}
\beta ~=~ \frac{1}{k\,T}    \,.
\label{temp}
\end{equation} 
For for static black holes one therefore continues $t \to i \tau$ and then examines the periodicity of the geometry in $\tau$ \cite{Gibbons:1976pt}.  For the Reissner-Nordstr\"om metric one has
\begin{equation}
ds^2 ~=~  r^{-2} \, \big((r-r_+)(r-r_-)  \big)  \, d\tau^2  ~+~ r^2\, \big((r-r_+)(r-r_-)  \big) ^{-1} \, dr^2  ~+~  r^2\, \big(d\theta^2 + \sin^2 \theta\,d \phi^2) \,.
\label{EuclRNmet1}
\end{equation} 
where  $r_\pm$  is given by (\ref{horlocation}).  For $r \sim r_+$ one has 
\begin{align}
ds^2 &~=~  r_+^{-2} \, (r_+ - r_-) \, \rho^2 \,  d\tau^2  ~+~4\, r_+^2\,  (r_+-r_-)^{-1} \, d\rho^2  ~+~  r_+^2\, \big(d\theta^2 + \sin^2 \theta\,d \phi^2) \nonumber  \\ 
& ~=~4\, r_+^2\,  (r_+-r_-)^{-1} \, \Big( d\rho^2~+~ \coeff{1}{4} \, r_+^{-4} \, (r_+ - r_-)^2 \, \rho^2 \,  d\tau^2\Big)  ~+~   r_+^2\, \big(d\theta^2 + \sin^2 \theta\,d \phi^2) \nonumber  \\ 
& ~=~4\, r_+^2\,  (r_+-r_-)^{-1} \, ( d\rho^2~+~ \rho^2 d\chi^2)  ~+~   r_+^2\, \big(d\theta^2 + \sin^2 \theta\,d \phi^2)\,.
\label{EuclRNmet2}
\end{align} 
where 
\begin{equation}
\rho ~\equiv~ \sqrt{r- r_+}  \,, \qquad \chi ~\equiv~   \coeff{1}{2} \, r_+^{-2} \, (r_+ - r_-) \,\tau\,.
\label{psirhodefn}
\end{equation} 
To be smooth, $\chi$ must have a period of $2\pi$, which means one must have:
\begin{equation}
\tau ~\equiv~\tau ~+~ \frac{4\pi \,  r_+^{2}}{ (r_+ - r_-)}  
\label{period}
\end{equation} 
and hence (taking $k=1$) the Hawking temperature is given by:
\begin{equation}
T_H ~=~   \frac{(r_+ - r_-)}{ 4\pi \,  r_+^{2}}  ~=~   \frac{\hbar\, c^3\,\sqrt{m^2 -q^2}}{ 2\pi \,  G\, k\,\big(m+ \sqrt{m^2 -q^2}\,\big)^2} \,,
\label{RNtemp}
\end{equation} 
where I have restored Planck's, Newton's and Boltzmann's constants as well as the speed of light.

For our purposes,  the most important aspect of this is that the Hawking temperature vanishes for extremal black holes.  Thus, not only are these solutions classically stable and static, but they are also time-independent in quantum mechanics. 

%%%%%%%%%%%%%%%%%%%%%%%%
\subsection{Supersymmetric ``branes'' in four dimensions}
\label{ss:Susy}
%%%%%%%%%%%%%%%%%%%%%%%%

Another extremely important property of the extreme Reissner-Nordstr\"om metric is that it is supersymmetric, and it is this property that allows one to construct explicit and remarkable generalizations in the form of solitons.

To exhibit the supersymmetry, and how it fixes the solution, we consider a generalization of  (\ref{extRNmet})  with  more general choices of functions:
\begin{equation}
ds^2 ~=~  -  H(\vec y)^{-2}  \, dt^2  ~+~ H(\vec y)^2 \, d \vec y \cdot d \vec y  \,.
\label{gen4harm}
\end{equation} 
where $H(\vec y)$ is an arbitrary function on the $\IR^3$ defined by $\vec y$.  We similarly consider a more general electrostatic field 
\begin{equation}
A ~=~ \Phi(\vec y) \, dt    \,, \qquad  F ~=~  dA ~=~  -  \partial_ i \Phi(\vec y)  \, dt \wedge dy^i     \,.
\label{FForm4}
\end{equation} 
Introduce the frames
\begin{equation}
e^0 ~=~ H(\vec y)^{-1} \, dt \,, \qquad  e^i ~=~ H(\vec y) \, dy^i   \,, \quad i =1,2,3 \,.
\label{4frames}
\end{equation} 
The spin connection, $\omega_{\mu\, ab} = -\omega_{\mu\, ba}$, is  then defined so to make the frames covariant constant:
\begin{equation}
0~=~ \nabla_\mu \,{e^a}_\nu ~\equiv~ \partial_\mu  {e^a}_\nu~+~  {{\omega_{\mu}}^a}{}_b \, {e^b}_\nu ~-~ \Gamma^\rho_{\mu \nu}  \, {e^a}_\rho \,.
\label{spincdefn}
\end{equation} 
where the frame indices, $a,b,c ...$ are raised and lowered using the Minkowski metric $\eta_{ab} = \eta^{ab} = {\rm diag}(-1,1,1,1)$.  The frame components and connection $1$-forms are  defined via:
\begin{equation}
e^a ~\equiv~{e^a}_\nu\, d x^\nu \,, \qquad \omega_{ab} ~\equiv~ \omega_{\mu\, ab} \, d x^\mu \,,
\label{c1forms}
\end{equation} 
and (\ref{spincdefn}) implies:
\begin{equation}
d \, e^a ~=~ - {{\omega}^a}{}_b \wedge e^b  \,.
\label{CM1}
\end{equation} 
This equation is sufficient to determine the connection $1$-forms, $\omega_{ab}$. 

A straightforward calculation using (\ref{4frames}) yields:
\begin{equation}
\omega_{0i} ~=~ -\omega_{i0} ~=~  H^{-2} (\partial_i H) \, e^0 \,, \qquad \omega_{ij} ~=~ -\omega_{ji} ~=~ - H^{-2} \, \big( (\partial_i H) \, e^j  -  (\partial_j H) \, e^i \big)  \,.
\label{spinconn1}
\end{equation} 

Supersymmetry transforms bosons into fermions, and {\it vice versa}:
\begin{equation}
\delta_\epsilon {\rm boson} ~=~ \epsilon \, {\rm fermions} \,, \qquad \delta_\epsilon {\rm fermion} ~=~ \epsilon \, {\rm bosons} \,.
\label{susy0}
\end{equation} 
and a supersymmetric background is defined to be one that is invariant under such a supersymmetry variation. Thus the variations in (\ref{susy0}) are required to vanish.  Supersymmetric backgrounds almost universally involve non-trivial bosonic fields and  have all the fermions set to zero.  This means that the non-trivial supersymmetry constraints come from solving the variations of the fermions. A solution, $\epsilon$, to $\delta_\epsilon {\rm fermions} ~=~ 0$ is called a {\it Killing spinor}.

It is  instructive to do some dimensional analysis in (\ref{susy0}).  Bosonic actions involve two derivatives whereas fermionic actions only involve one derivative.  This means that the dimensions of fermions  exceed the dimensions of the corresponding bosons by $\half$.  This is consistent with the bosonic variation if we assign $\epsilon$ to have a dimension of $-\half$.  To make the dimensions work in the fermion variation one must have a derivative on the right-hand side.  Thus the fermionic variations generically involve derivatives of bosons and the supersymmetry equations are typically first-order differential equations on the bosons.  

Strictly speaking, the foregoing argument is only valid in the absence of gravity because Newton's constant  has dimension $2$ and so can also be used to make up any discrepancy in dimensions.  One can refine the argument above to describe how and where  Newton's constant can appear and how it can rescale fields.  I am not going to do this because I am simply trying to set up expectations rather than prove theorems:  the BPS equations should involve first order derivatives of fundamental bosons and sometimes there will also be bosonic terms without derivatives, but these will come with factors of the square-root of Newton's constant.  

In a supergravity theory there are necessarily gravitini, $\psi_\mu^i$, and these transform at least into the graviton and, in extended supergravity theories, they also transform into  the Maxwell field strengths. Thus the primary set of supersymmetry equations come from setting the gravitino variations to zero.     More complicated supergravity theories can also have spin$-\frac{1}{2}$ fields and their variations lead to further constraints on the bosonic background.  In the simplest four-dimensional supergravity ($\Neql{2}$) that combines gravitons and Maxwell fields, the gravitino variation may be written as \cite{Gibbons:1982fy}:
\begin{equation}
\delta \psi_\mu ~=~ \nabla_\mu \epsilon ~-~ \coeff{1}{4} \,F_{\rho \sigma} \gamma^\rho  \gamma^\sigma  \gamma_\mu \epsilon  \,.
\label{gravvar1}
\end{equation} 
Thus supersymmetry requires us to solve the {\it first order system}: 
\begin{equation}
 \nabla_\mu \epsilon ~-~ \coeff{1}{4} \,F_{\rho \sigma} \gamma^\rho  \gamma^\sigma  \gamma_\mu \, \epsilon ~\equiv~  \partial_\mu \epsilon ~+~ \coeff{1}{4} \omega_{\mu a b} \,\gamma^a  \gamma^b\, \epsilon ~-~  \coeff{1}{4} \,F_{\rho \sigma} \gamma^\rho  \gamma^\sigma  \gamma_\mu \epsilon ~=~ 0 \,.
\label{susyeqn0}
\end{equation} 
It is simplest to write everything in terms of frame components:
\begin{equation}
{e^\mu}_c \,  \partial_\mu \epsilon ~+~ \coeff{1}{4} \omega_{c  a b} \,\gamma^a  \gamma^b\, \epsilon ~-~  \coeff{1}{4} \,F_{a b} \gamma^a  \gamma^b  \gamma_c \epsilon ~=~ 0 \,.
\label{susyeqn1}
\end{equation} 
where ${e^\mu}_c$ are the inverse frames (defined via ${e^\mu}_c {e_\mu}^a = \delta_c^a$) and 
\begin{equation}
F ~=~   -  \partial_ i \Phi(\vec y)  \, dt \wedge dy^i   ~=~  -( \partial_ i \Phi)  \, e^0 \wedge e^i   \,.
\label{Fframes}
\end{equation} 
The components of these four equations are:
\begin{align}
&H\, \partial_t \epsilon ~+~\coeff{1}{2} \, H^{-2} (\partial_i H) \,\gamma^0 \gamma^i \epsilon   ~+~\coeff{1}{2} \,  (\partial_i \Phi) \,\gamma^0 \gamma^i \gamma_0 \, \epsilon ~=~ 0  \,,\label{susyeqn2a} \\ 
&H^{-1}  \, \partial_i \epsilon ~+~\coeff{1}{4} \, H^{-2} (\partial_j H) \,(\gamma^j \gamma_i - \gamma_i  \gamma^j ) \epsilon   ~-~\coeff{1}{2} \, (\partial_j \Phi) \,\gamma^0 \gamma^j \gamma_i \, \epsilon  ~=~ 0   \,.
\label{susyeqn2b}
\end{align} 

There is a relatively obvious family of solutions to this.  Since the metric is time-independent, it is natural to seek solutions with $\partial_t \epsilon =0$.  The first equation then gives: 
\begin{equation}
\gamma^0 \gamma^i \Big[ - \partial_i \big( H^{-1}\big)   ~+~  (\partial_i \Phi) \,\gamma_0 \,  \Big] \, \epsilon  ~=~ 0  \,.
\label{susyeqn2c}
\end{equation} 
Since $\gamma^0 = - \gamma_0$ has eigenvalues $\pm 1$, this means we must take:
\begin{equation}
\gamma^0  \, \epsilon ~=~ \pm \epsilon \,,  \qquad\qquad \Phi  ~=~ \mp   H^{-1} ~+~ {\rm const.}
\label{proj1a}
\end{equation} 
Using  (\ref{proj1a}) this in  (\ref{susyeqn2b}) gives:
\begin{align}
0 ~=~   & H^{-1}  \, \partial_i \epsilon ~-~ \Big[\, \coeff{1}{4} \,  \big(\partial_j \big( H^{-1}\big)\big)\,(\gamma^j \gamma_i - \gamma_i  \gamma^j )    ~-~\coeff{1}{2} \,\gamma^j \gamma_i \,\Big]\, \epsilon  \\
 ~=~   & H^{-1}  \, \partial_i \epsilon ~-~  \coeff{1}{2} \,  \big(\partial_i \big( H^{-1}\big)\big)    \,  \epsilon   \,,
\label{susyeqn2d}
\end{align} 
and hence 
\begin{equation}
\epsilon ~=~  H^{-\frac{1}{2}} \, \epsilon_0 \,, 
\label{susyspinor}
\end{equation} 
where $\epsilon_0$ is a constant spinor.

Finally, we observe that the Maxwell equation, $d*F =0$ yields: 
\begin{equation}
\delta^{ij} \, \partial_i  \big( H^2 \partial_j  \Phi \big)  ~=~  0  \,, 
\label{Max3}
\end{equation} 
and using (\ref{proj1a}), this becomes
\begin{equation}
\delta^{ij} \, \partial_i   \partial_j \, H ~=~  0  \,.
\label{Max4}
\end{equation} 
Thus $H$ is a harmonic function on the $\IR^3$.

It is a tedious, but straightforward exercise that  (\ref{proj1a}) and  (\ref{Max4}) imply that the Einstein equations are satisfied.

%%%%%%%%%%
\subsubsection{Spinor conventions}
%%%%%%%%%%

Following \cite{Gibbons:1982fy}, I am taking $\gamma^a \gamma^b + \gamma^b \gamma^a  = -2 \eta^{ab}$, with explicit forms of $\gamma^a$ given by:
\begin{equation}
\gamma^0 ~=~  
\begin{pmatrix} 
0 &\oneone_{2\times 2}  \\ \oneone_{2\times 2} & 0
\end{pmatrix}  \,,  \qquad
\gamma^i  ~=~  
\begin{pmatrix} 
0 & \sigma_i  \\ - \sigma_i& 0
\end{pmatrix}  \,,  \qquad
\gamma^5  ~=~  
\begin{pmatrix} 
\oneone_{2\times 2}  & 0  \\ 0 & - \oneone_{2\times 2} 
\end{pmatrix}  \,.
\label{gammamats}
\end{equation} 
 Frame indices, $a, b, c.... $, are raised and lowered with  $\eta^{ab}$ and $\eta_{ab}$ and space-time indices, $\rho, \mu, \nu.... $, are raised and lowered with  $g^{\mu\nu}$ and $g_{\mu\nu}$.  The $\gamma$-matrices with space-time indices are defined by $\gamma^\mu \equiv {e^\mu}_a \gamma^a$.  

%%%%%%%%%%%%%%%%%%%%%%%%
\subsection{Some general observations about the supersymmetric solution}
\label{ss:SSsol}
%%%%%%%%%%%%%%%%%%%%%%%%

There are many important and  general lessons arising from this computation.

First, this is a huge generalization of the Reissner Nordstr\"om metric because $H$ is, {\it a priori}, a general harmonic function on $\IR^3$.  Indeed one can take  
\begin{equation}
 H ~=~  \varepsilon_0 ~+~ \sum_{j=1}^N \, \frac{q_j}{|\vec y - \vec y^{(j)} |}  \,.
\label{Hform1}
\end{equation} 
for some constant $\varepsilon_0$ and some charges $q_j$ sourced at the points $\vec y^{(j)}$.   Metric regularity requires that $H$ be strictly positive or  strictly negative.  We take the convention $H>0$, and then regularity means 
\begin{equation}
  \varepsilon_0 ~\ge~ 0  \,,\qquad   q_i ~\ge~ 0 \,,
\label{bounds1}
\end{equation} 
with at least one of them non-zero.   If $\varepsilon_0 \ne 0$, one can scale $t$ and the coordinates $\vec y$ so that $\varepsilon_0 = 1$. 

If $\varepsilon_0 = 1$ then $H \to 1$ as $r \equiv |\vec y| \to \infty$ and the metric (\ref{gen4harm}) is asymptotically flat.  If $\varepsilon_0 = 0$ then 
\begin{equation}
H ~\to~ \frac{Q}{r}  \,,\qquad  Q ~\equiv~ \sum_{j=1}^N \, q_j\,,
\label{Hasympinf}
\end{equation} 
and the metric is asymptotic to AdS$_2 \times S^2$ at infinity, as in (\ref{RBmet}).  

As one approaches the charge source at $\vec y^{(j)}$, one has 
\begin{equation}
H ~\sim~ \frac{q_j}{|\vec y - \vec y^{(j)} |} \,,
\label{Hasympj}
\end{equation} 
and the metric becomes exactly like that of extreme Reissner-Nordstr\"om, as described in Section \ref{ss:extRN}.   This solution is thus a static collection of extreme Reissner-Nordstr\"om black holes:  There is a perfect balance between the gravitational attraction ($\sim m$) and electrostatic repulsion  ($\sim q$).  The positions of the black holes, $\vec y^{(j)}$, are freely choosable ``moduli.''  This solution was first obtained over 70 years ago by  Majumdar and Papapetrou \cite{Majumdar:1947eu,Papapetrou:1948jw}.

%%%%%%%%%%%%%
\combox {There is a common belief that, in BPS solutions, the perfect balance of gravitational attraction and electrostatic repulsion leads to a large moduli spaces for the relative positions of such objects.  This belief is false:  as we will see in later lectures, there can be other interactions between BPS components and these interactions can create a ``potential'' that reduces the naive dimension of the moduli space.}
%%%%%%%%%%%%%

Turning to the supersymmetry, we note that the first constraint that arises is $\partial_i \Phi  ~=~ \mp  \partial_i( H^{-1})$ in (\ref{proj1a}).  The choice of sign is initially arbitrary, but it  determines which sign of the charge is to be viewed as supersymmetric (BPS) and which sign breaks the supersymmetry (is anti-BPS).  Indeed,  this sign choice is correlated with the sign of the projector: $\gamma^0  \epsilon ~=~ \pm \epsilon$.  Since $\gamma^0$ is traceless and $(\gamma^0)^2 = \oneone$, the matrix $\gamma^0$ has eigenvalues $+1$ and $-1$ each with degeneracy $2$.   The projection condition in  (\ref{proj1a})  thus cuts the number of supersymmetries in half and the two possible sets of supersymmetries (corresponding to $\pm$ in (\ref{proj1a})) are mutually incompatible.  For simplicity, we will take 
\begin{equation}
\gamma^0  \, \epsilon ~=~ - \epsilon \,,  \qquad\qquad \Phi  ~=~ +   H^{-1} \,,
\label{proj1final}
\end{equation} 
so that all the electric charges are positive.  Adding a negative charge would then break the supersymmetry and  make the metric (\ref{gen4harm}) singular ($H$ would vanish on some hypersurface). 

The underlying supersymmetric model has $\Neql 2$ supersymmetry in four dimensions.  This means that the supersymmetry parameter, $\epsilon$, can be taken to be a (complex) Dirac spinor\footnote{The designation, $\Neql 2$, is reflected in the fact that supersymmetry is usually counted in terms of Majorana (real) spinors, and a Dirac spinor has two Majorana pieces, that are essentially its ``real and imaginary parts.'' }.  There are thus eight independent supersymmetries because the total number of supersymmetries is determined by counting real parameters.  The projection condition (\ref{proj1final}) cuts this number to four supersymmetries, and thus, relative to the underlying model, one can view the multi-black-hole solution as \nBPS{2}.  However, these black holes are typically embedded in a M-theory, or in a ten-dimensional string theory, both of which have 32 supersymmetries.  Relative to such theories, the multi-black-hole solution is \nBPS{8}.  

%%%%%%%%%%%%%
\combox {Think of this as a warning: the only unambiguous way to express the amount of  supersymmetry is to specify the number of supersymmetry parameters.  Other ways of expressing the amount of supersymmetry depend on dimension and context.  In these lectures I will consider solutions with $4$ supersymmetries and take the macroscopic view that they are generically  \nBPS{8} states of M-theory or type II supergravity.}
%%%%%%%%%%%%%

The physical impact of the constraint $\partial_i \Phi  ~=~ \mp  \partial_i( H^{-1})$ should be evident from  (\ref{Newton1}) and (\ref{susyeqn1}):  supersymmetry sets the mass equal to the charge.  It is for this reason that  supersymmetry and BPS have almost become synonymous.  

The other fundamental feature of supersymmetry is that it ``squares to the hamiltonian.'' This has several important consequences for the analysis of the supersymmetry equations.  

First, consider the commutator of two covariant derivatives.  It is trivial to write this in terms of the curvature tensor:
\begin{equation}
\big[ \, \nabla_\mu,\nabla_\nu \, \big] \,\epsilon ~=~  \coeff{1}{4}\, R_{\mu \nu a b}\, \gamma^a \gamma^b \epsilon \,.
\label{commders}
\end{equation} 
On the other hand, one can simplify the left-hand side using the supersymmetry equations (\ref{susyeqn0}) to produce expressions in terms of the Maxwell field and its derivatives.  Indeed, the result is the ``integrability condition'' for the  supersymmetry equations (\ref{susyeqn0}) and it provides algebraic relationships between the curvature tensor and the Maxwell field  and its derivatives.  A careful analysis shows that these integrability conditions imply that almost all of the equations of motion are satisfied.  That is, almost all of the Maxwell equations and  Einstein equations are satisfied as a result of solving the supersymmetry conditions.  There are a number of theorems that cover this issue (see, for example, \cite{Gauntlett:2002nw,Gauntlett:2002fz,Gauntlett:2003wb}), and generically one needs to supplement the supersymmetry equations with just one (carefully chosen) component of the equations of motion so as to satisfy all of the equations of motion.  There are also many circumstances in which solving the supersymmetry equations actually solves all the equations of motion.   

One should also note that the converse is not true:  solving the equations of motion does not imply supersymmetry. One can easily see that the generic Reissner-Nordstr\"om black hole with $m \ne q$ is not supersymmetric. 

In the example above,  the supersymmetry equations related the electric potentials to metric coefficients, but did not not fully determine the underlying solution:  The fact that $H$ must be harmonic {\it only followed from using a particular equation of motion.}   We will find something similar with microstate geometries.

The other fundamental consequence of the supersymmetry ``squaring to the hamiltonian'' is that, if $\epsilon$ solves the supersymmetry equations then the vector field 
\begin{equation}
K^\mu  ~=~  \bar \epsilon \gamma^\mu \epsilon ~=~  {e^\mu}_a\, \bar \epsilon \gamma^a \epsilon   \,.
\label{KVec}
\end{equation} 
is a Killing vector (where $\bar \epsilon \equiv  \epsilon^\dagger \gamma^0$ is the Dirac conjugate).  The detailed proof depends on the specifics of the supersymmetry equation, but one simply computes  $\nabla_\mu K_\nu$ and then uses the supersymmetry equations to replace $\nabla_\mu \epsilon$ and   $\nabla_\mu  \bar \epsilon$.  Upon symmetrization, $\nabla_\mu K_\nu +\nabla_\nu K_\mu$, one usually finds that everything cancels.  This has been proven for many supergravity theories coupled to matter, and, in particular, for $M$-theory  (see, for example, \cite{Gauntlett:2002nw,Gauntlett:2002fz,Gauntlett:2003wb}).

One can indeed verify this for the solution above.  In particular, there is only one  Killing vector (namely,  $\frac{\partial}{\partial t}$) for the metric (\ref{gen4harm}). The time-like component of (\ref{KVec}) is given by 
\begin{equation}
K^0  ~=~  {e^{\mu =0} }_{a=0} \, \bar \epsilon \gamma^{0}  \epsilon ~=~ H \,  \epsilon^\dagger  \gamma^{0} \gamma^{0}  \epsilon~=~ -  H \,  \epsilon^\dagger \epsilon ~=~ -   \epsilon_0^\dagger \epsilon_0     \,.
\label{KVec0}
\end{equation} 
where I have used (\ref{susyspinor}) in the last step.  The important point is that the time-component of $K^\mu$ must be a constant (and all other componets must vanish) for $K^\mu$ to be the Killing vector.  

As a practical consideration, the fact that $K^\mu$, defined by (\ref{KVec}), must be a Killing vector is usually used in reverse to determine the norm of the spinor, as in (\ref{susyspinor}).

More generally, we know that  (\ref{KVec})  always defines a time-like or null Killing vector.  Indeed the proof that $K^\mu$ is time-like or null is relatively straightforward use of the Schwartz inequality on the spin space.  

To see this, consider the two spinors $\epsilon$ and $\eta$ where 
\begin{equation}
\eta ~\equiv~   v_i \gamma^0 \gamma^i \epsilon   \,,
\label{etadefn}
\end{equation} 
for some vector $v_i$ in $\IR^3$ and the indices, $i$, are spatial frame indices.  One can use the properties of $\gamma$-matrices to show that:
\begin{equation}
\eta^\dagger \eta~=~   |v|^2 \, \epsilon^\dagger \epsilon  \,.
\label{etanorm}
\end{equation} 
The Schwarz inequality implies 
\begin{equation}
| v_i \, (\epsilon^\dagger   \gamma^0 \gamma^i \epsilon) |^2  ~\le~  (\epsilon^\dagger \epsilon)\,(\eta^\dagger \eta) ~=~   |v|^2 \, (\epsilon^\dagger \epsilon)^2  \,,
\label{Schwartz1}
\end{equation} 
Choose $v_i = (\epsilon^\dagger   \gamma^0 \gamma^i \epsilon) =  (\bar \epsilon  \gamma^i \epsilon)$ and one gets 
\begin{equation}
|v|^4  ~\le~    |v|^2 \, (\epsilon^\dagger \epsilon)^2  \qquad \Leftrightarrow\qquad |(\bar \epsilon \gamma^i \epsilon)|^2  ~\le~    (\epsilon^\dagger \epsilon)^2 ~=~  (\bar \epsilon \gamma^0 \epsilon)^2  \,,
\label{Schwartz2}
\end{equation} 
The last inequality is precisely the inequality 
\begin{equation}
\eta_{ab} \, K^a \,K^b ~\le~   0  \,.
\label{Knonspacelike}
\end{equation} 
It is also amusing to note that there is equality in (\ref{Knonspacelike}) if and only if $\eta$  is proportional to $\epsilon$.  This means that $K^\mu$ is null if and only if 
\begin{equation}
v_i  \gamma^0 \gamma^i \epsilon ~=~  \lambda \, \epsilon   \,.
\label{nullproj}
\end{equation} 
for some vector $v_i$ and some number, $\lambda$.  This is a projection condition on $\epsilon$ and it is, in fact, incompatible with (\ref{proj1a}).  
Using this kind of argument, it is not very difficult to determine when the Killing vector (\ref{KVec}) is time-like or null.  In these lectures, $K^\mu$ will always be time-like.

%%%%%%%%%%
\subsubsection{A supersymmetric summary}
\label{ss:susysummary}
%%%%%%%%%%

The take-away messages here are 
\begin{itemize}
\item Supersymmetry involves solving {\it first order equations} for spinors on a manifold.  The solutions to these equations are called {\it Killing spinors.}
\item The supersymmetries, or Killing spinors, are usually confined to a subspace of all the spinors on the manifold.  This subspace is typically defined by projection conditions on the spinors and is sometimes called the {\it supersymmetry bundle.} 
\item Supersymmetry usually imposes a ``BPS condition'' in that the overall mass of the system is locked to (some of) the charges of the system 
\item {\bf Most important:} Solving the supersymmetry equations is usually much easier than solving all the equations of motion.  This is because the former are a first order system while the latter are a second order (non-linear) system.  Solving supersymmetry equations typically solves almost all of the equations of motion.
\item If $\epsilon$ is a supersymmetry, then the tensors, $T^{{\mu_1}  \dots {\mu_k} } = \bar \epsilon \gamma^{\mu_1} \dots  \gamma^{\mu_k} \epsilon$ have really interesting geometric properties, and in particular $K^\mu = \bar \epsilon \gamma^\mu \epsilon$ is generically a non-space-like Killing vector.   {\it The study of and classification of these rich tensor structures is a very active programme at Saclay, led by Mariana Gra\~na and Ruben Minasian.}
\item {\bf Most important:} Imposing supersymmetry is a ``Faustian Bargain'' for the physics.  Computations are much easier and the solutions are relatively simple, but the solutions are necessarily BPS and time-independent, both at the classical and quantum levels.  In particular, supersymmetric black-hole solutions have vanishing Hawking temperature.  The hope is that, like super-QCD, there is still important essential physics in the supersymmetric theory that gives valuable insight into the non-supersymmetric theories that underpin the natural world.
\end{itemize}
%

%%%%%%%%%%%%%%%%%%%%%%%%
\subsection{A final footnote}
\label{ss:footnote1}
%%%%%%%%%%%%%%%%%%%%%%%%

In  Section  \ref{ss:Susy} I considered a very simple metric Ansatz  (\ref{gen4harm}).  In principle, the most general supersymmetric metric is only required to be time-independent and so I could have started with the most general  time-independent metric, which has the form 
\begin{equation}
ds^2 ~=~  -  H^{-2}  \, (dt + \omega_i  d y^i) ^2  ~+~  \gamma_{ij} \, d y^i  \, d  y^j  \,.
\label{gen4met2}
\end{equation} 
where $H$, $\omega$ and $ \gamma_{ij}$ all depend upon the $y^k$.  One can also start with generic electric and magnetic fields. 

Gibbons and Hull \cite{Gibbons:1982fy} argue that this leads only to Majumdar-Papapetrou metrics obtained in  Section  \ref{ss:Susy}.  While I believe the result is correct, in preparing this course I discovered that their proof appears to have an error wrong.  There is a sentence where they say ``Each of the three terms in (27) is non-negative and so each must vanish separately.'' Unfortunately, the middle term is actually non-positive (and not non-negative).  If one sets the magnetic field to zero, then the offending term vanishes. It does, however, make me wonder if they missed some amusing magnetic generalization ... but I doubt it.  I leave a careful examination of this as an exercise for the reader.

%%%%%%%%%%%%%%%%%%%%%%%%%%%%%%%%%%%%%
\section{Solitons, horizons and topology}
\label{Sect:SHT}
%%%%%%%%%%%%%%%%%%%%%%%%%%%%%%%%%%%%%

The goal here is to understand how microstate geometries evade the ``No solitons without horizons'' theorems.  Such theorems were rigorously proved in $(3+1)$ dimensions and have generalizations (under implicit and sometimes unstated assumptions) to higher dimensional theories.  The first lesson is therefore that to find non-trivial microstate geometries, one must work in more than four space-time dimensions.   We are going to consider a simple version of the ``No solitons'' theorem in a basic class of supergravity theories in five dimensions.  This setting suffices to see how the theorems usually work, and how they can be evaded.  Before we start this, it is important to recall some basic facts about asymptotics and charges.

%%%%%%%%%%%%%%%%%%%
\subsection{Interlude:  Mass and conserved charges}
\label{AsympCharges}
%%%%%%%%%%%%%%%%%%%

To investigate the ``No Solitons'' theorem we are going to have to dissect the definitions of mass, angular momentum and charge for a generic solution in $D$ space-time dimensions.  My discussion here draws heavily on the excellent review by Peet  \cite{Peet:2000hn} and my work with Gary Gibbons \cite{Gibbons:2013tqa}.

%%%%%%%%%%%%%%%%%%%
\subsubsection{Expansions at infinity}
\label{ss:Asymptotics}
%%%%%%%%%%%%%%%%%%%

To determine the normalized asymptotic charges for an asymptotically flat metric in a $D$-dimensional space-time one should start from the canonically normalized action:
\begin{equation}
S ~=~  \int d^D x \, \sqrt{-g} \,  \bigg( \frac{R}{16\pi G_D}  ~+~{\cal{L}}_{\rm matter} \bigg) \,,
\label{canonact}
\end{equation}
where $G_D$ is the Newton constant.   The Einstein equations  are:
\begin{equation}
R_{\mu \nu}  - \frac{1}{2} \, R \, g_{\mu \nu} ~=~  8 \pi G_D  \, T_{\mu \nu} \,,
\label{Ein2}
\end{equation}
where $T_{\mu \nu}$ is the canonically normalized energy-momentum tensor.  The Einstein equations may be rewritten as   
\begin{equation}
R_{\mu \nu}  ~=~  8 \pi G_D \,  \Big(  T_{\mu \nu} - \frac{1}{(D-2)} \,  T \, g_{\mu \nu} \Big) \,,
\label{Ein3}
\end{equation}
where $T$ is the trace of $T_{\mu \nu} $. 

If one linearizes around a flat metric, using an expansion $g_{\mu \nu} = \eta_{\mu \nu} + h_{\mu \nu}$, then, in an appropriate harmonic gauge, one has
\begin{equation}
\eta^{\rho\sigma} \partial_\rho \partial_\sigma \,  h_{\mu \nu} ~\approx~ 16 \pi G_D \,  \Big(  T_{\mu \nu} - \frac{1}{(D-2)} \,  T \, g_{\mu \nu} \Big) \,.
\label{lingrav1}
\end{equation}
If one assumes that the matter is non-relativistic one can neglect time derivatives and write this as a solution to the Laplace equation on a  $(D-1)$-dimensional hypersurface, $\Sigma$: 
\begin{equation}
 h_{\mu \nu}(\vec x)  ~\approx~ \frac{16 \pi G_D}{A_{D-2}}  \, \int_\Sigma\, d^{D-1} \vec y \, \bigg(\,  \frac{1}{|\vec x - \vec y|^{D-3}}\, \Big(T_{\mu \nu}(|\vec x - \vec y| ) - \frac{1}{(D-2)} \,  T(|\vec x - \vec y| )  \, g_{\mu \nu} \Big)\,\bigg) \,,
\label{lingrav2}
\end{equation}
where $A_{D-2}$ is the volume of a unit $(D-2)$ sphere and is an inherent part of the relevant Green function. For future reference we note that $A_3 = 2 \pi^2$.

One then recalls that the momentum and angular momentum of the configuration are obtained my various integrals of $T_{\mu \nu}$ over the space-like hypersurface: 
\begin{equation}
P^\mu  ~=~  \int_\Sigma d^{D-1} x  \, T^{\mu 0}  \,,  \qquad 
J^{\mu \nu}  ~=~     \int_\Sigma d^{D-1} x  \, \big( x^\mu  T^{\nu 0} - x^\nu  T^{\mu 0} \big) \,.
\end{equation} 
By expanding (\ref{lingrav2}) one obtains  \cite{Myers:1986un, Peet:2000hn}:
\begin{eqnarray}
g_{00} &=&  -1 ~+~  \frac{16\pi G_D} {(D-2)\, A_{D-2}}  \frac{M}{\rho^{D-3} }~+~ \dots  \,,  \label{asympg1} \\
g_{ij}   &=&  1  ~+~  \frac{16\pi G_D} {(D-2)\, (D-3)\, A_{D-2}}   \frac{M}{\rho^{D-3}} ~+~ \dots  \,,\label{asympg2} \\ 
g_{0i} &=&   \frac{16\pi G_D} { A_{D-2}}   \frac{x^j J^{ji} }{\rho^{D-1}}~+~ \dots  \label{asympg3}\,,
\end{eqnarray}
where $\rho$ is the radial coordinate. 

More generally, the expansions (\ref{asympg1})--(\ref{asympg3}) are used to define asymptotic charges of a generic, asymptotically-flat metric

For Maxwell fields,  $F_{\mu \nu}$, one can generalize Gauss' law and integrate $*F$ over the $(D-2)$-sphere at infinity on $\Sigma$.  However, a standard normalization that is commonly used in the literature, and we will use here, is to take the gauge potential for time-independent solutions to be:
\begin{equation}
A  ~\sim~     \frac{Q}{\rho^{D-3} } \,dt \,, \qquad F   ~\sim~    (D-3)\, \frac{Q}{\rho^{D-2} } \, dt \wedge d\rho \qquad \rho \to \infty \,.
\label{asympQ}
\end{equation}
%

%%%%%%%%%%%%%%%%%%%
\subsubsection{Killing vectors and Komar integrals}
\label{KVKIs}
%%%%%%%%%%%%%%%%%%%

A Killing vector, $K^\mu$, defines a symmetry of the metric (an isometry).  It satisfies the Killing equation 
\begin{equation}
\nabla_\mu K_\nu ~+~\nabla_\nu K_\mu     ~=~     0 \,.
\label{Killing1}
\end{equation}
Any single vector can be locally integrated to that it is tangent to a coordinate axis: $K^\mu \frac{\partial}{\partial x^\mu} = \frac{\partial}{\partial v} $, where $v$ is one of the coordinates.  In this coordinate frame,  (\ref{Killing1}) is simply equivalent to:  
\begin{equation}
 \frac{\partial}{\partial v} \, g_{\mu \nu}  ~=~     0 \,.
\label{Killing2}
\end{equation}
It follows that, in this frame, the curvature tensors are all independent of $v$ and so 
\begin{equation}
K^\mu \nabla_\mu  \, R     ~=~     0 \,,
\label{RSinv}
\end{equation}
where $R$ is the Ricci scalar.  Note that this equation is  independent of  the choice of coordinates.

Indeed one can also show, using (\ref{Killing1}) that a Killing vector, satisfies some rather more general identities
\begin{equation}
 \nabla_\mu  \nabla_\nu  \, K_\rho     ~=~    {R^\sigma}_{ \mu \nu \rho}\, K_\sigma    \,, \qquad  \nabla^\rho   \nabla_\rho  \, K_\mu     ~=~     -R_{\mu \nu}\, K^\nu   \,,
\label{KVeceqns}
\end{equation}
where ${R^\sigma}_{ \mu \nu \rho}$ is the Riemann tensor and $R_{\mu \nu}$ is the Ricci tensor.

A Killing vector can be used to define conserved currents using the energy momentum tensor.  That is, if $K^\mu$ is a Killing vector, then the current 
\begin{equation}
 \hat  J_\mu ~\equiv~  K^\nu \, T_{\mu \nu}  \,,
\label{cons1}
\end{equation}
is necessarily conserved:
\begin{equation}
\nabla^\mu \hat J_\mu ~=~ (\nabla^\mu  K^\nu) \, T_{\mu \nu}  ~+~ K^\nu  (\nabla^\mu  \, T_{\mu \nu} ) ~=~  0 \,,
\label{divcons1}
\end{equation}
where the first term vanishes because of the symmetry of $T_{\mu \nu}$ and (\ref{Killing1}) and the second term vanishes because of the conservation of $T_{\mu \nu}$.  For smooth space-times with Killing vectors one can use this to define a corresponding conserved quantity.  However, there is a more convenient refinement given by the Komar integral.

\begin{figure}
\leftline{\hskip 1.8cm \includegraphics[width=5in]{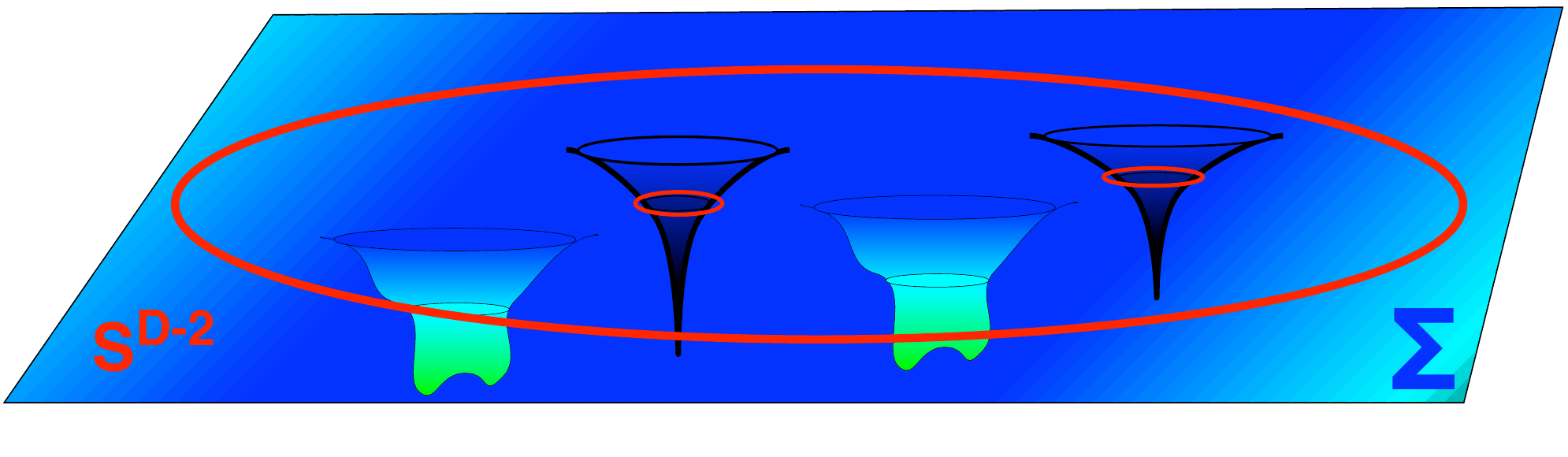}}
\setlength{\unitlength}{0.1\columnwidth}
\caption{\it The space-like hypersurface $\Sigma$ slicing through black holes and solitonic lumps.  The sphere``at infinity'' is taken to be large enough to encompass all the sources and only be sensitive to the leading asymptotics of all fields.}
\label{fig:Hyperslice}
\end{figure}

Consider a space-like hypersurface, $\Sigma$, in an asymptotically flat space-time. Let $S^{(D-2)}$ be the $(D-2)$-sphere at infinity on $\Sigma$. (See Fig. \ref{fig:Hyperslice}.) Consider the Komar integral
\begin{equation}
\cI_K ~=~ \int_{S^{(D-2)}} \, *d K \,,
\label{IntChg}
\end{equation}
where $K = K_\mu d x^\mu$ and $*$ is the Hodge dual in $D$ dimensions.

%%%%%%%%%%%%%%   Figure 1   %%%%%%%%%%%%%%%
%%%%%%%%%%%%%%   End Figure 1   %%%%%%%%%%%%%%%

If $\Sigma$ is smooth, and one applies Stokes' theorem, one obtains 
\begin{equation}
\cI_K ~=~ \int_{\Sigma} \, d*d K~=~  -2 \,\int_{\Sigma} \, R_{\mu \nu}  K^\mu n^\nu \, d \Sigma\,,
\label{cons2}
\end{equation}
where $n^\nu$ is the unit normal to $\Sigma$ and I have used  (\ref{KVeceqns}).

We can now use Einstein's equations, (\ref{Ein2}) to write this as 
\begin{equation}
\cI_K ~=~  -16 \pi G_D \,  \,\int_{\Sigma} \,  \Big(  T_{\mu \nu} - \frac{1}{(D-2)} \,  T \, g_{\mu \nu} \Big) K^\mu n^\nu \, d \Sigma\,,
\label{cons3}
\end{equation}
If it were not for the trace term, $T$, this would be precisely the charge associated with $\hat J^\mu$, defined in (\ref{cons1}) 

Instead, we can actually define the current associated with $K^\mu$ as: 
\begin{equation}
J_\mu ~\equiv~  K^\nu \, R_{\mu \nu} ~~=~ 8 \pi G_D \,   \Big(  T_{\mu \nu} - \frac{1}{(D-2)} \,  T \, g_{\mu \nu} \Big) K^\nu  \,.
\label{current2}
\end{equation}
It is easy to verify that this is also conserved:
\begin{equation}
\nabla^\mu J_\mu ~=~ (\nabla^\mu  K^\nu) \, R_{\mu \nu}  ~+~ K^\nu  (\nabla^\mu  \, R_{\mu \nu} ) ~=~  \coeff{1}{2} \,K^\nu  \, \nabla_\nu  \, R ~=~ 0\,,
\label{cons5}
\end{equation}
where the first term vanishes, one again, because of the symmetry of $R_{\mu \nu}$ and (\ref{Killing1}).  The second term is simplified using the Bianchi identity:  
\begin{equation}
\nabla^\mu R_{\mu \nu} ~=~ \coeff{1}{2} \, \nabla_\nu  \, R \,,
\label{bianchi0}
\end{equation}
and the final equality in (\ref{cons5}) follows from (\ref{RSinv}).  Equivalently, one can use the energy-momentum  expression in (\ref{current2}) and then one needs to know that $ K^\nu  \, \nabla_\nu T =0$, but this follows from (\ref{RSinv}) and the trace of Einstein's equations.

The reason why one uses the ``improved'' current (\ref{current2})  is that it can be written in terms of a surface integral (\ref{cons1}) and therefore can be generalized to situations in which the interior geometry is not smooth.

So far we have not specified any other properties of the Killing vector, $K^\mu$, and so the Komar integral can give any type of conserved quantity.

Suppose that $K^\mu$ is time-like.  The Komar mass is then defined by:
\begin{equation}
M ~=~  -  \frac{1}{16\pi G_D} \,  \frac{(D-2)}{(D-3)}  \,   \int _{S^{D-2}}   \, * d K   ~=~  -  \frac{1}{16\pi G_D} \, \frac{(D-2)}{(D-3)}  \,    \int _{S^{D-2}}  \,\big(\partial_\mu K_\nu - \partial_\nu K_ \mu  \big)  d \Sigma^{\mu \nu} ~ \,.
 \label{Komar1}
\end{equation}
Note that if we take $K^\mu \frac{\partial}{\partial x^\mu} = \frac{\partial}{\partial t}$, then the $1$-form $K$, is given by $K = g_{0 \nu} dx^\nu$ and 
\begin{equation}
 *  dK  ~=~ (\partial_\mu  g_{0 \nu})\,  * ( dx^\mu \wedge dx^\nu) ~\to~ (\partial_\rho  g_{0 0}) \, \rho^{D-2} \,  {\rm Vol}_{S^{D-2}}
\label{dKform1}
\end{equation}
Using (\ref{asympg1}), one sees that the normalized formula   (\ref{Komar1}) does indeed yield the correct asymptotic ``Keplerian''/ADM mass.

Also note that in the non-relativistic limit in which $T_{00} \gg |T_{ij}|$ and $|g_{00}| \gg |g_{ij}|$, one has $T \equiv g^{\mu \nu} T_{\mu \nu} \approx g^{00} T_{00}  \approx (g_{00})^{-1} T_{00}$, and hence

\begin{equation}
\cI_K ~=~  -16 \pi G_D \,  \,\int_{\Sigma} \,  \Big(  T_{0 0} - \frac{1}{(D-2)} \,  T \, g_{0 0} \Big)  \, d \Sigma ~=~ -16 \pi G_D \,  \,\int_{\Sigma} \,  \Big( 1- \frac{1}{(D-2)}\Big)\,T_{0 0}   \, d \Sigma \,.
\label{conssimp}
\end{equation}
Thus, in the non-relativistic limit one also has 
\begin{equation}
M ~=~  \,\int_{\Sigma} \,T_{0 0}   \, d \Sigma \,.
 \label{Komarnonrel}
\end{equation}
as one would expect.

%%%%%%%%%%%%%%%%%%%
\subsection{A relatively simple supergravity theory}
\label{fiveDsugr}
%%%%%%%%%%%%%%%%%%%

We will work with ungauged, $\Neql2$ supergravity coupled to two vector multiplets in five dimensions. This theory contains three vector fields, $A^I$,  with field strengths, $F^I  \equiv d A^I$, and  two independent scalars,  which may   conveniently be parametrized by   the fields, $X^I$, $I=1,2,3$ satisfying the constraint $X^1 X^2 X^3  = 1$.   The bosonic action is 
\begin{eqnarray}
  S ~=~ \int\!\sqrt{-g}\,d^5x \Big( R  -\coeff{1}{2} Q_{IJ} F_{\mu \nu}^I   F^{J \mu \nu} - Q_{IJ} \partial_\mu X^I  \partial^\mu X^J -\coeff {1}{24} C_{IJK} F^I_{ \mu \nu} F^J_{\rho\sigma} A^K_{\lambda} \bar\epsilon^{\mu\nu\rho\sigma\lambda}\Big) \,,
  \label{5daction}
\end{eqnarray}
with $I, J =1,2,3$. The  structure constants are given by $C_{IJK} \equiv |\epsilon_{IJK}|$ and the metric for the kinetic terms is 
\begin{equation}
 Q_{IJ} ~=~    \frac{1}{2} \,{\rm diag}\,\big((X^1)^{-2} , (X^2)^{-2},(X^3)^{-2} \big) \,.
\label{scalarkinterm}
\end{equation}
One should note the new feature, the Chern-Simons term $F \wedge F \wedge A$, which will be critical to the construction of solitons. 

We could, in fact, work with ``minimal supergravity,'' that is, with pure $\Neql2$ supergravity and no extra vector multiplets.  This corresponds to taking $A^1 = A^2 =A^3$ and $X^1 = X^2 =X^3 = 1$.  Note that, with this choice, the action still contains a non-trivial Chern-Simons term.  Going to the minimal theory  simplifies the computations a little, but having three vector fields makes the role and structure of the Chern-Simons interaction all the more transparent.  

In this section we will not use the supersymmetry explicitly: we will simply consider solutions to the equations of motion obtained from the action (\ref{5daction}).

We are going to seek solitons with a time-independent metric {\it and} time-independent matter.  (There are well-known families of solitons, called Q-balls,    \cite{Coleman:1985ki,Friedberg:1986tq} that have time-dependent matter  but the energy-momentum tensor, and hence the metric, are  time independent.) The most general form of such a metric on a five-dimensional Lorentzian, stationary space-time, $\cM_5$, is: 
\begin{equation}
ds_5^2 ~=~ -Z^{-2} \,(dt + k)^2 ~+~ Z \, ds_4^2  \,, 
\label{metform}
\end{equation}
where $ds_4^2$ is a general Riemannian metric on a four-dimensional spatial base manifold,  $\cB$.   The ``warp factor,'' $Z$, is a function and $k$ is a vector (one-form) field on $\cB$. For later convenience I have added  $Z$ as a warp factor in front of the $ds_4^2$.

The Maxwell fields are time independent and so may be decomposed into  electric and magnetic components:
\begin{equation}
A^I   ~=~  -  Z_I^{-1}\, (dt +k) ~+~ B^{(I)}  \,,
\label{Aform}
\end{equation}
where $B^{(I)}$ is a one-form on $\cB$.   It will prove  convenient to define magnetic field strengths:
\begin{equation}
\Theta^{(I)}    ~\equiv~  d B^{(I)}    \,,
\label{Thetadefn}
\end{equation}
which is a $2$-form on $\cB$.

The Einstein equations coming from   (\ref{5daction}) are:
\begin{equation}
R_{\mu \nu} -\coeff{1}{2} g_{\mu \nu} R  ~=~   Q_{IJ}\,\Big[  F^{I}_{\,\mu \rho}  \, {{F^J}_\nu \,}^\rho   - \coeff{1}{4} \, g_{\mu \nu}  \,  F^{I}_{\, \rho \sigma} F^{ J\, \rho \sigma} +  \partial_\mu X^I \,  \partial_\nu X^J  
 -  \coeff{1}{2} \, g_{\mu \nu}  \, g^{\rho \sigma}  \,  \partial_\rho  X^I \,  \partial_\sigma X^J   \Big]\,.  \label{Einstein1} 
\end{equation}
Taking traces and rearranging gives the equation:
\begin{equation}
R_{\mu \nu} ~=~  Q_{IJ}\,\Big[  F^{I}_{\, \mu \rho}  \, {{F^J}_\nu \,}^\rho   - \coeff{1}{6} \, g_{\mu \nu}  \,  F^{I}_{\, \rho \sigma} F^{ J\, \rho \sigma} +  \partial_\mu X^I \,  \partial_\nu X^J  
 \Big]\,.  \label{Einstein2} 
\end{equation}

The Maxwell equations coming from    (\ref{5daction})  are:  
\begin{equation}
\nabla_{\rho} \big(Q_{IJ}  {{F^J}^{\rho}}_\mu \big)  ~=~  J^{CS}_{I \,\mu}  \,, \label{Max5} 
\end{equation}
where the Chern-Simons currents are given by:
\begin{equation}
J^{CS}_{I \,\mu}  ~\equiv~  \coeff {1}{16} \, C_{IJK}\, \epsilon_{\mu \alpha \beta \gamma \delta } \, F^{J\, \alpha \beta}  \, F^{k\,\gamma \delta}   \,. \label{CScurrents} 
\end{equation}
One can also easily obtain the  equations of motion for the scalars.

%%%%%%%%%%%%%%%%%%%
\subsection{``No solitons without topology''}
\label{Sect:NSwoT}
%%%%%%%%%%%%%%%%%%%

In addition to assuming that metric is stationary and that the matter is time-independent, I  now assume that the five-dimensional space-time, $\cM_5$,  is smooth, horizonless and asymptotic to Minkowski space at infinity.  In particular, I will require the space-time to be sectioned, at fixed times, by smooth, space-like hypersurfaces, $\Sigma$.  The goal is to study the properties of the Komar mass, and the strategy of the ``No Solitons'' theorem is to argue that the mass, $M$, must be zero.  One then leverages this to show that all the dynamical fields must be trivial and the the solution must, in fact, be  Minkowski space globally.
 
 The first step is to massage the equations of motion into a more useful form.  Define the dual of the Maxwell fields, $G^I$, via
\begin{equation}
G_{I\, \rho \mu \nu}  ~\equiv~  \coeff {1}{2} \, Q_{IJ} \, F^{J\, \alpha \beta}   \, \epsilon_{\alpha \beta \rho \mu  \nu  }     \label{Gdefns} 
\end{equation}
and introduce the inverse, $Q^{IJ}$ of $Q_{IJ}$:
\begin{equation}
Q^{IJ}   \, Q_{JK} ~= ~ \delta^I_K  \,. \label{QIJinv} 
\end{equation}
If follows from the Bianchi identities ($d*F^J =0$) for $ F^{J}_{\mu  \nu}$ that $G_{J}$ satisfies:
\begin{equation}
\nabla_{\rho} \big(Q^{IJ} {G_{J}}^{\,\mu  \nu \rho  }   \big)  ~=~ 0 \,, \label{MaxG} 
\end{equation}
Similarly, from the equations of motion (\ref{Max1}) for $ F^{J}_{\mu  \nu}$  one has 
\begin{equation}
\nabla_{[ \lambda}  G_{| J | \, \rho \mu  \nu  ] }  ~=~ + \coeff{3}{8} \, C_{IJK} \,  F^{J}_{[\lambda \rho} \, F^{K}_{\mu  \nu]} \qquad \Leftrightarrow \qquad d G_I ~=~ + \coeff{1}{4} \, C_{IJK} \,  F^{J} \wedge F^{K}  \,.\label{BianchiG} 
\end{equation}
where $| J |$ means that the index $J$ is not involved in the skew-symmetrization bracket $[\dots]$.

One can  easily verify that
\begin{equation}
Q^{IJ}   \, G_{I\, \mu \rho  \sigma} \,{G_{J}}^{\nu \rho  \sigma}     ~=~  Q_{IJ} \, \big( 2\, F^{I}_{\, \mu  \rho} \,    F^{J\,  \nu  \rho}   -   \delta_\mu^\nu \, F^{I}_{ \, \rho  \sigma}   \,   F^{J \,\rho  \sigma}    \big)      \label{GGFF} 
\end{equation}
and so we may rewrite the Einstein equation (\ref{Einstein2}) as
\begin{equation}
R_{\mu \nu} ~=~  Q_{IJ}\,\Big[  \coeff{2}{3}\, F^{I}_{\, \mu \rho}  \, {{F^J}_\nu \,}^\rho  +  \partial_\mu X^I \,  \partial_\nu X^J  
 \Big] ~+~  \coeff{1}{6}\, Q^{IJ} \,  G_{I\, \mu \rho  \sigma} \,{G_{J\, \nu}}^{ \rho  \sigma}       \,.  \label{Einstein3} 
\end{equation}

Since I am assuming that the matter is time independent, this means that the Lie derivatives of all the fields along  $K^\mu$ must vanish: 
\begin{equation}
{\cal L}_K  F^I ~=~0 \,, \qquad   {\cal L}_K  G_I ~=~0 \,, \qquad {\cal L}_K  X^I ~=~0  \,,   \label{invariances1}  
\end{equation}
Cartan's formula states that for a $p$-form, $\alpha$, one has
\begin{equation}
{\cal L}_K  \alpha ~=~d (i_K( \alpha) ) ~+~  i_K(d \alpha)  \,.  \label{Cartan}  
\end{equation}
Taking $ \alpha  =  F^I $ one has, locally, 
\begin{equation}
 K^\rho  F^I_{ \rho \mu} ~=~  \partial_\mu  \lambda^I  \,, \label{KdotF}  
\end{equation}
for some functions $ \lambda^I$.

If the space-time manifold were not simply connected one could, in principle, encounter jumps in value of $\lambda^I$ if one were to integrate (\ref{KdotF}) around a closed curve.  To avoid this issue I will, from now on, assume that our space-time manifold is simply connected.  With this assumption, the arbitrary constants in the definitions of the functions, $\lambda^I$, may be fixed by requiring that the $\lambda^I$ vanish at infinity.  Physically, the functions, $\lambda^I$, are electrostatic potentials of the $2$-forms, $F^I$.   

Taking $ \alpha  =  G_I$ one has 
\begin{align}
d(i_K (G_I)) ~=~ &   -i_K (d G_I) ~=~   -\coeff{1}{4}  \, C_{ILM} \, i_K (  F^{L}\wedge F^{M})   ~=~   -\coeff{1}{2}  \, C_{ILM} \,  d \lambda^L \wedge  F^{M}  \nonumber  \\ ~=~ &   -\coeff{1}{2}  \,C_{ILM} \,  d( \lambda^L \, F^{M} )   \label{dKdotG}  
\end{align}
where I have used (\ref{BianchiG}) and  (\ref{KdotF}). The assumption of simple connectivity is a weak one because we could pass to a covering space. However, one cannot assume that the $H^2(\cM_5)$ is trivial. Indeed, {\it this is the crucial issue that makes solitons possible}.  For the moment, however, I will assume that $H^2(\cM_5)$ is trivial and hence
\begin{equation}
K^\rho  G_{I\,  \rho \mu \nu}  ~+~ \coeff{1}{2} \, C_{IJK}  \, \lambda^J  F^{K}_{\mu  \nu} ~=~   \partial_\mu  \Lambda_{I\, \nu} - \partial_\nu  \Lambda_{I\, \mu} \,,   \label{KdotG}  
\end{equation}
where $\Lambda_I$ are globally defined one-forms. 

Using (\ref{KdotF}) and (\ref{KdotG}) one finds that:
\begin{align}
 K^\mu \big( Q_{IJ}\,   F^{I}_{\, \mu \rho}  \, {{F^J}_\nu \,}^\rho  \big) ~=~&   - \nabla_\rho \big(  Q_{IJ}\, \lambda^I \, {F^J}^{ \rho\nu}  \big) ~+~   \coeff{1}{16} \, C_{IJK}  \,  \epsilon^{\nu \alpha \beta \gamma \delta} \, \lambda^I \, {F^J}_{  \alpha \beta} \,{F^K}_{\gamma \delta} \\
  K^\mu \big( Q^{IJ}\, G_{I \, \mu  \rho \sigma}  \, {G_J}^{\, \nu  \rho \sigma}  \big) ~=~&  - 2\, \nabla_\rho \big(  Q^{IJ}\, \Lambda_{I \, \sigma} \, {G_J}^{\,   \rho \nu \sigma} \big) ~-~   \coeff{1}{4} \, C_{IJK}  \,  \epsilon^{\nu \alpha \beta \gamma \delta} \, \lambda^I \, {F^J}_{  \alpha \beta} \,{F^K}_{\gamma \delta}
\end{align} 
and hence,  Einstein's equations (\ref{Einstein3}) imply:
\begin{equation}
 K^\mu R_{\mu \nu} ~=~   - \coeff{1}{3}\, \nabla^\mu \, \big[ \, 2\,  Q_{IJ}\, \lambda^I \, {F^J}_{ \mu\nu}  
 ~+~   Q^{IJ}\, {\Lambda_{I }}^{\sigma} \, {G_J}_{\,   \mu \nu \sigma} \, \big] \,,  \label{KdotR} 
\end{equation}
where I have used $K^\mu \partial_\mu X^I =  \cL_K X^I  =0$.  Note that the $\lambda (* F\wedge F)$  terms  have canceled in (\ref{KdotR}).  

The whole point is that the mass of the solution is given by 
\begin{equation}
M~=~  {\rm const.} \,  \,\int_{\Sigma} \, R_{\mu \nu}  K^\mu n^\nu \, d \Sigma ~=~  {\rm const.} \,  \,\int_{\Sigma} \,  \nabla_\mu  \cX^\mu \, d \Sigma\,,
\label{Mass1}
\end{equation}
where $\cX^\mu$ can be read off from (\ref{KdotR}).  This means that the mass is given by a pure boundary term.  Moreover,  all the fields $\lambda^I$, ${F^J}$ and $G_J$ fall off at infinity too fast for there to be a finite boundary term and thus one finds that 
\begin{equation}
M~=~ 0\,.
\label{Massvan}
\end{equation}

Now consider the volume integral over $\Sigma$ in (\ref{Mass1}).  Since $M=0$, one now has 
\begin{equation}
 0 ~=~   \,\int_{\Sigma} \, R_{0 0}   d \Sigma  \,.
\label{zeroint}
\end{equation}
Now remember that the metric on $\Sigma$ is positive definite.  This means that not only is the measure in  (\ref{zeroint}) positive definite, but that all the terms in the $00$-component of (\ref{Einstein3}) are positive definite:
\begin{equation}
R_{0 0} ~=~  Q_{IJ}\,\Big[  \coeff{2}{3}\, F^{I}_{\, 0 \rho}  \, {{F^J}_0 \,}^\rho   
 \Big] ~+~  \coeff{1}{6}\, Q^{IJ} \,  G_{I\, 0 \rho  \sigma} \,{G_{J\, 0}}^{ \rho  \sigma}   ~\ge~ 0    \,.  \label{Einstein3a} 
\end{equation}
(In writing this I have dropped all the terms involving $\partial_0 X^I = \partial_t X^I =0$ because $X^I$ is time-independent.)

As a result, one learns that 
\begin{equation}
G_{I\, 0 \rho  \sigma}  ~=~  0 \,, \qquad F^{I}_{\, 0 \rho}  ~=~  0   \,.  \label{0compsvan} 
\end{equation}
Since $G_I$ is the dual of $F^I$, the first equation means that $F^I_{ij} =0$ and hence
\begin{equation}
 F^{I}_{\mu \nu}  ~\equiv~  0   \,.  \label{Fzero} 
\end{equation}
Finally,  (\ref{5daction}) reveals that only the $F^{I}$ source the scalars, and since the $F^{I}$ vanish, one has 
\begin{equation}
 \partial_\mu \Big[\sqrt{|g|} \, g^{\mu \nu}\, Q_{IJ}\, \partial_\nu X^I \, \Big] ~=~0  \,. 
  \label{scaleqn}
\end{equation}
Since the fields are time independent, this is a negative definite scalar Laplacian on $\Sigma$.  The only non-singular solutions that go to a constant at infinity are thus the solutions with  
\begin{equation}
X^I  ~=~  {\rm const.}   \label{Xconst} 
\end{equation}
One therefore concludes that all the dynamical fields are trivial and the space-time is simply Minkowski space.

\bigskip
%%%%%%%%%%%%%%%%%%%
\leftline{\bf Comments:}
%%%%%%%%%%%%%%%%%%%
\noindent (i) The basic structure of all these theorems is to argue that if $\Sigma$ is smooth, then the Komar mass density is always a total derivative and hence the mass vanishes. One can then invoke positive-mass theorems that tell us that if the matter satisfies the dominant energy condition and the space-time is asymptotic to Minkowski space and has $M=0$  then it can only be a global Minkowski space.  Rather than invoking the sledge-hammer of positive-mass theorems, I established the result by  using the details of the equations of motion. 

\noindent (ii) The ``No solitons'' theorem also depends upon $\Sigma$ being smooth, with a positive definite induced metric.  If there are  event horizons then the hypersurface becomes null at the horizons and singularities form in the interior of the horizons.  One can repair the ``no solitons'' theorem by excising the horizons and introducing interior boundaries on the horizons.  The integral for $M$ then gets boundary contributions from these horizons.  When this is unpacked one gets an expression for M (the internal energy of the system)  in terms of the ``classical thermodynamic variables'' for each black hole: the horizon area (entropy), surface gravity (temperature), angular momentum, angular velocity of the horizon, charge, electrostatic potential ...  .  For more details, see \cite{Carter:1973rla,Carter2}.  This was believed to be the only way to get solitons and hence the original mantra of ``No solitons without horizons.'' Our purpose here is to obtain smooth, horizonless solitons and so the only boundary of $\Sigma$ is at infinity.

%%%%%%%%%%%%%%%%%%%
\subsection{Supporting mass with topology}
%%%%%%%%%%%%%%%%%%%

From the way I presented the theorem, it is pretty evident how one gets around ``No solitons without horizons:'' There can only be massive solitons when $H^2(\cM_5)$ is non-trivial.

Indeed the correct form of (\ref{KdotG}) is 
\begin{equation}
K^\rho  G_{I\,  \rho \mu \nu}  ~+~ \coeff{1}{2} \, C_{IJK}  \, \lambda^J  F^{K}_{\mu  \nu}~=~   \partial_\mu  \Lambda_{I\, \nu} - \partial_\nu  \Lambda_{I\, \mu}  ~+~ H_{I \, \mu \nu}\,,   \label{KdotGcorr}  
\end{equation}
where $\Lambda_I$ are globally defined one-forms and $H_I$ are closed but not exact two forms. That is, one cannot write $H_I = d \nu_I$ where $\nu_I$ are globally well-defined one-forms.

This then means that (\ref{KdotR}) has an extra term:
\begin{equation}
 K^\mu R_{\mu \nu} ~=~   - \coeff{1}{3}\, \nabla^\mu \, \big[ \, 2\,  Q_{IJ}\, \lambda^I \, {F^J}_{ \mu\nu}  
 ~+~   Q^{IJ}\, {\Lambda_{I }}^{\sigma} \, {G_J}_{\,   \mu \nu \sigma} \, \big] ~+~ \coeff{1}{6}\,  Q^{IJ}\, H_I^{\rho \sigma} \, {G_J}_{\rho \sigma \nu } \,,  \label{KdotRcorr} 
\end{equation}
The last term in  (\ref{KdotRcorr}) may be expressed as 
\begin{equation}
\coeff{1}{6}\,  Q^{IJ}\, H_I^{\rho \sigma} \, {G_J}_{\rho \sigma \nu }  ~=~  \coeff {1}{12} \, \epsilon_{\alpha \beta \rho \sigma  \nu  }   F^{I\, \alpha \beta}\,H_I^{\rho \sigma}       \,.  \label{HdotG} 
\end{equation}

This means that rather than finding $M=0$, one finds that $M$ can be supported by a topological integral:
\begin{align}
M  ~=~    & \frac{1}{32 \pi G_5} \,  \int _{ \Sigma}   \Big[  Q^{IJ}\, H_{I \rho \sigma} \, {G_J}^{\rho \sigma \nu }\Big]  \, d \Sigma_{\nu}~=~     \frac{1}{64 \pi G_5} \,  \int _{ \Sigma}    \epsilon^{\alpha \beta \rho \sigma  \nu  }   F^I_{\alpha \beta}\,H_{I\, \rho \sigma}  \, d \Sigma_{\nu}  \nonumber \\
   ~=~    &  \frac{1}{16 \pi G_5} \,  \int _{ \Sigma}    F^I \wedge  H_I   \label{Manswer} \,.
\end{align}
%
 
%%%%%%%%%%%%%%%%%%%
\subsection{The BPS equations}
\label{ss:BPSeqns}
%%%%%%%%%%%%%%%%%%%

Before leaving the discussion of this particular supergravity theory, I will summarize the BPS equations that arise from requiring supersymmetry.  The first detailed analysis of the BPS equations was done in  \cite{Gauntlett:2002nw}, however this first work was incomplete in that it missed a major, and essential simplification that was subsequently discovered in \cite{Bena:2004de,Bena:2005va}.  

Supersymmetry necessarily makes the metric stationary and the other fields time independent. One can therefore  take the metric to have the form (\ref{metform}).  One can similarly decompose the Maxwell fields according to (\ref{Aform}) and use the definition (\ref{Thetadefn}).

If one seeks the solutions that possess four supersymmetries,  one first finds that the scalars and  warp factors are directly related to the electrostatic potentials:  
\begin{equation}
Z ~\equiv~ \big( Z_1 \, Z_2 \, Z_3  \big)^{1/3}\,,\quad    X^1    =\bigg( \frac{Z_2 \, Z_3}{Z_1^2} \bigg)^{1/3} \,, \quad X^2    = \bigg( \frac{Z_1 \, Z_3}{Z_2^2} \bigg)^{1/3} \,,\quad X^3   =\bigg( \frac{Z_1 \, Z_2}{Z_3^2} \bigg)^{1/3}  \,.
\label{XZrelns}
\end{equation}
This is, once again, a BPS constraint and the expression for $Z$ means that the mass, $M$, is given by the sum of the electric charges $M =Q_1 + Q_2 +Q_3$.

Requiring four supersymmetries also imposes that the  metric, $ds_4^2$, on the base, $\cB$,   be hyper-K\"ahler.  Finally,  the complete system of BPS equations and field equations can then be reduced to solving the following system \cite{Bena:2004de,Bena:2005va}:
\begin{eqnarray}
 \Theta^{(I)}  &~=~&  \star_4 \, \Theta^{(I)} \label{BPSeqn:1} \,, \\
 \nabla^2  Z_I &~=~&  \coeff{1}{2} \, C_{IJK} \, \star_4 (\Theta^{(J)} \wedge
\Theta^{(K)})  \label{BPSeqn:2} \,, \\
 dk ~+~  \star_4 dk &~=~&  Z_I \,  \Theta^{(I)}\,,
\label{BPSeqn:3}
\end{eqnarray}
where $\star_4$ is the Hodge dual taken with respect to the  four-dimensional metric, $ds_4^2$ and $\nabla$ is the covariant derivative in this metric.  Solving this system on a hyper-K\"ahler base, $\cB$, yields the most general solutions to the supergravity action with four supersymmetries.

While we will only need the foregoing details, it is interesting to note how this comes about in solving the supersymmetry conditions.  In particular, the solution requires the supersymmetries to satisfy: 
\begin{equation}
 \Theta^{(I)}_{a b}\, \gamma^a \,  \gamma^b \,  \epsilon ~=~ 0 \,, \qquad \hat R_{ \alpha \beta ab}\gamma^a \,  \gamma^b  \epsilon  ~=~ 0  \,, 
\label{proj0}
\end{equation}
where where $a,b,c, d ... $  and  $\alpha,\beta  ...$ are, respectively, frame and  tangent-space indices on $\cB$ and $\hat R_{ \alpha \beta a b}$ is the Riemann tensor on $\cB$.

One can solve these equations by imposing the projection condition: 
\begin{equation}
\gamma^1 \,  \gamma^2\, \gamma^3 \, \gamma^4 \, \epsilon ~=~ \epsilon \quad \Leftrightarrow \quad  \gamma^a \,  \gamma^b \, \epsilon ~=~ - \coeff{1}{2}\, \epsilon^{abcd} \,\gamma^c \,  \gamma^d \,  \epsilon\,,
\label{proj1}
\end{equation}
and imposing (\ref{BPSeqn:1}) and
\begin{equation} 
\hat R_{a b}{}^{c d} ~=~ \coeff{1}{2}\, \epsilon^{cdef} \, \hat R_{ a b e f }\,.
\label{sdual1}
\end{equation}
Note that $\gamma^1 \,  \gamma^2\, \gamma^3 \, \gamma^4$ is traceless and satisfies:
\begin{equation}
(\gamma^1 \,  \gamma^2\, \gamma^3 \, \gamma^4)^2 ~=~  \oneone 
\label{gammaprops}
\end{equation}
Thus (\ref{proj1}) reduces the supersymmetries by half:  The solutions of  (\ref{BPSeqn:1})--(\ref{BPSeqn:3}) thus lead to BPS solutions with $4$ supersymmetries (we started with $\cN=2$  supersymmetry, which, in five dimensions, means eight real supercharges).

Let $\hat \nabla_\alpha$ be  the covariant derivative on $\cB$.   For any spinor one has:
\begin{equation} 
\big[\,\hat \nabla_\alpha \,, \hat \nabla_\beta \,\big]  \,    \epsilon ~=~ \coeff{1}{4}\, \hat R_{ \alpha \beta \gamma \delta}\gamma^\gamma \,  \gamma^\delta  \epsilon \,.
\label{susy02}
\end{equation}
and so, for the spinors satisfying (\ref{proj1}), and hence (\ref{proj0}), the right-hand side vanishes.  This is the integrability equation condition for 
\begin{equation} 
\hat \nabla_\alpha  \,    \epsilon  ~=~ 0 \,,
\label{susy01}
\end{equation}
and hence the Killing spinors are covariant constant on $\cB$.

Recall that the Riemman curvature measures the monodromy for parallel transport around closed loops, and on the $4$-manifold, $\cB$, the generic monodromy is $SO(4) = (SU(2) \times SU(2))/\ZZ_2$.  The self-duality constraint on the curvature means that the metric, $ds_4^2$, must be ``half-flat,'' that is, have trivial monodromy on one $SU(2)$ factor. This allows the solution of (\ref{susy01}) on the spin bundle with trivial monodromy.  If there is trivial monodromy on both $SU(2)$ factors then the Riemann tensor vanishes and the manifold is flat.  More generally, one can allow non-trivial monodromy in one $SU(2)$ factor and such metrics are hyper-K\"ahler.  

There are many things to note about the BPS system.  First observe that it is, in fact, a linear system of equations.  (It was this fact was not discovered in \cite{Gauntlett:2002nw} but was shown with some more deconstruction in \cite{Bena:2004de,Bena:2005va}.). The form of the BPS equations (\ref{BPSeqn:1})--(\ref{BPSeqn:3}) descends from the non-linear form of the Maxwell equations, $d*F \sim F\wedge F$.  However,  (\ref{BPSeqn:1}) is linear and homogeneous; while  (\ref{BPSeqn:2}) and (\ref{BPSeqn:3}) are linear but with sources that are quadratic in the solutions to the preceding equations.  The entire system is basically some variant of four-dimensional, Euclidean electromagnetism.

Next observe that equations (\ref{BPSeqn:1}) and (\ref{BPSeqn:1}) are first order while (\ref{BPSeqn:2}) is second order.  The former come from imposing the supersymmetry conditions while the latter is actually one of the equations of motion.

Observe that one can take $ \Theta^{(I)}  \equiv 0$ and then one has $ \nabla^2  Z_I  =0$, which means that the $Z_I$ are harmonic.  Similarly, $k$, then reduces to a harmonic vector field.  If these are to fall off at infinity, then one must have singular sources on $\cB$ and this will lead to multi-black-hole solutions.   The choice of harmonic $k$ means that one can add angular momentum to these five-dimensional black holes while still preserving supersymmetry.  There are thus no solitons if $\Theta^{(I)}  \equiv 0$.

The magnetic fluxes, by definition (see (\ref{Thetadefn})), satisfy $d \Theta^{(I)}  = 0$  and so (\ref{BPSeqn:1}) implies $d   \star_4 \Theta^{(I)}  = 0$.  Thus the $\Theta^{(I)}$ are harmonic, and if they are to be smooth, they must be cohomological, that is belong to $H^2(\cB)$.   It is then through  (\ref{BPSeqn:2}) that they contribute quadratically to the $Z_I$, and then through (\ref{XZrelns}) to $Z$ and hence to $g_{00}$ and the mass, $M$.  One thus sees, very explicitly how  (\ref{Manswer}) is being implemented in the BPS equations.  Conversely, one sees that if $H^2(\cB)$ is trivial, then there are no smooth magnetic sources, which means $\Theta^{(I)}  \equiv 0$, which leads back to either multi-black-holes or empty space.

One also sees,  in  (\ref{BPSeqn:2}), how the Chern-Simons interaction in $d*F \sim F\wedge F$ enables a pair of magnetic fluxes to combine to source an electric charge.  Thus the BPS solutions based on smooth cohomological fluxes do not have any point-source electric charges.  Instead, the electric charge sources are smoothly distributed into cohomological fluxes.  

The last BPS equation, (\ref{BPSeqn:3}), also has a very interesting physical meaning.  The vector $k$ contains the information about the angular momentum of the solution (see, (\ref{asympg3})).  The source is the product of electric potentials, $Z_I$, and magnetic fluxes, $\Theta^{(I)}$ and so should be thought of as analogues of $\vec E \times \vec B$ in electrodynamics.  Thus the angular momentum is generated by a three-way interaction: two smooth magnetic fields, $\Theta^{(J)}$ and $\Theta^{(K)}$ creating a smooth electrostatic field, $Z_I$, which in turn creates angular momentum when it interacts with the third smooth magnetic field,  $\Theta^{(I)}$.  The mechanism is the same as the $\vec E \times \vec B$ interaction that generates angular momentum for an electron in the presence of a magnetic monopole.

To conclude, it is important to note that I may have just built a ``castle in the air'' in that it appears that all this beautiful BPS structure cannot lead to smooth solitonic solutions. 

The starting point of BPS solutions is to choose a hyper-K\"ahler metric, $ds_4^2$, on $\cB$.  To obtain a solution that is asymptotic, at infinity, to Minkowski space, one must therefore require  $ds_4^2$ to be asymptotic to the flat metric on $\IR^4$.  In the early 1980's there was a goal to classify all Riemannian metrics in four dimensions with self-dual curvature:  these metrics were known as gravitational instantons.  One of the by-products of the proof of the positive mass theorems  \cite{Schon:1979rg,Schon:1979uj,Schon:1981vd,Witten:1981mf} was to establish that there were no asymptotically Euclidean gravitational instantons.  That is, the only smooth, hyper-K\"ahler, Riemannian metric that is asymptotic to $\IR^4$ is $\IR^4$ itself with its flat metric.  Thus there can be no topology and seemingly no solitons .... 

However, we will see that the mathematical universe has much richer possibilities ... and even in the face of such a discouraging theorem, there are indeed solitons.  Like many ``no go'' theorems, one can evade them by weakening some of the assumptions.

%%%%%%%%%%%%%%%%%%%%%%%%%%%%%%%%%%%%%
\newpage
\section{Microstate geometries in five dimensions}
\label{Sect:GG5}
%%%%%%%%%%%%%%%%%%%%%%%%%%%%%%%%%%%%%

We need non-trivial hyper-K\"ahler metrics the four-dimensional base manifold, $\cB$, and to that end we will start with one of the most useful and explicit examples of such metrics in four dimensions: the Gibbons-Hawking ALE metrics \cite{Gibbons:1979zt}.  These were created as part of the gravitational instanton program and they are not ruled out by the theorems of Schoen, Yau and Witten because, while they are flat at infinity, they are not Euclidean at infinity.  Indeed, as we will discuss, the asymptotic structure is $\IR^4/\ZZ_n$.  

My presentation here will draw heavily on the review article \cite{Bena:2007kg}.

%%%%%%%%%%%%%%%%%%%
\subsection{Gibbons-Hawking metrics}
%%%%%%%%%%%%%%%%%%%

 Gibbons-Hawking (GH) ALE spaces are non-trvial $U(1)$ fibrations over a flat $\IR^3$ base:
\begin{equation}
ds_4^2 ~=~ V^{-1} \, \big( d\psi + \vec{A} \cdot d\vec{y}\big)^2  ~+~ V\, (dy_1^2 +  dy_2^2 +   dy_3^2) \,,
\label{GHmetric}
\end{equation}
where $V$ is harmonic on the flat $\IR^3$: 
\begin{equation}
\nabla^2 V  ~=~ 0\,.
\label{Vharm}
\end{equation}
while the connection, $A = \vec A \cdot d\vec{y} $, is related to $V$ via
\begin{equation}
\vec \nabla \times \vec A ~=~ \vec \nabla V \,.
\label{AVreln}
\end{equation}

The scaling transformation: $V \to \lambda^2 V$, $A \to \lambda^2 A$, $y_i \to \lambda^{-1} y_i$ and $\psi \to \lambda  \psi$   preserves (\ref{GHmetric})--(\ref{AVreln}).  We will fix this choice of scaling by fixing the period of the $\psi$ coordinate:
\begin{equation}
\psi ~\equiv~ \psi ~+~ 4\,\pi \,.
\label{psiperiod}
\end{equation}

This family of metrics is the unique  class of four-dimensional hyper-K\"ahler metrics with a tri-holomorphic $U(1)$ isometry\footnote{Tri-holomorphic means that the $U(1)$ isometry preserves all three complex structures of the hyper-K\"ahler metric.}.  Moreover,  a four-dimensional hyper-K\"ahler manifold with a $U(1)\times U(1)$ symmetry must, at least locally, have the Gibbons-Hawking form with an extra $U(1)$ symmetry around an axis in the $\IR^3$ \cite{Gibbons:1987sp}.  

Perhaps, more usefully,  these GH spaces have a very explicit and easily analyzed family of hyper-K\"ahler metrics.  The standard form is to take $V$ to be sourced at discrete points, $\vec{y}^{(j)}$,  in the $\IR^3$:
\begin{equation}
 V ~=~\varepsilon_0 ~+~ \sum_{j=1}^N \,  \frac{q_j}{r_j} \,, \qquad r_j ~\equiv~ |\vec{y}-\vec{y}^{(j)}| \,.
\label{Vform}
\end{equation}
For the metric to be Riemannian (positive definite), one must take $\varepsilon_0 \ge 0$ and $q_j \ge 0$. 

%%%%%%%%%%%%%%%%%%
\exbox{Compute the curvature on this GH metric and show that it satisfies (\ref{sdual1}). (The sign will only work out correctly if you choose the correct orientation on the manifold.)}
%%%%%%%%%%%%%%%%%%

To determine $\vec A$, we need  the vector fields, $\vec v_i$, that satisfy:
\begin{equation}
\vec \nabla \times \vec v_{i} ~=~  \vec \nabla\, \bigg( {1\over r_i} \bigg) \,.
\label{vieqn}
\end{equation}
Choose coordinates, $\vec y ~=~ (x,y,z)$, so that $\vec y^{(i)} ~=~ (0,0,a)$  and let $\phi$ denote the polar angle in the  $(x,y)$-plane, then:
\begin{equation}
\vec v_{i} \cdot d \vec y~=~ \Big( {(z -a) \over r_i} ~+~ c_i \Big) \, d \phi \,,
\label{vzdefns}
\end{equation}
where $c_i$ is a constant.  The  vector field, $\vec v_i$, is regular away from the $z$-axis, but has a Dirac string along the $z$-axis.  By choosing $c_i$ we can cancel the string along the positive or negative $z$-axis, and by moving the axis we can arrange these strings  to run in any  direction we choose, but they must start or finish at some $\vec y^{(i)}$, or run out to infinity.

There appear to be singularities in the metric at $r_j =0$. However, if one changes to polar coordinates, $(\hat r, \hat \theta, \hat \phi)$, centered at $r_j =0$  the metric limits to the form:
\begin{align}
ds_4^2 ~\sim~&  q_j^{-1} \, \hat r \, \big( d\psi + q_j  \cos \hat \theta \, d\hat \phi \big)^2  ~+~q_j \, \hat r^{-1} \,   (d \hat r^2 ~+~ \hat r^2 \,  d\hat \theta^2 ~+~ \hat r^2\, \sin^2 \hat \theta  \,d\hat \phi^2) \nonumber \\
~=~&  q_j \,\Big[ \hat r \, \big( q_j^{-1} d\psi +   \cos \hat \theta \, d\hat \phi \big)^2  ~+~ \hat r^{-1} \,   (d \hat r^2 ~+~ \hat r^2 \,  d\hat \theta^2 ~+~ \hat r^2\, \sin^2 \hat \theta d\hat \phi^2) \Big] \,,  
\label{GHmetricpt}
\end{align}
where we have used the fact that near $r_j =0$ one has $V \sim \frac{q_j}{r_j} + const.$ for which the solution to (\ref{AVreln})  gives:
\begin{equation}
 A ~=~q_j  \cos \hat \theta \, d\hat \phi   \,.
\label{Apt}
\end{equation}
Now choose a new radial coordinate
\begin{equation}
\rho ~=~ 2\,  \sqrt{\hat r} ~=~ 2\, \sqrt{|\vec{y}-\vec{y}^{(j)}|}   \,.
\label{newradcoord}
\end{equation}
then the metric is locally of the form:
\begin{equation}
ds_4^2 ~\sim~ q_j \,\big( d \rho^2 ~+~ \rho^2  \, d \Omega_{3, q_j}^2 \big) \,,
\label{asympmet}
\end{equation}
where 
\begin{equation}
d \Omega_{3, q_j}^2 ~\equiv~ \big( q_j^{-1} d\psi +   \cos \hat \theta \, d\hat \phi \big)^2 ~+~   d\hat\theta^2 ~+~ \sin^2 \hat \theta \,d\hat \phi^2   \,.
\label{Lensmet}
\end{equation}
Define $\chi =  q_j^{-1} d\psi$ and observe that 
\begin{equation}
d \Omega_{3}^2 ~\equiv~ \big( d \chi +   \cos  \theta \, d \phi \big)^2 ~+~   d\theta^2 ~+~ \sin^2  \theta \,d \phi^2  
\label{S3met}
\end{equation}
is the metric on the $S^3$ defined by $|\zeta_1|^2 + |\zeta_2|^2 = 1$ in $\IC^2$ with
\begin{equation}
\zeta_1~=~ e^{\frac{i}{2}( \chi - \phi)} \cos \frac{\theta}{2}  \,, \qquad   \zeta_2~=~ e^{\frac{i}{2}( \chi + \phi)} \sin \frac{\theta}{2} \,.
\label{cplxcoords}
\end{equation}
To fully cover the sphere one must have:
\begin{equation}
0 \le \chi \le 4 \, \pi\,, \qquad 0 \le \theta \le \pi  \,, \qquad  0 \le \phi \le 2 \,\pi  \,.
\label{Ranges}
\end{equation}
However,  (\ref{Lensmet}) and   (\ref{psiperiod}) imply that a precise matching to (\ref{S3met}) requires
\begin{equation}
\chi ~\equiv~ q_j^{-1} \psi  ~\equiv~ q_j^{-1}( \psi + 4\, \pi)  ~\equiv~ \chi ~+~ \frac{ 4\, \pi}{q_j}   \,.
\label{Lensperiod}
\end{equation}
 Thus the metric (\ref{Lensmet})  is  that of $S^3/\ZZ_{q_j}$ in which the quotient is taken on the Hopf fiber, $\chi$, according to (\ref{Lensperiod}).
 
 %%%%%%%%%%%%%%%%%%
\bigskip \bigskip
\centerline{\exboxx{ Check the claims underlying (\ref{S3met})--(\ref{Ranges}).}}
%%%%%%%%%%%%%%%%%%

For such a quotient to be well-defined, one must  have:
\begin{equation}
q_j ~\in~ \ZZ  \,.
\label{qjint}
\end{equation}
If $q_j = \pm 1$ then there are no identifications on the $S^3$  and the metric in region around $r_j \to 0$ reduces to that of flat $\IR^4$.   If the $|q_j | > 1$ then $r_j = 0$ is an orbifold point. Such singularities are nicely resolved in string theory and thus acceptable singularities of manifolds. 

Also note that at infinity one has: 
\begin{equation}
V ~\sim~\varepsilon_0 ~+~ \frac{q_0}{r} \,, \qquad  q_0 ~\equiv~  \sum_{j=1}^N \, q_j \,.
\label{Vinf}
\end{equation}
If $\varepsilon_0 \ne 0$ then the metric is asymptotic to $\IR^3 \times S^1$ and  the GH metrics are known as multi-Taub-NUT.  If $\varepsilon_0 = 0$ then the metric is asymptotic to $\IR^4/\ZZ_{q_0}$ where the $\ZZ_{q_0}$ is modded out of the $S^3$ at infinity exactly as in (\ref{Lensmet}).  If $|q_0| =1$ then the metric is asymptotic to  $\IR^4$ and if $|q_0| > 1$ then the non-trivial identifications at infinity mean that it is an ALE (Asymptotically Locally Euclidean) space.  

Henceforth, (largely for simplicity of the asymptotics)  we take
\begin{equation}
\varepsilon_0 ~=~ 0\,.
\label{varepsilonzero}
\end{equation}

One should also remember that for the metric to be Riemannian then one must  $q_j  \in \ZZ_+ \cup \{0\}$.   This means that if you want the  metric to be asymptotic to  $\IR^4$ then $q_0 \equiv \sum_{j=1}^N \, q_j  =1$ and hence  one must have $q_i =1$  and $q_j =0, j \ne i$  for some $i$.  Then $V = \frac{1}{r_i}$ and the space is globally $\IR^4$.  Thus the only GH metric that is Riemannian and asymptotic to flat $\IR^4$  must be  flat $\IR^4$  everywhere.  So it seems we have to work with ALE spaces.  

%%%%%%%%%%%%%%%%%%%
\subsection{Harmonic forms on GH metrics}
\label{GHtopology}
%%%%%%%%%%%%%%%%%%%

Our goal is to solve the BPS equations and we start with the first layer,  (\ref{BPSeqn:1}), and seek harmonic magnetic fluxes on a GH space.  We start by identifying the dual homology cycles.

\begin{figure}
\leftline{\hskip 1.8cm \includegraphics[width=5in]{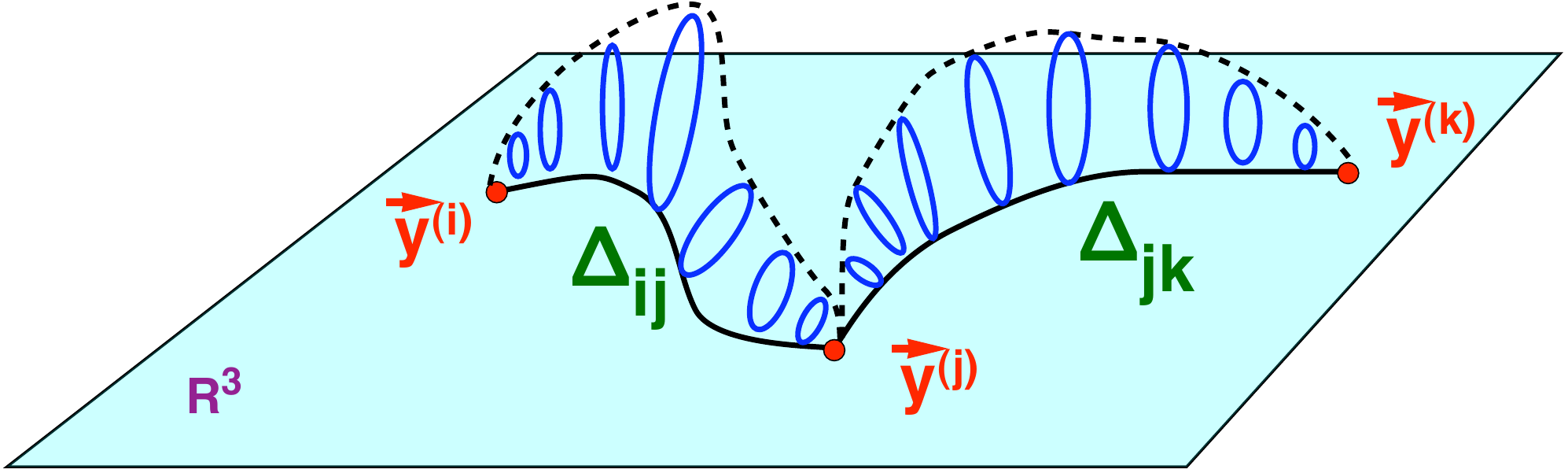}}
\setlength{\unitlength}{0.1\columnwidth}
\caption{\it This figure depicts some non-trivial cycles of the Gibbons-Hawking geometry.  The behaviour of the $U(1)$ fiber is shown along curves between the sources of
the potential, $V$. Here the fibers sweep out a pair of intersecting homology spheres.}
\label{GHcycles}
\end{figure}

 A generic GH metric has ${1 \over  2} N(N-1)$ topologically non-trivial two-cycles, $\Delta_{ij}$, that. run between the GH centers.  These two-cycles can be defined by taking any curve, $\gamma_{ij}$, between $\vec{y}^{(i)}$ and $\vec{y}^{(j)}$ and considering the $\psi$-fiber of (\ref{GHmetric}) along the curve. Because of the factor of $V^{-1}$ in (\ref{GHmetric}), the fiber collapses to zero at the GH centers, and so the curve and the fiber sweep out a $2$-sphere (up to $\ZZ_{|q_j|}$ orbifolds).  See Fig. \ref{GHcycles}.  These spheres intersect one another at the
common points $\vec{y}^{(j)}$.  There are $(N-1)$ linearly independent homology two-spheres, and the set $\Delta_{i\, (i+1)}$ represents a basis\footnote{If one has $q_j =1$ at every GH center, then the integer homology corresponds to the root lattice of $SU(N)$ with an intersection matrix given by the inner product of the roots.}.

%%%%%%%%%%%%%%   Figure 2  %%%%%%%%%%%%%%%
%%%%%%%%%%%%%%   End Figure 2   %%%%%%%%%%%%%%%

To define the dual cohomology, it is convenient to introduce a set of frames
\begin{equation}
\hat e^1~=~ V^{-{1\over 2}}\, (d\psi ~+~ A) \,,
\qquad \hat e^{a+1} ~=~ V^{1\over 2}\, dy^a \,, \quad a=1,2,3 \,.
\label{GHframes}
\end{equation}
and two associated sets of two-forms:
\begin{equation}
\Omega_\pm^{(a)} ~\equiv~ \hat e^1  \wedge \hat
e^{a+1} ~\pm~ \coeff{1}{2}\, \epsilon_{abc}\,\hat e^{b+1}  \wedge
\hat e^{c+1} \,, \qquad a =1,2,3\,.\
\label{twoforms}
\end{equation}
The two-forms, $\Omega_-^{(a)}$, are anti-self-dual,  harmonic and
non-normalizable  and they define the
hyper-K\"ahler  structure on the base.  The forms, $\Omega_+^{(a)}$, are
self-dual and can be used to construct harmonic fluxes that are dual to the
two-cycles.  Consider  the self-dual two-forms:
\begin{equation}
\Theta^{(I)}~  \equiv~ - \sum_{a=1}^3 \, \big(\partial_a \big( V^{-1}\, K^I \big)\big) \,
\Omega_+^{(a)} \,.
\label{harmtwoform}
\end{equation}
Then $\Theta^{(I)}$ is closed (and hence co-closed and harmonic) if and only if $K^I$ is harmonic in $\IR^3$, {\it i.e.}  $\nabla^2 K^I =0$. 

 %%%%%%%%%%%%%%%%%%
\exbox{ Compute the spin connection for the GH metric using the frames  (\ref{GHframes}) and solve the equation (\ref{susy01}) using the projection condition  (\ref{proj1}).}
%%%%%%%%%%%%%%%%%%

We now have the choice of how to distribute sources of $K^I$ throughout the $\IR^3$ base of the GH space.  One can make all sorts of black objects by allowing singular sources, but we want smooth, cohomological fluxes. Indeed, the $\Theta^{(I)}$  will be smooth if and only if $K^I/V$ is smooth; this occurs if and only if
$K^I$ has the form:
\begin{equation}
K^I ~ =~ k^I_0 ~+~ \sum_{j=1}^N \,  {k^I_j \over r_j} \,.
\label{KIdefn}
\end{equation}
For some constants $k^I_0$ amd $k^I_j$.  Also note that the ``gauge transformation:''
\begin{equation}
K^I ~ \to~ K^I  ~+~ c^I \, V \,,
\label{gaugetrf}
\end{equation}
for some constants, $c^I$, leaves $\Theta^{(I)}$ unchanged, and so there are only $N$ independent parameters in each $K^I $.  In addition, since $\varepsilon_0 =0$ then one must take $ k^I_0  =0$ for $\Theta^{(I)}$ to remain finite at infinity.  The remaining $(N-1)$  parameters then describe harmonic forms that are dual to the non-trivial two-cycles\footnote{If $\varepsilon_0 \ne 0$ then  the extra parameter is that of a Maxwell field whose gauge potential gives the Wilson line around  the $S^1$ at infinity.}.

%%%%%%%%%%%%%%%
\exbox{ Show that the two-forms, $\Theta^{(I)}$, defined by (\ref{harmtwoform}) and
(\ref{KIdefn}) are normalizable on standard GH spaces (with  $V>0$ everywhere).
That is, show that the $\Theta^{(I)}$ square integrable:
\begin{equation}
\int  \Theta^{(I)} \wedge \Theta^{(I)} ~<~ \infty \,, \qquad ({\rm with\ no\ sum\ on }\  I)
\end{equation}
where the integral is taken of the whole GH base space.}
%%%%%%%%%%%%%%%

It is straightforward to find a {\it local} potential  such that $\Theta^{(I)} = dB^{(I)}$:
\begin{equation}
B^{(I)} ~\equiv~  V^{-1}\,   K^I  \, (d\psi ~+~ A) ~+~  \vec{\xi}^{(I)} \cdot d \vec y \,,
\label{Bpot}
\end{equation}
where
\begin{equation}
\vec  \nabla \times \vec \xi^{(I)}  ~=~ - \vec \nabla K^I \,.
\label{xidefn}
\end{equation}
Hence, the $\vec \xi^{(I)}$ are the vector potentials for magnetic monopoles located at the singular points of the $K^I$.

One can use these local potentials to compute the fluxes, $\flux^{(I)}_{ij}$, of $\Theta^{(I)}$ through the cycles:
\begin{equation}
\flux^{(I)}_{ij} ~\equiv~  {1 \over 4\, \pi}\, \int_{\Delta_{ij}} \, \Theta^{(I)} ~=~   \bigg( { k^I_j  \over q_j} ~-~
 { k^I_i \over q_i} \bigg) \,.
 \label{basicflux}
\end{equation}
I have normalized these periods for later convenience.

%%%%%%%%%%%%%%%
\exbox{Excise the points $r_i =0$ and $r_j =0$ from $\Delta_{ij}$ and then use  (\ref{Bpot}) on this punctured cycle to prove (\ref{basicflux}).}
%%%%%%%%%%%%%%%

%%%%%%%%%%%%%%%
\subsection{Solving the BPS equations}
\label{BPSequations}
%%%%%%%%%%%%%%%

We have solved the first layer of the BPS equations (\ref{BPSeqn:1}) and our task now is to solve the remaining two, (\ref{BPSeqn:2}) and (\ref{BPSeqn:3}).   Such solutions were derived in \cite{Gauntlett:2002nw, Gauntlett:2004qy} for Riemannian Gibbons-Hawking metrics  and the result is relatively simple.

%%%%%%%%%%%%%%%
\exbox{Substitute the two-forms  (\ref{harmtwoform}) into (\ref{BPSeqn:2}) and show that the resulting equation has the
solution:
\begin{equation}
Z_I ~=~ \coeff{1}{2}  \, C_{IJK} \, V^{-1}\,K^J K^K  ~+~ L_I \,,
\label{ZIform}
\end{equation}
where the $L_I$ are independent harmonic functions.}
%%%%%%%%%%%%%%%

Now write the one-form, $k$, as:
\begin{equation}
k ~=~ \mu\, ( d\psi + A   ) ~+~ \omega
\label{kansatz}
\end{equation}
and then (\ref{BPSeqn:3}) becomes:
\begin{equation}
\vec \nabla \times \vec \omega ~=~  ( V \vec \nabla \mu ~-~
\mu \vec \nabla V ) ~-~ \, V\, \sum_{I=1}^3 \,
 Z_I \, \vec \nabla \bigg({K^I \over V}\bigg) \,.
\label{roteqn}
\end{equation}
Taking the divergence yields the following equation for $\mu$:
\begin{equation}
\nabla^2 \mu ~=~ \, V^{-1}\, \vec \nabla \cdot
\bigg( V \sum_{I=1}^3 \, Z_I ~\vec \nabla {K^I \over V} \bigg) \,,
\label{mueqn}
\end{equation}
which is solved by:
\begin{equation}
\mu ~=~ \coeff {1}{6} \, C_{IJK}\,  {K^I K^J K^K \over V^2} ~+~
{1 \over 2 \,V} \, K^I L_I ~+~  M\,,
\label{mures}
\end{equation}
where $M$ is yet another harmonic function on $\IR^3$.  Indeed, $M$
determines the anti-self-dual part of $dk$ that cancels out of
 (\ref{BPSeqn:3}). Substituting this result for $\mu$ into
 (\ref{roteqn})  we find that $\omega$ satisfies:
\begin{equation}
\vec \nabla \times \vec \omega ~=~  V \vec \nabla M ~-~
M \vec \nabla V ~+~   \coeff{1}{2}\, (K^I  \vec\nabla L_I - L_I \vec
\nabla K^I )\,.
\label{omegeqn}
\end{equation}
The integrability condition for this equation is obtained by taking
 the divergence of both sides.  The left-hand side trivially vanishes, while the right-hand side vanishes because
$K^I, L_I, M$ and $V$ are harmonic.

The  harmonic functions in (\ref{omegeqn}) will involve constants and point sources at $r_j =0$. Thus the right-hand  side of (\ref{omegeqn}) will have two kinds of terms:
\begin{equation}
{1\over r_i}\, \vec \nabla\, {1\over r_j} ~-~ {1\over r_j}\, \vec \nabla\,
{1\over r_i}   \qquad {\rm and } \qquad \vec \nabla \, {1\over r_i} \,.
\label{rhsterms}
\end{equation}
To determine the pieces that make up $ \vec \omega$, introduce coordinates with the $z$-axis running through $\vec y^{(i)}$ and  $\vec y^{(j)}$ so that $\vec y^{(i)} ~=~ (0,0,a)$ and $\vec y^{(j)} ~=~ (0,0,b)$ with $a > b$.   Define
\begin{equation}
\omega_{ij} ~\equiv~ - {(x^2 +  y^2 + (z-a+ r_i)(z-b - r_j)) \over (a-b)  \, r_i  \, r_j } \, d \phi  \,.
\label{omegaijdefn}
\end{equation}
One can then easily verify that these vector fields satisfy:
\begin{equation}
  \vec \nabla \times \vec \omega_{ij} ~=~  {1\over r_i}\, \vec
\nabla\, {1\over r_j} ~-~ {1\over r_j}\, \vec \nabla\, {1\over r_i}~+~
 {1 \over r_{ij} } \, \bigg(\vec \nabla\,  {1\over r_i} ~-~ \vec \nabla\,
 {1\over r_j} \bigg) \,,
\label{omijeqn}
\end{equation}
where
\begin{equation}
r_{ij} ~\equiv~ |\vec y^{(i)} ~-~ \vec y^{(j)}  |
\label{dijdefn}
\end{equation}
is the distance between the $i^{\rm th}$ and $j^{\rm th}$ center in the Gibbons-Hawking metric.

We then see that the general solution for $\vec \omega$ may be written
as:
\begin{equation}
\vec \omega ~=~  \sum_{i,j}^ N \, a_{ij} \, \vec \omega_{ij} ~+~
 \sum_{i}^ N \, b_{i} \, \vec v_{i}\,,
\label{omsol}
\end{equation}
for some constants $a_{ij}$, $b_i$, where the $\vec v_i$ are defined in  (\ref{vzdefns}).

The important point about the choice of solution, $\omega_{ij}$, is that they have {\it no string singularities whatsoever.}  They can be used to solve (\ref{omegeqn}) with the first set of source terms in (\ref{rhsterms}), without introducing Dirac-Misner strings, but at the cost of adding new source terms of the form of the second term in (\ref{rhsterms}).  Using the $\omega_{ij}$ shows the string singularities in $\vec \omega$ can be reduced to those associated with the second set of terms in (\ref{rhsterms}) and so there are at most $N$ possible string singularities and these can be arranged to run in any direction from each of the points $\vec{y}^{(j)}$.

%%%%%%%%%%%%%%%
\subsection{Removing singularities}
\label{ss:Singularities}
%%%%%%%%%%%%%%%

The functions, $L_I$ and $M$ should be chosen to ensure that metric is regular as $r_j \to 0$, which means that the warp factors, $Z_I$, and the function, $\mu$, must be regular as $r_j \to 0$.   From (\ref{ZIform}) and (\ref{mures}), one can easily see that regularity requires:
\begin{equation}
L^I ~=~ \ell^I_0 ~+~  \sum_{j=1}^N \, {\ell_j^I \over r_j} \,, \quad
M ~=~ m_0 ~+~  \sum_{j=1}^N \, {m_j \over r_j} \,,
\label{LMexp}
\end{equation}
with
\begin{eqnarray}
\ell_j^I  &~=~& -  \coeff{1}{2}\,  C_{IJK} \,
{ k_j^J \, k_j^K  \over q_j} \,,  \qquad j=1,\dots, N \,;\\
m_j &~=~&   \coeff {1}{12}\,C_{IJK} {k_j^I \, k_j^J \, k_j^K \over q_j^2}  ~=~
\coeff{1}{2}\,  {k_j^1 \, k_j^2 \, k_j^3 \over q_j^2} \,,  \qquad j=1,\dots, N \,.
\label{lmchoice}
\end{eqnarray}

As I noted above, in order to obtain solutions that are (locally) asymptotic to five-dimensional Minkowski space, $\IR^{4,1}$, (possibly divided by $\ZZ_{q_0}$), one
must take $\varepsilon_0 = 0$ in (\ref{Vform}), and $k_0^I =0$ in (\ref{KIdefn}).  Moreover, $\mu$ must vanish at infinity, and this fixes $m_0$.  For simplicity
I will also take $Z_I \to 1$ as $r \to \infty$. Hence, the
solutions that are asymptotic to five-dimensional Minkowski space
have:
\begin{equation}
\varepsilon_0 = 0 \,, \qquad  k_0^I =0\,, \qquad   l_0^I =1\,, \qquad
 m_0  = -\coeff{1}{2}\, q_0^{-1} \, \sum_{j=1}^N\, \sum_{I=1}^3 k_j^I \,.
\label{fiveDsol}
\end{equation}
It is straightforward to generalize these results to solutions with different asymptotics, and in particular to Taub-NUT.

We have now created a solution to the BPS equations that is based on harmonic fluxes and has no divergent behaviour in the metric.

%%%%%%%%%%%%%%%
\subsection{The bubble equations and closed time-like curves}
\label{ss:BubbleEqns}
%%%%%%%%%%%%%%%

While we have removed obvious divergent behaviour in the fields and in the metric, we have not yet created a non-singular physical five-dimensional metric.   In particular, we must make sure that there are no closed time-like curves (CTC's).    \

The easiest way to expose the potential problem is to consider the constant time, $t$, slices of the metric (\ref{metform}) and the resulting metric induced in the hypersurfaces, $\cB$:
\begin{equation}
d\hat s_4^2 ~=~ -Z^{-2} \ k^2 ~+~ Z \, ds_4^2  \,.
\label{inducedmet}
\end{equation}
The danger arises when, along some closed curve, the first (negative) term wins out over the second factor and thus produces a CTC.  A physically acceptable metric requires us to eliminate this possibility.  There is another,  more stringent,  highly desirable (but not strictly essential) constraint:  requiring that the metric  (\ref{inducedmet}) be positive definite.  If this is true then the function, $t$, represents a globally-defined time function and the five-dimensional metric, (\ref{metform}), is called {\it stably causal}.  This means that not only is the metric causal, but it also remains causal under small perturbations.

One begins the investigate at the obvious danger points: $r_j \to 0$.  The functions, $Z_I$, and $\mu$ are finite, however $V \to \infty$ and so  the $\psi$-circle has vanishing size when measured using $ds_4^2$.  Thus the $\psi$-circles become CTC's unless we impose the further conditions that: 
\begin{equation}
 \mu(\vec y = \vec y^{(j)} ) ~=~ 0\,, \quad j=1,\dots,N \,.
\label{mucond}
\end{equation}
This ensures that the $\psi$-circles pinch off in the metric  (\ref{inducedmet}).

There is a second danger coming from Dirac strings in $\omega$:  If one goes to the axis between two GH points, $r_i \to 0$ and $r_j \to 0$, then the azimuthal angle, $\phi$, around that axis can become a CTC if $\omega$ has a Dirac string.  One therefore needs to collect all the terms in the sources for $\omega$ that give rise to  Dirac strings and set them to zero.  It is relatively easy to show (see \cite{Bena:2007kg}) that (\ref{roteqn}) implies that if one imposes the constraint   (\ref{mucond}), then it also removes the Dirac strings from $\omega$.

Performing the expansion of $\mu$ using (\ref{mures}), (\ref{KIdefn}), (\ref{LMexp}) and (\ref{lmchoice}) around each Gibbons-Hawking point one finds that (\ref{mucond}) becomes the {\it  Bubble Equations}:
\begin{equation}
 \sum_{{\scriptstyle j=1} \atop {\scriptstyle j \ne i}}^N \,
  {\Gamma_{ij}  \over  |\vec y^{(j)} - \vec y^{(i)} |  } ~=~
-2\, \Big(m_0 \, q_i ~+~  \coeff{1}{2} \sum_{I=1}^3  k^I_i \Big) \,,
\label{BubbleEqns}
\end{equation}
where 
\begin{equation}
\Gamma_{ij} ~\equiv~  q_i \, q_j\, \Pi^{(1)}_{ij} \,   \Pi^{(2)}_{ij} \,  \Pi^{(3)}_{ij} \,, \qquad r_{ij} \equiv |\vec y^{(i)} -\vec y^{(j)}|  \,.
\label{Gammadefn}
\end{equation}
If one adds together all of the bubble equations, then the left-hand side vanishes identically (because $\Gamma_{ji} =-\Gamma_{ij}$), and one obtains the
condition on $m_0$ in (\ref{fiveDsol}).  This is simply the condition $\mu \to 0$ as $r \to \infty$ and also means that there is no Dirac-Misner string running out to infinity.  Thus there are only $(N-1)$ independent bubble equations.

We refer to (\ref{BubbleEqns}) as the {\it bubble equations} because they relate the flux through each bubble to the physical size of the bubble, represented by $r_{ij}$.  Note that for a generic configuration, a bubble size can only be non-zero if and only if {\it all three} of the fluxes are non-zero.  Thus the bubbling transition will only be generically possible for the three-charge system.    If all the fluxes are fixed then there are $3N$ moduli, $\vec y^{(j)}$, but really this is $3(N-1)$ because one can translate the centroid of the $\vec y^{(j)}$ without changing the metric.  Once we impose the bubble equations, there  are $3(N-1) - (N-1) =2(N-1)$ moduli remaining.  

It is for this reason that I made the remark in Section \ref{ss:SSsol} that the relative positions of BPS objects are not always free parameters.  

Solving the bubble equations guarantees that there are no CTC's in the immediate vicinity of the GH points.  This does not, however, guarantee that there are not CTC's elsewhere, and the solution may also have other serious pathologies, particularly if one of the $Z_I$'s changes sign. 

There is also one last, crucial, and surprising, ingredient that appears to be essential to the construction of viable microstate geometries.

%%%%%%%%%%%%%%%
\subsection{Ambi-polar GH metrics}
\label{ss:Ambipolar}
%%%%%%%%%%%%%%%

We concluded Section \ref{ss:BPSeqns} by noting that {\it the only} smooth, hyper-K\"ahler, Riemannian metric that is asymptotic to $\IR^4$ is $\IR^4$ itself with its flat metric.  This creates an obvious challenge to microstate geometries.  But it gets worse:  If one starts with a Riemannian GH metric, it is almost impossible to find solutions to the bubble equations that result in globally smooth metrics.  The physical problem is that magnetic create an expansion force on the bubbles and  there is no counterbalancing force to create the equilibrium that the bubble equations seem to require.

The solution is extremely simple, and seemingly radical:  drop the requirement that the metric of $\cB$ be Riemannian.  More specifically, we will allow the metric on $\cB$ be ambi-polar: that is, the signature of the base metric, $ds_4^2$, can change from $+4$ to $-4$ with apparently singular intervening surfaces.  Away from these singular surfaces, we require that the metric still be smooth and hyper-K\"ahler.  For the GH metrics, this means  that we are going to allow $V$ to change sign and, in particular, allow the GH charges, $q_i \in \ZZ$, to have any sign.  This now leads to vast families of ambi-polar GH metrics that are asymptotic to flat $\IR^4$:  One simply imposes 
\begin{equation}
\varepsilon_0 ~=~0  \,, \qquad  q_0 ~\equiv~  \sum_{j=1}^N \, q_j ~=~ +1 \,.
\label{asympflat}
\end{equation}
The apparent disaster is the horribly singular behaviour at the surfaces on which $V = 0$.

The miracle is that even though the base metric is extremely singular,  the ‘warp factor,’ $Z$, of (\ref{metform}) changes sign (by passing through a pole)  at the $V = 0$ surfaces in precisely the right way to create a smooth, Lorentzian five-manifold.   

Indeed, (\ref{XZrelns}) and  (\ref{ZIform}) imply that, near $V = 0$, $Z \sim V^{-1}$.  This cancels the factor of $V$ in front of  $d\vec y \cdot d\vec y$ in (\ref{GHmetric}) and produces a finite limit for the metric along $\IR^3$.  On the other hand, the same factor of $V^{-1}$ appears to create a double pole along the $\psi$-fiber. However, the explicit form of $\mu$ in (\ref{mures}) can be used to show that this double pole, and the sub-leading single pole, are  exactly cancelled by poles coming from the $-Z^{-2}(dt+k)^2$ term.  It is also evident from (\ref{XZrelns}) that the factors of $V$ cancel in the expressions for the scalars, $X^I$. It is also very straightforward to show that in the Maxwell potential  (\ref{Aform}), the  singularity at $V=0$ in (\ref{Bpot}) cancels against the singularity coming from $Z_I^{-1}\, k$ and thus $A^{(I)}$ is smooth in the neighbourhood of $V=0$.

%%%%%%%%%%%%%%%%%%
\exbox{
\begin{itemize}
\item[(i)] Consider what happens to the homology $2$-cycles defined in Section \ref{GHtopology} but now in the full physical metric,  (\ref{metform}), or, equivalently,  in the  metric (\ref{inducedmet}) on $\cB$.  Show that these cycles remain compact and well defined if and only if one imposes (\ref{mucond}).
\item[(ii)] Observe that the divergence in $\Theta^{(I)}$ at $V=0$ makes the period integral,  (\ref{basicflux}), completely ill defined.    Show that (\ref{Aform}) is regular in the neighbourhood of $V=0$ surfaces. Consider $F^{(I)}   = d A^{(I)}$ and prove the  period integral of $F^{(I)}$ is well-defined and is given by:
\begin{equation}
\flux^{(I)}_{ij} ~\equiv~  {1 \over 4\, \pi}\, \int_{\Delta_{ij}} \, F^{(I)} ~=~   \bigg( { k^I_j  \over q_j} ~-~
 { k^I_i \over q_i} \bigg) \,,
 \label{FIflux}
\end{equation}
\end{itemize}
}
%%%%%%%%%%%%%%%%%%

%%%%%%%%%%%%%%%%%%
\exbox{
Write the metric (\ref{inducedmet})  in the form:
\begin{equation}
\begin{aligned}
{d\hat s}^2_4   =~ & - Z^{-2}\, \big( \mu  (d \psi+ A ) + \omega  \big)^2 \nonumber \\
& ~+~ {Z V^{-1}}\big( d\psi + A \big)^2 + Z V \big(dr^2 +
r^2 d\theta^2 + r^2 \sin^2 \theta \, d\phi^2\big)   \\
=~ & {\cQ \over Z^2 V^2} \Big( d\psi + A  - {\mu \, V^2 \over \cQ }\, \omega \Big)^2 ~+~
Z  V \Big( r^2 \sin^2 \theta \, d \phi^2 -{\omega^2  \over \cQ} \Big) 
~+~  Z  V (dr^2 + r^2 d\theta^2) \,,
\end{aligned}
 \label{dtzero}
\end{equation}
where 
\begin{equation}
Z~\equiv~(Z_1\, Z_2\, Z_3)^{1/3} \,, \qquad \cQ ~\equiv~    Z_1 Z_2 Z_3 V ~-~ \mu^2 \, V^2 \,.
\label{ZQdefn}
\end{equation}
Show that one can write $\cQ$ in the form:
\begin{equation}
\begin{aligned}
\cQ ~=~&  - M^2\,V^2   - \coeff{1}{3}\,M\,C_{IJK}{K^I}\,{K^J}\,{K^K} ~-~ M\,V\,{K^I}\,{L_I}
~-~ \coeff{1}{4} \,(K^I L_I)^2 \nonumber \\
&  +~\coeff{1}{6} \, V C^{IJK}L_I L_J L_K ~+~ \coeff{1}{4} \,C^{IJK}
C_{IMN}L_J L_K K^M K^N
\end{aligned}
\label{QasEseven}
\end{equation}
Observe that $Z  V$ and $\cQ$ are smooth (and generically non-vanishing) near the surfaces where $V=0$, and hence conclude that  (\ref{metform}) is smooth across these surfaces.
}
%%%%%%%%%%%%%%%%%%

The surfaces at which one $V=0$ pose no inherent problems with smoothness in ambi-polar metrics:  (\ref{metform})  can indeed be a smooth Lorentzian metric.  Moreover, everything we discussed about magnetic fluxes remains valid when translated to the full, five-dimensional metric and the  full, five-dimensional Maxwell fields.

There is one rather interesting physical feature of the $V=0$ surfaces:  the Killing vector, $\frac{\partial}{\partial t}$, is time-like everywhere except at the $V=0$ surfaces, on which $\frac{\partial}{\partial t}$ becomes a null vector.  The $V=0$ surfaces are thus known as {\it evanescent ergosurfaces}.  Because the time-like Killing vector is going null on these surfaces   they  potentially have a very important role in collecting and storing information   \cite{Bena:2008nh,Eperon:2016cdd,Keir:2016azt,Marolf:2016nwu,Eperon:2017bwq}

%%%%%%%%%%%%%%%
\subsection{Two centres:  $AdS_3 \times S^2$}
\label{AdSS}
%%%%%%%%%%%%%%%
  
It is extremely instructive to look at a very simple example \cite{Denef:2007yt,deBoer:2008fk,Bena:2010gg} that represents a ``local model'' of how the full geometry is resolved around a pair of GH charges with opposite signs.  

For simplicity we take all the $ K^I$'s to be equal and set:
\begin{equation}
V ~=~ q\,  \Big({ 1 \over r_+} ~-~ {1 \over r_-} \Big)\,, \qquad  K^I ~\equiv~ K~=~ k\, \Big({ 1 \over r_+} ~+~ {1 \over r_-} \Big)  \,,  
\end{equation}
where
\begin{equation}
r_\pm ~\equiv~   \sqrt{\rho^2 ~+~ (z\mp a)^2 } \,,
\end{equation}
in  cylindrical polar coordinates, $(z, \rho, \phi)$, on the $\IR^3$ of the GH base.

Regularity of the functions $Z_I$ and $\mu$ at the GH points determines the functions $L_I$ and $M$ up to additive constants.  Since we do not want any rotation at infinity we need $\mu$ to vanish at infinity and for simplicity we  set the constants in $L_I$ to zero.  Thus we find:
\begin{equation}
\qquad  L_I ~\equiv~ L ~=~  -  {k^2 \over q}  \, \Big({ 1 \over r_+} ~-~ {1 \over r_-} \Big) \,, \qquad  M ~=~ - {2\, k^3 \over a\, q^2}~+~ \frac{1}{2} \,{k^3 \over q^2} \Big({ 1 \over r_+} ~+~ {1 \over r_-} \Big)\,.
\end{equation}

The vector potentials for this solution are then:
\begin{equation}
A ~=~  q\,  \Big({(z -a) \over r_+} - {(z +a) \over r_-} \Big) \, d\phi  \,, \qquad \omega ~=~ -{2\, k^3 \over a\, q} \,
{\rho^2 + (z-a +r_+)(z+a - r_-)  \over r_+ \, r_-}  \, d\phi  \,.
\end{equation}
The five-dimensional metric is then:
\begin{equation}
ds_5^2 ~\equiv~ - Z^{-2} \big(dt+ \mu (d\psi+A) + \omega\big)^2 ~+~
Z\, \big(   V^{-1} (d\psi+A)^2 ~+~ V(d\rho^2 + \rho^2 d\phi^2 + dz^2)  \big)\,,
\label{fivemetric}
\end{equation}
where
\begin{eqnarray}
Z &=&  V^{-1} K^2 + L ~=~ - {4\, k^2 \over q}  \, { 1 \over (r_+ - r_- )}\,,  \\
\mu &=&   V^{-2} K^3 +  \coeff{3}{2}\, V^{-1} K \, L + M ~=~ {4\, k^3 \over q^2}  \,
{  (r_+ + r_- ) \over (r_+ - r_- )^2} ~-~  {2\, k^3 \over a\, q^2} \,. \nonumber
\end{eqnarray}

To map this onto a more familiar metric,  one must make a  transformation to oblate spheroidal coordinates like those employed in \cite{Prasad:1979kg} to map the positive-definite, two-centered GH space onto the Eguchi-Hanson form:
\begin{equation}
 z =  a\, \cosh 2\xi \,\cos \theta \,, \qquad  \rho =  a\, \sinh 2\xi \, \sin \theta \,, \qquad
 \xi \ge 0\,, \ \ 0 \le \theta \le \pi \,.
  \label{coordsa}
 \end{equation}
In particular, one has $r_\pm =  a (\cosh 2\xi  \mp \cos \theta)$.  One then rescales and shifts the remaining variables according to:
\begin{equation}
 \tau ~\equiv~   \coeff{a\, q}{8\, k^3}\,  t \,, \qquad \varphi_1 ~\equiv~   \coeff{1}{2\,q} \, \psi -
 \coeff{a\, q}{8\, k^3}\,  t   \,, \qquad \varphi_2 ~\equiv~ \phi -   \coeff{1}{2\,q} \, \psi +
 \coeff{a\, q}{4\, k^3}\,  t   \,,
 \label{coordsb}
 \end{equation}
and the five-dimensional metric takes the standard $AdS_3 \times S^2$ form:
\begin{equation}
ds_5^2 ~\equiv~ R_1^2 \big[ - \cosh^2\xi \,  d\tau^2 + d\xi^2 +  \sinh^2 \xi \, d\varphi_1^2 \big] ~+~  R_2^2 \big[   d \theta ^2 + \sin^2\theta  \, d\varphi_2^2 \big]  \,,
 \label{AdS3S2}
 \end{equation}
with
\begin{equation}
  R_1~=~  2 R_2 ~=~ 4 k \,.
 \label{Radii}
 \end{equation}
Note that the first factor in the metric is {\it global} $AdS_3$ with $-\infty < \tau < \infty$.  One should also recall that the GH fiber coordinate has period $4 \pi$ and therefore, for $|q|=1$, the angles, $\varphi_j$, both have periods $2 \pi$.  For $q \ne 1$, the $\ZZ_{|q|}$ orbifold associated with the GH points emerges as a simultaneous $\ZZ_{|q|}$ quotient on the longitudes of the $AdS_3$ and $S^2$.

This metric is completely smooth and the  ``bubble,'' or non-trivial topology, is simply the $2$-sphere.   It is a ``local model'' in that if the two GH points are far from the other GH points, one can ``zoom-in''  on that pair and the metric locally reduces to the construction here.

One can also check that the time-like Killing vector, $T \equiv \frac{\partial}{\partial t} $ transforms as follows:
\begin{equation}
\frac{\partial}{\partial t} = \frac{1}{R_1} \, \bigg [\, \frac{\partial}{\partial \tau} ~-~\frac{\partial}{\partial \varphi_1} ~+~ 2 \,\frac{\partial}{\partial \varphi_2}  \,\bigg ]
\end{equation}
whose norm is
\begin{equation}
T^\mu \, T_{\mu} = -R_1^{-2}\, \cos ^2 \theta. 
\end{equation}
One may think of the integral curves of $T$ as  world-lines of  non-space-like ``observers'' that are time-like everywhere except for $\theta =   \frac{\pi}{2}$.  From this perspective, there is nothing unusual about the evanescent ergosurface.

%%%%%%%%%%%%%%%
\subsection{Regularity}
\label{ss:RegularityMGs}
%%%%%%%%%%%%%%%

We have examined smoothness on evanescent ergosurfaces and the absence of CTC's near the GH points, $\vec y^{(j)}$.  
There is now the much broader question of whether the metric is regular globally and whether it is globally free of CTC's and perhaps even stably causal.  

First we observe that writing the metric in the form (\ref{dtzero}) means that one must have  $Z V > 0$ and $\cQ >0$.  However the parametrization  of the scalar fields in  (\ref{XZrelns}) requires that all the functions, $Z_I$, have the same sign.  We therefore must require 
\begin{equation}
Z_I \, V ~>~ 0 \,, \qquad I=1,2,3\,,
\label{ZVpos}
\end{equation}

For the five dimensional metric to be stably causal, $\cQ$ must not only be positive but must satisfy \cite{Berglund:2005vb}:
\begin{equation}
- g^{\mu\nu} \partial_{\mu}t \, \partial_{\nu} t = - g^{tt} =  (Z V)^{-1}
(\cQ -  \omega^2) > 0\,,
\label{stabcausal}
\end{equation}
where $\omega$ is squared using the $\IR^3$ metric.

First we note that for generic fluxes, satisfying the bubble equations is nowhere near sufficient to guarantee  (\ref{ZVpos}) and (\ref{stabcausal}).  This is  because generic fluxes can produce positive electric charge contributions from some collections of bubbles and negative  electric charge contributions from other collections of bubbles.  Such combinations of localizable positive and negative charges creates a very pathological solution that typically fails to satisfy (\ref{ZVpos})  in some region.  While there are no theorems, if one satisfies the bubble equations and makes sure that  (\ref{ZVpos}) is satisfied then the five-dimensional metric usually satisfies (\ref{stabcausal}). (I know of no counterexamples and have constructed many, many families of such microstate geometries.)

So the bottom line is that one must solve the bubble equations and make sure that (\ref{ZVpos}) and (\ref{stabcausal}) are satisfied.  Satisfying (\ref{ZVpos}) seems to guarantee  (\ref{stabcausal}) but one should always explore the metric numerically to make sure that  (\ref{stabcausal}) is indeed true.

%%%%%%%%%%%%%%%
\subsection{The asymptotic charges}
\label{ss:AsympChGH}
%%%%%%%%%%%%%%%

As described in Section \ref{AsympCharges}, one can obtain the electric charges, mass and  angular momenta of bubbled geometries by expanding $Z_I$ and $k$ at infinity.   It is, however, more convenient to translate the asymptotics into the standard coordinates of the Gibbons-Hawking spaces. In particular, one should remember that the GH radial coordinate, $r$, is related to the radial coordinate $\rho$ on flat $\IR^4$ via  (\ref{newradcoord}), that is, $r = {1 \over 4} \rho^2$.  One then finds
\begin{equation}
Z_I ~\sim~ 1 ~+~ \, {Q_I \over 4\, r }~+~ \dots  \,, \qquad  \rho \to \infty \,,
\label{ZIexpGH}
\end{equation}
and from (\ref{ZIform}) one easily obtains
\begin{equation}
Q_I ~=~ -2 \, C_{IJK} \, \sum_{j=1}^N \, q_j^{-1} \,
\tilde  k^J_j \, \tilde  k^K_j\,,
\label{QIchg}
\end{equation}
where
\begin{equation}
\tilde  k^I_j ~\equiv~ k^I_j ~-~    q_j\, N  \,  k_{avg}^I  \,,
\qquad {\rm and} \qquad k_{avg}^I ~\equiv~{1 \over N} \, \sum_{j=1}^N k_j^I\,.
\label{ktilde}
\end{equation}
Note that $\tilde  k^I_j$ is gauge invariant under (\ref{gaugetrf}).  

This may be recast in a more suggestive form that reflects the origin of the charges as coming from magnetic fluxes via the Chern-Simons interaction: 
\begin{equation}
Q_I  ~=~   \sum_{i, j=1}^N \,  {Q_I}_{ij}   \,, \qquad {Q_I}_{ij}   ~\equiv~ -  \coeff{1}{4} \, C_{IJK} \,  q_i \, q_j \, \Pi^{(J)}_{ij} \,   \Pi^{(K)}_{ij}    \,.
\label{QIanswer}
\end{equation}
where the ${Q_I}_{ij}$ may be thought of as the contribution to the charge coming from each bubble.

Expanding $g_{00}$, one has:
\begin{equation}
-g_{00}~=~  (Z_1 Z_2 Z_3)^{-\frac{2}{3}} ~\sim~ 1 ~-~ \frac{2}{3} \, \sum_{I=1}^3  \, {Q_I \over 4\, r }  \,,
\label{g00expansion}
\end{equation}
and comparing this with (\ref{asympg1})  one finds 
\begin{equation}
M ~=~ \frac{\pi}{4 G_5} \,( Q_1 ~+~ Q_2 ~+~ Q_3) \,. 
\label{MeqlQ}
\end{equation}
It is fairly common to go to a system of units in which the five-dimensional Planck length, $\ell_5$, is unity and this means (see, for example, \cite{Peet:2000hn,Elvang:2004ds}):
\begin{equation}
G_5~=~ \frac{\pi}{4} \,.
\label{NiceUnit}
\end{equation}
In particular, this means that the solution BPS condition takes the simpler standard form:
\begin{equation}
M ~=~ Q_1 ~+~ Q_2 ~+~ Q_3 \,. 
\label{BPSstandard}
\end{equation}

One can also recast the expressions for angular momenta in terms of the GH formulation:  
\begin{equation}
k ~\sim~ {1 \over 4 \,\rho^2} \, \big((J_1+J_2) ~+~ (J_1-J_2) \, \cos \theta   \big) \,
d\psi ~+~ \dots \,.,
\label{angmomform}
\end{equation}
which means one can obtain both commuting angular momenta from an expansion of  $\mu$.  There are two types of such terms, the simple ${1 \over r}$
terms and the dipole terms arising from the expansion of $V^{-1} K^I$.  Following \cite{Berglund:2005vb}, define the dipoles
\begin{equation}
\vec D_j ~\equiv~  \, \sum_I  \, \tilde k_j^I \, \vec y^{(j)} \,,
\qquad \vec D ~\equiv~ \sum_{j=1}^N \, \vec D_j \,.
\label{dipoles}
\end{equation}
and then the expansion of $k$ takes the form (\ref{angmomform})
if one takes $\vec D$ to define the polar axis from which $\theta$ is measured.
One then arrives at
\begin{equation}
J_R ~\equiv~ J_1 + J_2 ~=~ \coeff{4}{3}\, \, C_{IJK} \, \sum_{j=1}^N q_j^{-2} \,
\tilde  k^I_j \, \tilde  k^J_j \,  \tilde  k^K_j  \,,
\label{Jright}
\end{equation}
\begin{equation}
 J_L ~\equiv~ J_1 - J_2 ~=~ 8 \,\big| \vec D\big|  \,.
\label{Jleft}
\end{equation}
While we have put modulus signs around $\vec D$ in (\ref{Jleft}), one
should note that it does have a meaningful orientation, and so we
will sometimes consider $\vec J_L = 8 \vec D$.

One can use the bubble equations to obtain another, rather more
intuitive expression for $J_1 -J_2$.  One should first note  that the right-hand side
of the bubble equation, (\ref{BubbleEqns}), may be written as $-  \sum_I  \tilde k_i^I$.
Multiplying this by $\vec y^{(i)}$ and summing over $i$ yields:
\begin{eqnarray}
\vec J_L &~\equiv~&  8\, \vec D  ~=~  -  \coeff{4}{3}\, C_{IJK} \,
 \sum_{{\scriptstyle i, j=1} \atop {\scriptstyle j \ne i}}^N \,  \,  \Pi^{(I)}_{ij} \,
  \Pi^{(J)}_{ij} \,  \Pi^{(K)}_{ij} \ {q_i \, q_j \,  \vec y^{(i)}  \over r_{ij} } \nonumber \\
 & ~=~ & -  \coeff{2}{3}\, C_{IJK} \, \sum_{{\scriptstyle i, j=1} \atop
 {\scriptstyle j \ne i}}^N \,  q_i \, q_j \,  \Pi^{(I)}_{ij} \,   \Pi^{(J)}_{ij} \,  \Pi^{(K)}_{ij} \
{(\vec y^{(i)} - \vec y^{(j)}) \over \big|\vec y^{(i)} - \vec y^{(j)}\big| }\,,
\label{JLnice}
\end{eqnarray}
where we have used the skew symmetry $\Pi_{ij} = - \Pi_{ji}$ to obtain
the second identity.  This result suggests that one should define an angular
momentum flux vector  associated with the $ij^{\rm th}$ bubble:
\begin{equation}
\vec J_{L\, ij} ~\equiv ~ -  \coeff{4}{3}\,q_i \, q_j \, C_{IJK} \,
\Pi^{(I)}_{ij} \,   \Pi^{(J)}_{ij} \,  \Pi^{(K)}_{ij} \, \hat y_{ij} \,,
\label{angmomflux}
\end{equation}
where $\hat y_{ij}$ are {\it unit} vectors,
\begin{equation}
\hat y_{ij} ~\equiv ~  {(\vec y^{(i)} - \vec y^{(j)}) \over
 \big|\vec y^{(i)} - \vec y^{(j)}\big| } \,.
 \label{unitvecs}
\end{equation}
This means that the flux terms on the left-hand side of the bubble equation
actually have a natural spatial direction, and once this is incorporated, it yields the
contribution of the bubble to $J_L$.

This ``angular momentum associated with a bubble'' is a great significance to the quantization of bubbled geometries.  If one quantizes the moduli space then the corresponding symplectic form treats each bubble as a distinct spin system whose individual angular momentum must be quantized  \cite{Bena:2006kb,Bena:2007qc,deBoer:2008zn,deBoer:2009un}. 

%%%%%%%%%%%%%%%
\subsection{Summary}
\label{ss:GHSummary}
%%%%%%%%%%%%%%%

One can make a bubbled solution as follows
\begin{itemize}
\item Start with a GH metric  (\ref{GHmetric}) with (\ref{Vform}).  In particular, {\it choose} the $q_j\in \ZZ$ and  $\vec{y}^{(j)} \in \IR^3$ as one wishes. For metrics  asymptotic   to  $\IR^{4,1}$ impose the constraint (\ref{asympflat}) on the $q_i$. Compute $\vec A$ from (\ref{AVreln}). (To do this,  (\ref{vzdefns}) is useful.)
\item  {\it Choose} the magnetic fluxes, $\flux^{(I)}_{ij}$, by  {\it choosing} the flux parameters, $k_i^I$  in (\ref{KIdefn}).
\item  Fix the functions $L_I$ and $M$ according to (\ref{lmchoice}) and (\ref{fiveDsol})  and thereby determine the functions $Z_I$ and $\mu$ using (\ref{ZIform}) and (\ref{mures}). 
\item Solve (\ref{omegeqn}) to determine $\vec \omega$.  In doing this, (\ref{omegaijdefn}) and (\ref{omijeqn})  are very useful.
\item Solve the {\it Bubble Equations}  (\ref{BubbleEqns}):  This will put $(N-1)$ constraints on the separations of the $\vec{y}^{(j)}$.
\item Check for global absence of CTC's/causal stability:   Make sure (\ref{ZVpos}) and (\ref{stabcausal}) are satisfied
\item Compute the global charges: $Q_I$, $J_1$, $J_2$.
\end{itemize}

One can easily count the moduli in a solution with $N$ centers:  There are the $3(N-1)$ components of the $\vec{y}^{(j)} \in \IR^3$ minus $(N-1)$ bubble equations for a total of $2(N-1)$ geometric moduli.  There are then $3(N-1)$ flux parameters coming from $3N$ values of the $k_i^I$ minus the ``gauge transformations'' (\ref{gaugetrf}).  This gives $5(N-1)$ parameters.  Finally, there are also three global rotations on $\IR^3$ to be subtracted  to reduce this to $5(N-1)-3$.  If one fixes the $3$ electric charges, $Q_I$, and the angular momenta, $J_1$, $J_2$, then there remain $5(N-2)-3$ free parameters.

There are obviously a large number of degrees of freedom as $N$ becomes large.  As we will discuss in the next lecture, this is only the tip of an ice-berg:  the bubble can have shape modes, which means infinitely many Fourier coefficients. 

There are many issues I have ignored.  First, the flux parameters are necessarily quantized in string theory: see, for example, \cite{Berglund:2005vb}. As we noted above, the geometric moduli must also be quantized \cite{Bena:2006kb,Bena:2007qc,deBoer:2008zn,deBoer:2009un}.  This latter fact led to a remarkable triumph in the holographic field theory dual to these geometries.

%%%%%%%%%%%%%%%
\subsection{Final comment: Geometric transitions}
\label{ss:GHcomments}
%%%%%%%%%%%%%%%

One of the motivations behind the discovery of microstate geometries is the important physical idea of geometric transitions.  A geometric transition typically involves a topology change in which one kind of source is replaced by another.  Here we are thinking of black holes, based on singular electric charge sources, being replaced by magnetic fluxes on $2$-cycles.    The beautiful thing about ambi-polar GH metrics is that it can give this picture a natural realization.  

One can start from a black ring in flat space:  
\begin{equation}
V  ~= ~  \frac{1}{r} \,, \qquad  \Theta^{(I)}   ~= ~ 0\,, \qquad Z_I   ~= ~  \frac{Q_I}{r_b}\,, \qquad r_b   ~= ~  \sqrt{x^2 + y^2 + (z-b)^2}  \,.
 \label{BHfive}
\end{equation}
It is a black ring because it is singular around the entire $\psi$ fiber.

\begin{figure}
\leftline{\hskip 0.5cm \includegraphics[width=6in]{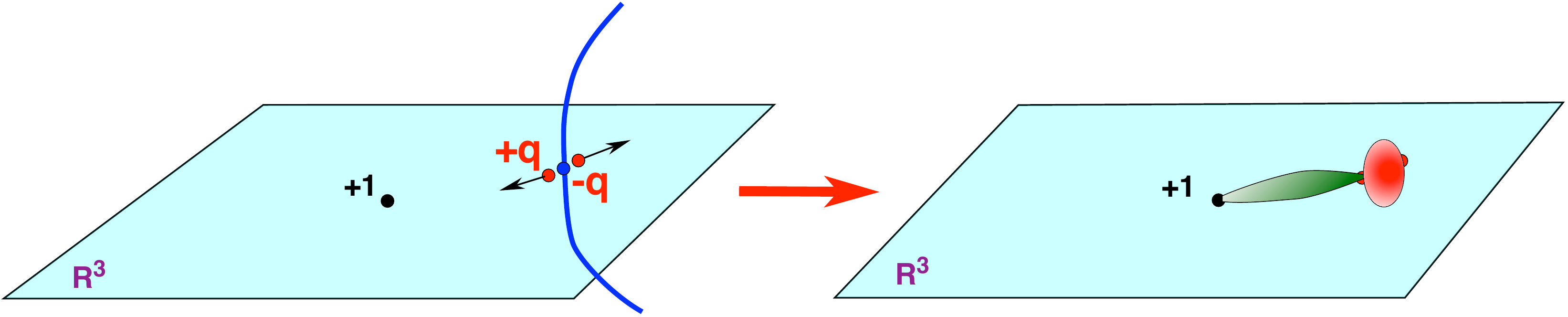}}
\setlength{\unitlength}{0.1\columnwidth}
\caption{\it  Bubbling a black ring by blowing up new topological cycles on a GH space.}
\label{fig:Transition}
\end{figure}

One can then imagine ``creating a pair'' of GH points (see Fig. \ref{fig:Transition}) in the neighborhood of the locus of the original black ring:
\begin{equation}
V  ~= ~  \frac{1}{r} ~+~ \frac{q}{r_+} ~-~ \frac{q}{r_-}\,, \qquad r_\pm   ~= ~  \sqrt{x^2 + y^2 + (z-b \pm a)^2}  \,.
 \label{bubbling}
\end{equation}
and then replacing the singular source by fluxes on the bubbles.  In this way one has ``blown up a cycle'' underneath the original black hole and created a microstate geometry instead.

 %%%%%%%%%%%%%%   Figure 3  %%%%%%%%%%%%%%%
%%%%%%%%%%%%%%   End Figure 3  %%%%%%%%%%%%%%%

The new solution has to satisfy the bubble equation in order to be time independent and BPS. Thus the separation, $2a$,  of $r_\pm$ is fixed by the fluxes.  However, one can imagine a dynamical process that nucleates such a bubble which then grows to an equilibrium size set by the bubble equations.  One can also blow down the bubble and even do this while preserving the BPS property. If one takes $q$ to be large, while keeping the fluxes (and hence the charges) fixed, then the bubble equations require  the separation, $2a$,  of $r_\pm$ to get smaller.  While this is far from a continuous process ($q \in \ZZ$ and $q\to \infty$ is a singular limit), this captures the essence of a transition from a microstate geometry to a black object.

Physically, we think of the geometric transition  to form a microstate geometry as a phase transition in the state of the matter system.  This idea is also the basis of the holographic descriptions of many infra-red phases of strongly coupled quantum field theories.

%\newpage
%%%%%%%%%%%%%%%%%%%%%%%%%%%%%%%%%%%%%
\section{Scaling  microstate geometries}
\label{Sect:scaling}
%%%%%%%%%%%%%%%%%%%%%%%%%%%%%%%%%%%%% 

We have realized our initial goal of creating solitonic solutions but the ultimate goal of the Microstate Geometry programme is to generate solitonic solutions that look like black holes to arbitrary precision and then determine to what extent they can capture the microstate structure of a black hole.  (Again, we are only considering BPS black holes here.)  In this respect, the most important classes of microstate geometries are the so-called {\it scaling geometries}.

Scaling solutions arise whenever there is a set of points, $\cS$,  for which the bubble equations admit homogeneous solutions  \cite{Denef:2000nb,Bates:2003vx,Bena:2006kb, Denef:2007vg, Bena:2007qc}: 
\begin{equation}
 \sum_{{\scriptstyle j \in \cS } \atop {\scriptstyle j \ne i}} \,   {\Gamma_{ij}  \over  |\vec y^{(j)} - \vec y^{(i)} |  } ~=~  0\,,  \qquad i \in \cS \,.\label{HomBubbleEqns}
\end{equation}
It then follows that such a cluster of points can be scaled: 
\begin{equation}
 \vec y^{(j)} - \vec y^{(i)}  ~\to ~  \lambda  \, ( \vec y^{(j)} - \vec y^{(i)} )    \,, \label{scaling}
\end{equation}
for  $\lambda  \in  \IR$, and one can then examine the limit in which $\lambda \to 0$.  

The geometries are, or course, required to satisfy  (\ref{BubbleEqns}) and not (\ref{HomBubbleEqns}), however, given a solution of  (\ref{HomBubbleEqns}) one can easily make infinitessimal perturbations of the points, $ \vec y^{(i)}$, and if $| \vec y^{(j)} - \vec y^{(i)} |$ is sufficiently small this will generate finite terms on the right-hand side of  (\ref{HomBubbleEqns}) and these can be used to generate solutions to the full bubble equations (\ref{BubbleEqns}).  In this way, the moduli space of physical solutions  that satisfy (\ref{BubbleEqns}) can contain scaling solutions in which a set of points, $\cS$, can approach one another arbitrarily closely. 

The simplest example of this kind of behaviour comes from scaling triangles.  Suppose that $|\Gamma_{ij}|$, $i,j =1,2,3$, satisfy the triangle inequalities:
\begin{equation}
|\Gamma_{13}|   ~<~ |\Gamma_{12}|  ~+~  |\Gamma_{23}|  \qquad { and \ \ cyclic}\,, \label{triangles1}
\end{equation}
which means that we may arrange the points so that 
\begin{equation}
| \vec y^{(j)} -  \vec y^{(i)} |  ~=~  \lambda \, |\Gamma_{ij}|\,, \label{triangles2}
\end{equation}
 for $\lambda \in \IR^+$.  The fluxes  can then usually be arranged so that the homogeneous bubble equations, (\ref{HomBubbleEqns}), are trivially satisfied since they amount to $\pm \lambda^{-1} \mp  \lambda^{-1} =0$.   When the triangle has infinitessimal size, making  infinitessimal deformations of the angles can be used to generate solutions to the original bubble equations  (\ref{BubbleEqns}).  In particular, in a physical solution to (\ref{BubbleEqns}) with three fluxes that obey  (\ref{triangles1}),  one can make the three points approach one another arbitrarily closely by adjusting the angles in the triangle so that they approach the angles in the triangle defined by (\ref{triangles2}).
 
The existence of scaling solutions to the bubble equations was first noted in  \cite{Denef:2000nb,Bates:2003vx}, in which the bubble equations emerged as {\it integrability conditions}.  However, from the four-dimensional perspective of \cite{Denef:2000nb,Bates:2003vx}, this appeared to be a  singular limit of a multi-black-hole solution. It was subsequently shown that, from the perspective of five-dimensional supergravity, this limit is not only  non-singular but also defines perhaps the most important class of physical solutions \cite{Bena:2006kb, Denef:2007vg, Bena:2007qc}.

 %%%%%%%%%%%%%%   Figure 4  %%%%%%%%%%%%%%%
\begin{figure}
\leftline{\hskip 0.5cm \includegraphics[width=6in]{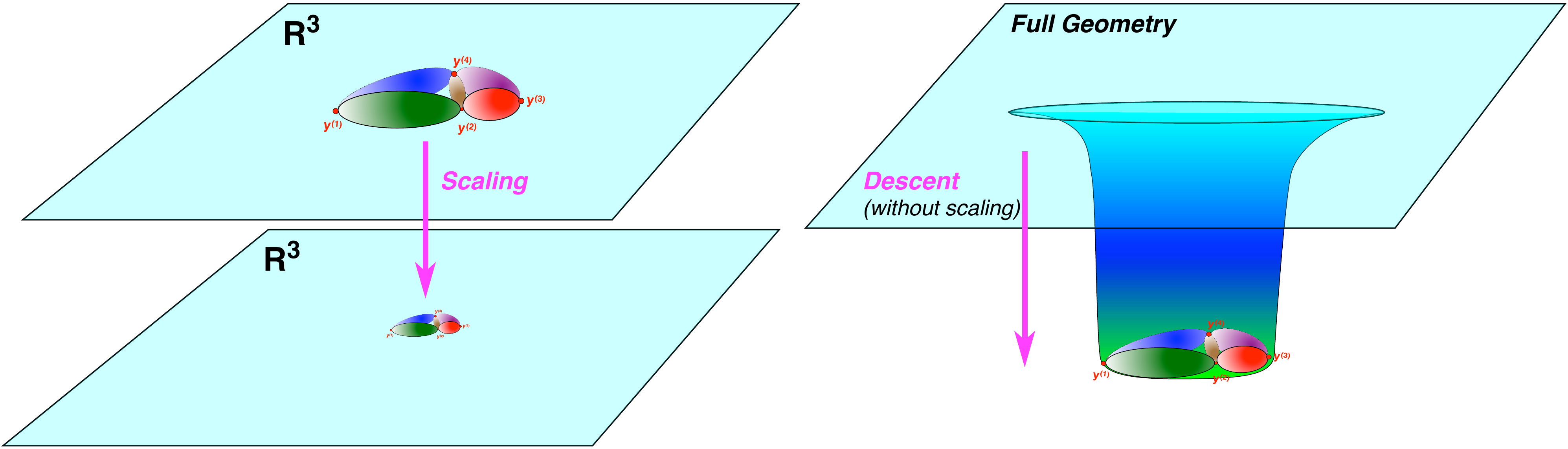} }
\setlength{\unitlength}{0.1\columnwidth}
\caption{\it  The effect of back-reaction on scaling geometries. As the configuration scales to zero size in $\IR^3$, it actually retains its physical size in the complete geometry while descending an AdS throat.}
\label{fig:Scaling}
\end{figure}
%%%%%%%%%%%%%%   End Figure 4  %%%%%%%%%%%%%%%

Suppose that we have a scaling cluster, $\cS$, that is centred on the origin, $r =0$.  Let $\epsilon$ be the largest separation (in $\IR^3$) between points in $\cS$ and let $\eta$ be the smallest distance from a point in $\cS$ and a point, $ \vec y^{(i)}$,  that not in $\cS$.  Assume that $\epsilon <\! <\! < \eta$ and,  for simplicity,  suppose that the total geometric charge of the cluster is unity:  $q_\cS  \equiv \sum_{i \in \cS} q_i = 1$.   In the intermediate range of $r$ in which, $\epsilon \ll r \ll \eta$, one has $V \sim \frac{1}{r}$ and all the other functions $K_I$ and $L_I$ behave as $\cO(r^{-1})$.   This means that, in the intermediate region, $Z_I \sim \frac{Q_{I , \cS}}{4 r}$. where  the $Q_{I, \cS}$ are the electric charges associated with the scaling cluster.  Using this in (\ref{metform}) and (\ref{GHmetric}) we see that the metric in the intermediate region becomes:
\begin{equation}
ds_5^2 ~=~  - \frac{16\,r^2}{a^4}  (dt +k)^2 ~+~      \frac{a^2}{4} \, \frac{dr^2}{r^2}   ~+~  \frac{a^2}{4} \, \big[ ( d \psi +  \cos \theta d \phi ) ^2 +  d\theta^2 + \sin^2 \theta d \phi^2  \big]   \,,
\label{Intermet}
\end{equation}
where $a =(Q_{1, \cS}Q_{2, \cS}Q_{3, \cS})^{1/6}$.   This is the metric of an $AdS_2 \times S^3$ throat of a rotating, extremal black hole.

There are several important consequences of this result.  First,  such scaling  clusters look almost exactly like extremal black holes  except that they ``cap off'' in a collection of bubbles just above\footnote{From the perspective of an infalling observer.} where the horizon would be for the extremal black hole (see Fig. \ref{fig:Scaling}).  Moreover, while it appears, from the perspective of the $\IR^3$ base, that the bubbles are collapsing in the scaling limit, they are, in fact, simply creating an $AdS$ throat and descending down it as it forms.  The physical size of the bubbles  approaches a large, finite value whose scale is set by the radius, $a$, of the $S^3$  of the throat, which corresponds to the horizon of the would-be black hole.  Thus the scaling microstate geometries represent deep bound states of bubbles that realize the goal of creating a smooth, solitonic solutions that look  like BPS black holes.     One obtains similar results for black rings from scaling clusters  whose net geometric charge, $q_\cS$, is zero.

The fact that one can adjust classical parameters so that the scaling points approach one another arbitrarily closely means that the $AdS$ throat can be made arbitrarily deep.  However, the angular momentum, (\ref{JLnice}), depends, via (\ref{angmomflux}), upon the details of locations of the points and  when angular momentum is  quantized this will lead to a discretization of the moduli space and will limit the depth of simple scaling solutions like those based on scaling triangles  \cite{Bena:2007qc}.  More generally, it was proposed in  \cite{Bena:2007qc} and then proven in \cite{deBoer:2008zn} that the individual contributions, $\vec J_{L\, ij}$ in (\ref{angmomflux}) must be separately quantized and so, upon quantization, the classical moduli space is completely discrete.  This has the very interesting physical consequence that even though very long, deep throats are macroscopic regions of space time in which the curvature length scale can be uniformly bounded to well above the Planck scale, quantum effects can wipe out such regions of space-time.

This also has very important implications for the the dual holographic theory, and, in particular leads to the correct holographic prediction of the energy gap of the typical sector of the conformal field theory \cite{Bena:2006kb,Bena:2007qc,deBoer:2008zn,deBoer:2009un}.

%\newpage
%%%%%%%%%%%%%%%%%%%%%%%%%%%%%%%%%%%%%
\section{Studying microstate geometries}
\label{Sect:study}
%%%%%%%%%%%%%%%%%%%%%%%%%%%%%%%%%%%%% 

So far I have taken a more pedagogical approach to these lectures.  For the remaining time, I will survey some of the broader ideas about microstate geometries.  This overview is necessarily incomplete and rather idiosyncratic.

%%%%%%%%%%%%%%%%%%%
\subsection{Stringy and M-theory realizations}
\label{ss:Strings-M}
%%%%%%%%%%%%%%%%%%%

To keep things simple, the focus so far has been on simple microstate geometries in five-dimensional supergravity coupled to vector multiplets.  While solitons in five dimensions are interesting in their own right, the primary goal of the microstate geometry program is to study solitons and associated microstructure  in string theory and M-theory.

First it should be remembered that one can obtain five-dimensional supergravity coupled to vector multiplets as part of dimensional reduction of M-theory on a Calabi-Yau complex $3$-fold.  Indeed, the simplest way to get to the model introduced in Section \ref{fiveDsugr} is to compactify $M$-theory on a $\IT^6$. Decomposing the $\IT^6$ into three $\IT^2$'s labelled by $(x_5,x_6)$,  $(x_7,x_8)$  and $(x_9,x_{10})$, one reduces the eleven-dimensional fields to the five-dimensional fields according to:
\begin{equation}
\begin{aligned}
 ds_{11}^2  ~=~ ds_5^2 & ~+~   \left(Z_2 Z_3  Z_1^{-2}  \right)^{1\over 3}
 (dx_5^2+dx_6^2) \nonumber \\
& ~+~\left( Z_1 Z_3  Z_2^{-2} \right)^{1\over 3} (dx_7^2+dx_8^2)    +
  \left(Z_1 Z_2  Z_3^{-2} \right)^{1\over 3} (dx_9^2+dx_{10}^2) \,,
 \end{aligned}
\label{elevenmetric}
\end{equation}
and
\begin{equation}
C^{(3)} ~=~   A^{(1)} \wedge dx_5 \wedge dx_6 ~+~  A^{(2)}   \wedge
dx_7 \wedge dx_8 ~+~ A^{(3)}  \wedge dx_9 \wedge dx_{10}  \,,
\label{Cfield:ring}
\end{equation}
Note how the five-dimensional scalars appear as the volumes of the $T^2$'s.  The three charges in five-dimensions now have the interpretation of $M2$-brane charges.  

Similarly, one can get the five-dimensional supergravity via a $\IT^5$ or $K3 \times S^1$ compactification of IIB supergravity.  

It is, of course, the reverse of this perspective that is important:  microstate geometries are a very natural part of string theory and  the deeper study of the properties of microstate geometries must necessarily be informed by string theory and what it tells us about the states of matter in such geometries.

%%%%%%%%%%%%%%%%%%%
\subsection{Holographic field theory}
\label{ss:Holography}
%%%%%%%%%%%%%%%%%%%

Holography has had a very great impact on our understanding of the physics of microstate geometries.   Basically, if there is string theory background that is sourced by  some brane charges  and the geometry, in some limit,  becomes that of AdS$_{p+1} \times S^{D-p-1}$, then the general ``stringy'' expectation is that there should be an underlying dual, strongly-coupled (quantum) CFT in $p$-dimensions lying on (some part of) the underlying branes.     In particular, an  AdS$_{3}$ should have a dual $(1+1)$-dimensional conformal field theory.  This is a very general principle in string theory and comes from the fact that strings can either be closed, and describe the gravity sector, or be open, ending on D-branes, and thus encoding the field theory on the branes.

The power and importance of this body of ideas is that it gives one two different ways to study the same problem: one can study gravity through the dual CFT, or study the CFT through the gravity.  A particularly important form of this arises in the study of microstate geometries in six dimensions.  Such geometries may be viewed as simple uplifts of the  five-dimensional geometries we have been studying here.  Specifically, one can work with the IIB compactification on a  $\IT^4$ to six-dimensions.  The five-dimensional  black holes and microstate geometries are then elementary $S^1$ compactifications of this six-dimensional theory and so it is easy to pass from one formulation to the other.   

The importance of the six-dimensional formulation is that supersymmetric black holes have infinitely long AdS$_{3}$ throats.  There is then a dual   $(1+1)$-dimensional CFT and the world-volume of this CFT lies along the $S^1$ of the compactification from six to five dimensions.  One can then study the microstates of the black hole by studying this CFT and its excitations.  This how we now understand the first detailed accounting of supersymmetric microstate structure achieved by Strominger and Vafa in 1996 \cite{Strominger:1996sh}:  The microstates are momentum excitations of the D1-D5 CFT.  The holographic duals of these states has given rise to another class of microstate geometries known as superstrata \cite{Bena:2015bea,Bena:2016ypk,Bena:2017xbt}.
  
The holography of microstate geometries raises a host of interesting questions.  For example,  what CFT states are described by the full range of multi-bubbled scaling microstate geometries? Or what are the CFT states described by small fluctuations around a scaling microstate geometry.  Using superstrata, we have made a great deal of progress on the latter question for the simplest scaling microstate geometries but the former question remains unanswered in anything other than very broad-brush ideas about renormalization group flows to tensor product theories.

%%%%%%%%%%%%%%%%%%%
\subsection{Excitations of microstate geometries}
\label{ss:Wiggling}
%%%%%%%%%%%%%%%%%%%

We have focussed entirely on microstate geometries that exist in five dimensions and whose dynamics is trivial in the compactified dimensions.  This raises two obvious questions:  Have we obtained the most general possible microstate geometry in five dimensions and are there new possibilities if we allow non-trivial dynamics in the extra dimensions.  The answers to these questions are no and yes.

%%%%%%%%%%
\subsubsection{Fluctuating  microstate geometries}
\label{sss:flucts}
%%%%%%%%%%

Most obviously, we chose a Gibbons-Hawking base metric and there are certainly more general hyper-K\"ahler  Riemannian base metrics and there are almost certainly even richer families of hyper-K\"ahler  ambi-polar base metrics.  The problem here is simply computational: outside the GH geometries, there are some explicitly known hyper-K\"ahler  Riemannian  metrics in four dimensions but they are either too symmetric to be very useful \cite{Bena:2007ju}, or too complicated to enable explicit computation (See, for example, \cite{Cherkis:1997aa,Cherkis:1998hi,Cherkis:1998xca,Cherkis:2001gm}).  Essentially nothing is known about ambi-polar hyper-K\"ahler metrics beyond the Gibbons-Hawking family.

Putting this issue aside, one can ask if one has the complete set of solutions to the BPS equations, (\ref{BPSeqn:1})--(\ref{BPSeqn:3}).  Again the answer is no.  As mentioned above, we now have superstrata, which are families of fluctuating {\it BPS} solutions to (\ref{BPSeqn:1}): there are closed, self-dual $2$-forms, $\Theta^{(I)}$, that depend on non-trivial modes along the $\psi$ and $\phi$.  Indeed these solutions depend on arbitrary Fourier modes of the form $n \phi + m \psi$ with $n \ge m$.    Thus the magnetic fluxes can fluctuate and through the back-reaction to these fluctuations, the bubbles can develop shape modes ... and still be BPS.   These solutions were first obtained \cite{Bena:2015bea,Bena:2016agb,Bena:2016ypk,Bena:2017xbt} from six-dimensional microstate geometries that fluctuate as functions of $(v,\psi,\phi)$, where $v$ is a coordinate along the extra $S^1$ of the six-dimensional theory.  The beauty of the six-dimensional formulation is that one can establish a precise correspondence between the fluctuations of the superstrata and specific excitations of the dual D1-D5 CFT.

So far, such fluctuating solutions have only been constructed on single bubble solutions.  It is an open problem as to how to construct such fluctuating solutions on multi-bubbled geometries and whether these fluctuations are constrained through matching conditions at intersection points of the bubbles.  There is, however, a very interesting proposal  \cite{Tyukov:2018ypq} for how one might generalize superstrata to multi-centered geometries.  Since the holography of multi-bubbled solutions is also an open problem, the entire holographic interpretation of fluctuating multi-bubbled solutions is completely new territory.

%%%%%%%%%%
\subsubsection{Other excitations}
%%%%%%%%%%

In string theory, once one has non-trivial topological cycles, it is very interesting to ask about the new fields and excitations that come from wrapping branes around the cycles. This was, for example, how Hull and Townsend \cite{Hull:1994ys} showed how to get the W-bosons of the $E_8 \times E_8$ string from the type II string compactified on $K3$.  

One can do the same kind of thing in microstate geometries and some of the corresponding supergravity solutions were investigated in \cite{Levi:2009az,Raeymaekers:2014bqa,Raeymaekers:2015sba,Tyukov:2016cbz}.

It is even more interesting to examine the classes of states and field theories that emerge from such brane wrapping of cycles in microstate geometries \cite{Martinec:2015pfa}. In particular, one finds vast numbers of massive states that are expected to become massless as the microstate geometry gets deeper and deeper.  It was shown in  \cite{Martinec:2015pfa} that the numbers of such states grows exponentially with the black-hole charges and, when they become massless, they will yield a leading contribution to the entropy.  From the perspective of holographic field theory, these states seem to be avatars of the Higgs branch of the dual CFT.   

The wrapped-brane states are extremely rich and complex because they also involve the physics of the compactified dimensions in very non-trivial ways.  The simplest such wrapped branes are point-like in the extra dimensions, but they feel the magnetic fluxes threading the extra dimensions and so are expected to settle into Landau orbits in the magnetic fields.  It is this structure that leads to their vast degeneracy.

%%%%%%%%%%%%%%%%%%%
\subsection{Underlying mathematical structure}
\label{ss:math}
%%%%%%%%%%%%%%%%%%%

Ambi-polar, hyper-K\"ahler metrics are a whole new world that mathematicians are just beginning to explore \cite{Hitchin1, Biquard1,Niehoff:2016gbi,Tyukov:2018ypq}.   The crucial observation that makes it all mathematically accessible is that the five-dimensional metric, (\ref{metform}), is smooth and Lorentzian, or, better, the four-dimensional metric,(\ref{inducedmet}), is smooth and Riemannian.

Perturbative computations strongly suggest that there should be new, rich families of ambi-polar  hyper-K\"ahler metrics that have yet to be discovered. Riemannian K\"ahler manifolds also have an extremely beautiful mathematical structure that relates moduli of the metric to harmonic analysis of $2$-forms.  The general theory of this for ambi-polar metrics has yet to be developed but the calculations  described in Section \ref{sss:flucts} show that there are infinite families of harmonic forms and so one expects an infinite dimensional families of metric moduli.  

The foregoing expectations of infinite families of harmonic forms and metric moduli fly in the face of ``Riemannian'' experience, but the evanescent ergosurfaces and their ``acceptable'' singular structure opens up a vast new set of possibilities and understanding what is ``acceptable,'' and what is not, remains to be understood.  In practice, the ultimate arbiter of acceptability is the regularity of the full Maxwell field (\ref{Aform}) and full metric (\ref{metform}).  

Finally, I cannot resist noting that the original Riemannian ALE spaces have an intersection form isomorphic to the Cartan matrix of the Lie algebra of $SU(n)$ and  the Weyl group of $SU(n)$ emerges as the monodromy group acting on the cycles as one moves around  moduli  space of the metric.  It would be extremely interesting to understand  the ambi-polar generalization of this story.  It was suggested in \cite{Tyukov:2018ypq} that this may involve an extension to super Lie algebras.

%%%%%%%%%%%%%%%%%%%
\subsection{Probing microstate geometries}
\label{ss:Probing}
%%%%%%%%%%%%%%%%%%%

There are many ways to try to probe microstate geometries, and even the simplest geodesic probes produce some extremely interesting results.  

Gravitational tidal forces are dictated by the Riemann tensor, and in ordinary black holes, one has, simply on dimensional grounds 
\begin{equation}
R^{\mu \nu \rho \sigma} \, R_{\mu \nu \rho \sigma}  ~\sim~ \frac{m^2}{r^6} \,,
  \label{Riemsq1}
\end{equation}
At the horizon one has $r \sim m$ and hence
\begin{equation}
R^{\mu \nu \rho \sigma} \, R_{\mu \nu \rho \sigma}  ~\sim~ \frac{1}{m^4} \,,
  \label{Riemsq2}
\end{equation}
This means that the tidal stresses at the horizon scale decrease as the mass of the black hole gets larger.  

A more careful analysis of this class of problem involves geodesic deviation.  That is, one considers a family of geodesics with proper velocities denoted by $V^\mu = \frac{dx^\mu}{d \tau}$.  One defines the ``deviation vector,'' $S^\rho$, to be a space-like displacement across the family of geodesics.  Indeed,  by appropriately synchronizing the proper time  between neighboring geodesics one can arrange $S^\rho V_\rho = 0$. This means that  $S^\rho$ is a space-like vector in the rest-frame of the geodesic observer.  One can also re-scale $S^\mu$ at any one point so that $S^\mu S_\mu =1 $ and it therfore  represents a unit displacement across the family.  
The tidal forces are then measured by the relative acceleration, $A^\mu ~\equiv~ \frac{D^2 S^\mu}{d \tau^2}$, of neighboring geodesics.  

A straightforward calculation leads to the equation of geodesic deviation:
\begin{equation}
A^\mu ~\equiv~ \frac{D^2 S^\mu}{d \tau^2}  ~=~ - {R^\mu}_{\nu \rho \sigma} \, V^\nu S^\rho   V^\sigma \,.
  \label{Geodev1}
\end{equation}
The skew-symmetry of the Riemann tensor means that $ A^\mu V_\mu  =0$ and so the tidal acceleration is similarly space-like, representing the tidal stress in the rest-frame of the infalling observer with velocity, $V^\mu$.   To find the largest stress one can maximize  the norm, $\sqrt{A^\mu A_\mu}$, of $A^\mu$  over all the choices of $S^\mu$, subject to the constraint $S^\mu S_\mu =1$.  

 To analyze the stress forces it is convenient to  introduce what is sometimes called the ``tidal tensor:'' 
\begin{equation}
{\cA^\mu}_\rho ~\equiv~ - {R^\mu}_{\nu \rho \sigma} \, V^\nu \, V^\sigma \,,
  \label{cAdefn}
\end{equation}
and consider its norm.  Indeed,  we define
\begin{equation}
|\cA| ~\equiv~  \sqrt{{\cA^\mu}_\rho\, {\cA^\rho}_\mu}  \,.
  \label{cAnorm}
\end{equation}
Note that since $V^\mu = \frac{dx^\mu}{d \tau}$ is dimensionless, $\cA$ has the same dimensions as the curvature tensor, $(\rm{length})^{-2}$.

The important thing about black holes is that they typically have only one scale: the mass, $m$, and so one finds the natural variant of  (\ref{Riemsq2})
\begin{equation}
|\cA| ~\sim~ \frac{1}{m^2}  \,,
 \label{Geodev2}
\end{equation}
which means that the tidal forces get very small as the mass grows larger.

The important thing about microstate geometries is that necessarily have multiple scales\footnote{This is also hugely important in the dual holographic field theory}.  In particular, there is the scale at the top of the black-hole throat, $r \sim b$, which is usually set by the mass, $m$, and there is the scale at the ``bottom'' of the  throat, $r \sim a$, at which one encounters the cap.   The scale, $a$, is determined by moduli, but, as noted above, these moduli are quantized and $a$ cannot be made arbitrarily small.  In this description, taking $a \to 0$ corresponds to the black-hole limit in which $r=0$ corresponds to the horizon.   

The ratio, $\frac{b}{a}$, determines the red-shift between the cap and the top of the throat, and for a ``typical'' microstate geometry it is extremely large.  

Given that there are now (at least) two scales, there are more possibilities for the right-hand side of (\ref{Geodev2}).  In particular, explicit calculations \cite{Tyukov:2017uig,Bena:2018mpb} show that there are generically terms of the form
\begin{equation}
|\cA| ~\sim~  \frac{a^2 \, b^2}{r^6}  \,,
 \label{Geodev3}
\end{equation}
This may be thought of as a higher multipole moment of the metric and it is induced by the presence of a cap. Note that it vanishes in the black-hole limit ($a \to 0$). 

As one gets near the cap, $r \to a$, this tidal tensor becomes extremely large, as one might expect: 
\begin{equation}
|\cA| ~\sim~  \frac{b^2}{a^4}  \,,
 \label{Geodev4}
\end{equation}
In this context, ``large'' means large compared to a mixture of the compactification scale and the string scale, and so other stringy or Kaluza-Klein modes will become important.  If the tidal forces become ``large,'' the probe will either discover that space-time is compactified or that it is made of strings.

However, there is an even more important transition: If the probe is dropped from the top of the throat then it will encounter the deviations from the black-hole metric while traveling at ultra-relativistic speeds.  Indeed, the  tidal tensor,  (\ref{cAdefn}), involves factors of the velocity, and for a particle released from rest at $r \sim b$, this adds a further factor of $b$ to the tidal force \cite{Tyukov:2017uig,Bena:2018mpb}.  As a result, one finds 
\begin{equation}
|\cA| ~\gtrsim~ 1\qquad \Leftrightarrow \qquad r ~\sim~ \sqrt{ab} \,,
 \label{Geodev5}
\end{equation}
In other words, the ultra-relativistic speeds create the transition to the stringy, or Kaluza-Klein phase long before the cap: in logarithmic terms, about ``half-way down'' the throat. 

This is expected to greatly influence our understanding of how matter ``scrambles'' into microstate structure.

%%%%%%%%%%%%%%%%%%%
\subsection{non-BPS microstate geometries}
\label{ss:nonBPS}
%%%%%%%%%%%%%%%%%%%

It would be remiss to finish these lectures without saying something about non-BPS microstate geometries.   Finding explicit non-BPS solutions has  become a very large enterprise.

The starting point of this enterprise was the JMaRT solution \cite{Jejjala:2005yu}.  More systematic methods were subsequently developed and they resulted in solutions in which the supersymmetry was broken in a controlled manner via gravitational holonomy \cite{Bena:2009qv,Bobev:2009kn,Bena:2013gma}.  In the last few years, an even more general systematic approach  has been developed   \cite{Bena:2015drs,Bena:2016dbw,Bossard:2017vii}  and this has produced a rich family of examples.  The basic idea is that with enough symmetry one can reduce a supergravity solution to an effectively two-dimensional problem. The dynamics can then be expressed in terms of a scalar sigma model. This can then be solved by an array of methods ranging from inverse-scattering methods through to exploiting special properties of nilpotent sub-algebras.  Ideally, one would like to develop inverse-scattering methods to find generic non-extremal microstate geometries.  While there has been some progress in this direction \cite{Katsimpouri:2013wka,Rocha:2014uqa,Roy:2018ptt}, the complexity of the underlying supergravity theory makes this approach extremely challenging.

The problem with the non-BPS and non-extremal microstate geometries constructed so far is that they seem to be very atypical outliers in the space of generic black hole microstates.  Correspondingly, they are dual to extremely exotic coherent states in the black-hole CFT.  I suspect that this limitation is caused by the limitations of analytic computation rather than the physics of microstate geometries.  It is possible that analytic methods may improve, but it also seems that numerical methods have progressed sufficiently far (see, for example, \cite{Dias:2015nua,Dias:2016eto}) that it will soon be feasible to start numerical searches for more generic non-extremal microstate geometries.

Finally, a promising and, as yet, relatively unexplored approach to non-extremal microstate geometries can be made using perturbation theory to construct near-BPS solutions.  Just as there are BPS fluctuations of bubbles in microstate geometries, there are supersymmetry breaking fluctuations of microstate geometries.  One also expects that the large scale, BPS bubbled geometries to be stable against such small fluctuations and so one might hope to find explicit solutions.  The challenge will be to handle gravitational and electromagnetic radiation that will be generated by such fluctuations, but with luck, the radiation field will be a second-order correction to such non-BPS solutions.  Such perturbations would provide a direct way to create solutions with small Hawking temperature and the radiation they generate will presumably be the microstate analog of  Hawking radiation.

%%%%%%%%%%%%%%%%%%%%%%%%%%%%%%%%%%%%%
\section{A last comment}
\label{Sect:final-comment}
%%%%%%%%%%%%%%%%%%%%%%%%%%%%%%%%%%%%% 

In Section \ref{Sect:NSwoT}, I established the five-dimensional version of the ``No Solitons without Topology'' theorem.    At least to this extent, microstate geometries represent the only viable gravitational mechanism that produce smooth, horizonless geometries that look just like black holes up until one is arbitrarily close to the horizon scale.  As a result, any classical, smooth, horizonless geometry must be some form of microstate geometry.  Moreover, if you try to replace a black hole with a strongly quantum pile of mush, then this pile of mush, should, near the horizon scale, have some form of semi-classical limit in order for it to be consistent with the General Relativistic  calculations that underpin the results of LIGO.  It is thus not much more of a stretch for the quantum pile of mush to have a semi-classical limit that takes one very close to the horizon scale.  If there is such a semi-classical limit, it must be described by a microstate geometry.  Similarly, if you want to study a quantum system near the horizon scale, then microstate geometries provide the only possible classical mechanism to support the quantum system you want to study.

It is also important to recall one of the lessons coming from 20 years of holographic field theory.  The gravity sector of the duality is extremely good at capturing the large-scale collective effects of the dual strongly-coupled  quantum system.  Supergravity is a fairly crude instrument when it comes to details of the microstructure but it excels in capturing the large-scale bulk expression of the quantum system.  It therefore seems that microstate geometries must emerge as one aspect of the universal, large-scale, strong-coupling expression of the  quantum systems that underpin the black-hole microstate structure.  In particular, this gives one hope that microstate geometries can capture the universal physics that will lead to a large-scale, hydrodynamic description of the quantum microstates of black hole much as holography provided hydrodynamic insight into the properties of quark-gluon plasmas. For example, one might reasonably hope to use microstate geometries to compute the effective viscosity created by horizon-scale microstructure. 

In short, whether or not you accept the entire microstate geometry/fuzzball paradigm, microstate geometries still provide one of the best ways to probe collective large-scale effects of black-hole microstructure and to study quantum systems close to the horizon scale.

%%%%%%%%%%%%%%%%%%%%%%%%%%%%%%%%%%%%%
\section*{Acknowledgments}
%%%%%%%%%%%%%%%%%%%%%%%%%%%%%%%%%%%%% 
\vspace{-2mm}
I am indebted to all my collaborators (especially Iosif Bena) and colleagues who were instrumental in developing many of the ideas described in these lectures.  I would like to thank Riccardo Guida and Marco Schiro for encouraging me to give these lectures and for handling the logistics. I would also like to thank Dominique Hirondel for his careful reading of these notes and for sending me very helpful corrections.  I am very grateful to the IPhT, CEA-Saclay for support and hospitality as well as for providing a forum for these lectures.   This work was supported in part by ERC Grant number: 787320 - QBH Structure  and by the DOE grant DE-SC0011687.

%%%%%%%%%%%%%%%%%%%%%%%%%%%%%%%%%%%%%
 \newpage
%%%%%%%%%%%%%%%%%%%%%%%%%%%%%%%%%%%%%

\begin{adjustwidth}{-1mm}{-1mm} % to adjust the L and R margins 
\bibliographystyle{utphys}      
\bibliography{microstates}       % calls file "microstates.bib"

\end{adjustwidth}

%%%%%%%%%%%%%%%%%%%

\end{document}